\documentclass[sigconf, nonacm]{acmart}

\usepackage{pvldb}

\renewcommand\vldbdoi{XX.XX/XXX.XX}
\renewcommand\vldbpages{XXX-XXX}
\renewcommand\vldbavailabilityurl{}

\newcommand{\out}[1]{}
\newcommand{\keys}[1]{{\mdseries #1}}
\newcommand{\hi}[1]{\textbf{#1}}             
\newcommand{\lo}[1]{\textcolor{gray}{#1}}    
\newcommand{\ex}[1]{\textit{#1}}             

\newcommand{\fid}{\mathrm{F}}      
\newcommand{\fidhat}{\hat{F}}        

\usepackage{alltt}
\usepackage{amsmath}
\usepackage{boxedminipage}
\usepackage{graphicx}
\usepackage{url}

\begin{document}

\title{Pre-Execution Plan Selection for Approximate Vector Search}

\author{Adithya Ramanatha}
\affiliation{
\institution{NIT Calicut}
\city{Calicut}
\country{India}
}
\email{adithyar2115@gmail.com}

\author{Sanskar S. Singh}
\affiliation{
\institution{IIT Delhi}
\city{New Delhi}
\country{India}
}
\email{ssanskar2705@gmail.com}

\author{Maya Ramanath}
\affiliation{
\institution{IIT Delhi}
\city{New Delhi}
\country{India}
}
\email{ramanath@cse.iitd.ac.in}

\begin{abstract}
Vector search is now a database operator with many physical implementations. Index families, parameter settings, and compression levels each trade cost against \emph{fidelity}, the fraction of a query's exact top-$k$ neighbors that a configuration returns. Deployments nevertheless choose one configuration offline and run every query on it. They are sized for their hardest queries and overpay for the rest. On 5.8M LAION image--caption pairs with four embedding models, running each query on the cheapest configuration that suffices for it would roughly halve the cost in a typical setting. Capturing this saving requires predicting, before any search runs, how each configuration will do on a query. This is hard, because fidelity is defined by the exact search that approximate indexes avoid. We show that a query's position in the embedding space predicts it, and that the prediction must be kept per configuration. We study two estimators trained once on a labeled sample. A small neural network maps the query vector to a probability per configuration. A \emph{fidelity synopsis} records, like an optimizer histogram, how often each configuration met the target in each region of the space. A budget rule turns either estimate into a per-query choice, tuned so that the workload misses the target no more often than a static deployment is allowed to. Against the best static choice among the 11 configurations we built, selection costs a median of 0.59$\times$ with the network and 0.70$\times$ with the synopsis. An advisor in the style of relational index advisors chooses which configurations to build. Three configurations keep the saving, at a fraction of the memory and labeling cost of the full catalog. Costs are estimated from per-configuration latencies measured with full batches.
\end{abstract}

\maketitle

\section{Introduction}
\label{sec:intro}

Dense embeddings have made nearest-neighbor search a core operation of text retrieval, cross-modal search, and retrieval-augmented generation~\cite{dpr,clip,siglip,qwen3-vl,rag}. Vector databases and relational extensions expose it as a query operator~\cite{milvus,analyticdbv,pgvector,vectordb}. Like any operator, it has many physical implementations. A collection can be indexed with a proximity graph such as HNSW~\cite{hnsw} or with an inverted file and product quantization (IVFPQ)~\cite{pq}. Each index has build-time and search-time parameters. We call an index family with fixed parameter values a \emph{configuration}. Configurations trade cost against \emph{fidelity}, the fraction of a query's exact top-$k$ nearest neighbors that the configuration returns.

\begin{figure}[t]
\centering
\includegraphics[width=0.92\linewidth]{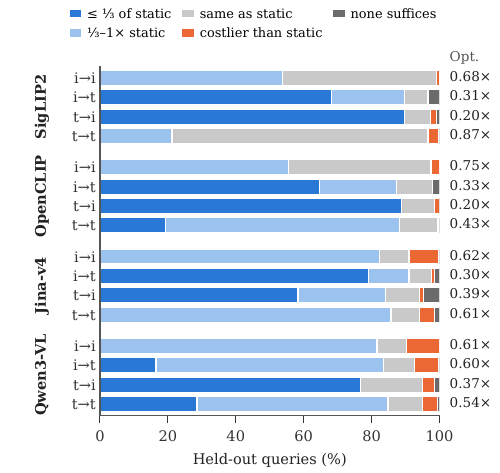}
\caption{The opportunity at a fidelity target of 0.7: held-out queries of each model and query direction, split by the cost of their cheapest sufficient configuration relative to the static plan. \keys{(i)~Dark blue: at most a third of the static plan's cost. (ii)~Light blue: between a third and the whole of it. (iii)~Grey: the static plan's own configuration. (iv)~Orange: a costlier configuration, for queries the static plan misses. (v)~Dark grey: no configuration suffices. (vi)~Right: the mean cost if every query ran on its cheapest sufficient configuration where that costs no more than the static plan, and on the static plan otherwise.}}
\label{fig:opportunity}
\end{figure}

Today, a configuration is chosen once, when the collection is indexed, and every query runs on it. In relational terms, the operator has several access paths but no optimizer that chooses among them. A fixed choice is costly because queries differ in how hard they are to approximate. A static deployment must ensure that most queries, say 90\%, meet a fidelity target. It is therefore sized for the hardest queries within that 90\%. Figure~\ref{fig:opportunity} shows the consequence on 5.8M LAION image--caption pairs~\cite{laion}, indexed under four embedding models and nine configurations. In many settings, most queries would meet the target on a configuration that costs at most a third of the static choice. Running every query on the cheapest configuration that suffices for it would cost about half as much in a typical setting.

Capturing this saving requires knowing, before any search runs, which configurations will meet the target on a given query. This is hard for a basic reason. Fidelity is defined relative to the exact top-$k$. The exact top-$k$ is precisely what approximate indexes exist to avoid computing.

We treat this as plan selection, the task a relational optimizer performs when it chooses among access paths~\cite{selinger}. Offline, a sample of queries runs once through exact search and through every configuration (Figure~\ref{fig:overview}). These labels train an \emph{estimator}, which predicts for a new query and every configuration the probability that the configuration meets the target. At query time, a \emph{budget rule} runs each query on a cheap configuration where that probability is high, and on a reliable one where it is low. It is tuned so that the share of queries that miss the target stays within the limit a static deployment must also meet. We study two estimators. A small multi-layer perceptron (MLP) maps the query vector to one probability per configuration. A \emph{fidelity synopsis} (Figure~\ref{fig:overview}) divides the embedding space into cells around \emph{landmarks} (representative points obtained by clustering the corpus). Like an optimizer histogram~\cite{histograms}, it records for every cell how often each configuration met the target. The MLP is more accurate and cheaper. The synopsis needs no training, is calibrated by construction, and can be inspected cell by cell. Both rest on two premises, which we test rather than assume. The first is \emph{smoothness}. Fidelity must vary slowly enough across the space that nearby queries fare alike. The second is \emph{configuration specificity}. The same region can be easy for one index family and hard for another, so estimates must be kept per configuration. Finally, selection can only choose among configurations that exist. We adapt relational index advisors to choose which ones to build.

Prior work leaves this per-query decision open. Index tuners and benchmarks~\cite{vdtuner,annbench,graphann-survey} find the best single configuration for a workload. They produce a better static plan, which plan selection can then use as one of its options. Early-termination and adaptive-parameter methods such as LAET~\cite{laet}, DARTH~\cite{darth}, Ada-ef~\cite{adaef}, and Quake~\cite{quake} decide how long a search runs inside an index that is already chosen. They cannot help when that index cannot reach the target, which our measurements show is common for cross-modal queries~\cite{roargraph,modality-gap}. Even within a single index, they tune each search while it runs, toward a target for that query. Plan selection decides before any search starts and holds the whole workload to a miss rate. It can therefore send a query to a cheap setting wherever a miss is unlikely, and spend the workload's allowance where it is safe. Hardness measures such as local intrinsic dimensionality~\cite{lid} and Steiner-hardness~\cite{steiner} explain why some queries are hard. They do not estimate whether a given configuration will meet a given target. Histograms and learned cardinality estimators~\cite{histograms,kipf-mscn} estimate selectivity, which depends on the data and the predicate but not on the physical plan. Fidelity depends on the plan as well, so its estimates must be kept per configuration. Index advisors~\cite{autoadmin} choose which indexes to build, but for plans that all return exact answers. Systems for approximate and semantic query processing trade accuracy for cost under explicit targets~\cite{blinkdb,thalamusdb,lotus,palimpzest,frugalgpt}. They choose models and plans, but they run vector search in one fixed way. MINT~\cite{mint} plans multi-vector queries across indexes under a recall threshold, but models recall from a data sample rather than from each query's surroundings. To our knowledge, no prior work estimates, before execution, how each candidate configuration will perform on a single query from where that query lies, and chooses among index families under a requirement on the whole workload.

\paragraph{Contributions.}
\begin{itemize}
  \item We quantify the opportunity. At the same level of service, the per-query optimum that serves the static plan's queries (the \emph{matched optimum}) costs a median of about half of the static plan (Section~\ref{sec:rq1}).
  \item We cast per-query choice among ANN configurations as plan selection under a requirement on the whole workload, with a budget rule tuned on a sample. We test the two premises that make it possible (Sections~\ref{sec:selection} and~\ref{sec:rq2}).
  \item We compare two estimators. An MLP over the query vector is the more accurate and cheaper one. A fidelity synopsis is a transparent, training-free alternative that never exceeded the requirement in the main comparison (Sections~\ref{sec:selection}, \ref{sec:rq2}, and~\ref{sec:rq3}).
  \item Against the best static choice among the 11 configurations we built, selection costs a median of 0.59$\times$ with the MLP and 0.70$\times$ with the synopsis (Section~\ref{sec:rq3}).
  \item We adapt index advisors to plan selection. Three configurations keep the saving, at a median of 56\,GB per collection and 30 CPU-hours of labeling per setting, and a 10\% sample usually finds them (Section~\ref{sec:rq4}).
\end{itemize}
This report is the extended version of our conference paper. It gives per-setting results for every experiment, two simpler selection rules, smaller catalogs, further targets, sensitivity to the synopsis's parameters, and statistical tests. Its sections match those of the paper.

Section~\ref{sec:prelim} defines fidelity and the plan-selection problem. Section~\ref{sec:selection} presents the two estimators, the selection rule, and the catalog advisor. Section~\ref{sec:setup} describes the evaluation setup, and Section~\ref{sec:experiments} answers five research questions. Section~\ref{sec:related} discusses related work, and Section~\ref{sec:conclusion} concludes.

\section{Problem Setting}
\label{sec:prelim}

This section fixes the setting and notation, defines fidelity, and states the plan-selection problem. Table~\ref{tab:notation} summarizes the notation.

\subsection{Corpus, Models, and Queries}
\label{sec:prelim:setting}

A corpus $C$ is a collection of items, each an image, a text, or both. An embedding model $M$ maps items and queries of either modality into a shared vector space, in which similarity is measured by a fixed function. In our experiments, embeddings are L2-normalized and compared by Euclidean distance, which ranks items exactly as cosine similarity does. Each model induces its own embedding space and hence its own nearest neighbors, so everything below is defined for a single model, which we leave implicit.

A query $q$ is an image or a text, embedded by the same model as the corpus. Searching a corpus of one modality with a query of one modality gives four \emph{query directions}: image-to-image (i$\to$i), text-to-text (t$\to$t), image-to-text (i$\to$t), and text-to-image (t$\to$i). The first two are \emph{uni-modal}, the last two \emph{cross-modal}. The distinction matters because an approximate index is built over the distribution of the corpus vectors. Uni-modal queries come from roughly that distribution. Cross-modal queries often do not, because contrastively trained models place the two modalities in partly separated regions of the embedding space~\cite{modality-gap}, so a query of one modality may lie far from the items of the other~\cite{roargraph}. The extent of this separation varies across models. We call a pair of model and query direction a \emph{setting}.

\subsection{Index Families and Configurations}
\label{sec:prelim:families}

Approximate nearest-neighbor (ANN) indexes avoid comparing a query with every item, at the price of sometimes missing true neighbors. We consider three widely deployed index families, which approximate search in different ways and therefore lose accuracy for different reasons.

\paragraph{HNSW} HNSW~\cite{hnsw} builds a multi-layer proximity graph over the corpus vectors and answers a query by greedy best-first traversal, keeping a candidate list of size \texttt{efSearch}. Its build-time parameters are the graph degree $M$ and \texttt{ef\-Con\-struc\-tion}. HNSW keeps full-precision vectors, so its losses come from navigation: the traversal can stop in a region that does not contain all true neighbors, particularly for queries far from the data the graph was built on~\cite{roargraph,steiner}.

\paragraph{IVFPQ} IVFPQ~\cite{pq} partitions the corpus around $n_{\mathrm{list}}$ centroids and compresses each vector with product quantization into a short code. A query probes the \texttt{nprobe} partitions whose centroids are closest to it and ranks their contents by approximate distance. It loses accuracy in two ways: true neighbors may lie in partitions that are not probed, and quantization may misorder those that are.

\paragraph{O-IVFPQ} O-IVFPQ applies a learned rotation (OPQ~\cite{opq}) before product quantization, which reduces quantization error for a given code length. Otherwise it behaves like IVFPQ.

\paragraph{Configurations, catalog, and cost.} A \emph{configuration} $A$ is an index family together with values for its build-time and search-time parameters. A \emph{catalog} $\mathcal{A} = \{A_1, \dots, A_m\}$ is the set of configurations available for a corpus and model. Configurations that differ only in search-time parameters share one built index, and we call them the \emph{search settings} of that index. All others need separate indexes. Each configuration has a per-query cost $c(A)$, such as its mean search latency on a given platform. Cost depends on the model, for instance through the embedding dimension, so the cost order of a catalog can differ between models. Memory is a per-index cost and is reported separately.

\subsection{Fidelity}
\label{sec:prelim:fidelity}

Let $\mathrm{Top}\text{-}k_{\mathrm{KNN}}(q)$ be the $k$ items of $C$ nearest to $q$, computed by exact search, and $\mathrm{Top}\text{-}k_{A}(q)$ the $k$ items returned by configuration $A$. The \emph{fidelity} of $A$ on $q$ is
\[
  \fid(q, A) \;=\; \frac{\big|\,\mathrm{Top}\text{-}k_{\mathrm{KNN}}(q) \,\cap\, \mathrm{Top}\text{-}k_{A}(q)\,\big|}{k} \;\in\; [0,1].
\]
Fidelity is what the ANN literature calls recall@$k$ against exact search. We use a separate name to distinguish it from recall against human relevance judgments. It isolates the loss caused by approximation: the model determines which items are nearest, and fidelity measures how many of them a configuration recovers. Computing it requires the exact top-$k$, which is affordable offline for a sample of queries but is exactly what approximate indexes avoid at query time.

A configuration is \emph{dominated} if another configuration is no more expensive and has at least its mean fidelity, with one of the two strictly better, for the same model and query direction. A dominated configuration is never the better choice for the workload as a whole, but it can still be the cheapest configuration that meets a target on individual queries.

\subsection{Problem: Per-Query Plan Selection}
\label{sec:prelim:problem}

An operator fixes a \emph{fidelity target} $\tau \in (0,1]$ and a \emph{reliability level} $\rho \in (0,1]$: a query should receive at least a fraction $\tau$ of its exact top-$k$, and at most a fraction $1 - \rho$ of the workload may fall short. We call the latter the \emph{requirement}. For a query $q$, the \emph{sufficient} configurations are those that meet the target,
\[
  S_\tau(q) \;=\; \{\, A \in \mathcal{A} \;:\; \fid(q, A) \ge \tau \,\}.
\]

\paragraph{Static plan.} A static deployment runs every query of a workload $W$ on one configuration. The \emph{static plan} is the cheapest configuration that meets the requirement on $W$,
\[
  A_{\mathrm{static}} \;=\; \arg\min_{A \in \mathcal{A}} \; c(A)
  \quad \text{s.t.} \quad \frac{|\{q \in W : A \in S_\tau(q)\}|}{|W|} \;\ge\; \rho ,
\]
or, if no configuration meets it, the one that meets the target on the largest fraction of $W$. Because a catalog offers only a few configurations, the static plan usually exceeds the requirement: it meets the target on more than a fraction $\rho$ of queries, because the next cheaper configuration meets it on fewer.

\paragraph{Per-query optimum.} If every query's fidelity under every configuration were known in advance, each query could run on its cheapest sufficient configuration,
\[
  A^\ast(q) \;=\; \arg\min_{A \in S_\tau(q)} c(A).
\]
This \emph{per-query optimum} is not deployable, since it requires every query's exact top-$k$, but its cost bounds what any per-query choice can save. Section~\ref{sec:setup:params} defines the variants we compare against.

\paragraph{An example.} Consider a catalog of two configurations. $A_1$ costs 1 and is sufficient for 85\% of queries. $A_2$ costs 4 and is sufficient for 98\%. With $\rho = 0.9$, the static plan must be $A_2$, because $A_1$ falls short of 90\%. Every query then costs 4, and only 2\% miss. Now suppose a selector could recognize the queries for which $A_1$ suffices. Running those on $A_1$ and the rest on $A_2$ costs $0.85 \cdot 1 + 0.15 \cdot 4 = 1.45$, just over a third of the static plan. The selector may even run a few more queries on $A_1$ and miss them, up to the 10\% that the requirement allows. A real selector cannot recognize these queries perfectly, so it gives up part of this saving. How much it keeps depends on how well it predicts which configurations suffice for each query.

\paragraph{Selection.} A deployable \emph{selector} chooses a configuration $\hat{A}(q)$ for each query from estimates that are available before any configuration runs. An \emph{estimator} provides them. For a query $q$ and every configuration $A_j$, it gives $\hat{\pi}_j(q)$, the estimated probability that $A_j$ is sufficient for $q$. Obtaining the estimates costs $c_{\mathrm{est}}$ per query. A \emph{miss} occurs when $\hat{A}(q) \notin S_\tau(q)$, that is, when the chosen configuration falls short of the target. The goal is the selector with the lowest mean cost $c_{\mathrm{est}} + c(\hat{A}(q))$ whose miss rate meets the requirement, at most $1 - \rho$: the same requirement the static plan is built to meet. As a more conservative comparison, we also hold the selector to the miss rate the static plan actually incurs, which is usually lower. The two goals trade off: a selector that always picks the cheapest configuration minimizes cost but misses often, while the static plan meets the requirement at the cost of serving every query as if it were among the hardest.

\begin{table}[t]
\footnotesize\centering
\begin{tabular}{@{}ll@{}}
\toprule
Symbol & Meaning \\
\midrule
$C$, $M$ & corpus; embedding model \\
$\mathcal{A} = \{A_1,\dots,A_m\}$ & catalog of configurations \\
$c(A)$ & per-query cost of configuration $A$ \\
$k$ & result depth \\
$\fid(q,A)$ & fidelity of $A$ on $q$ \\
$\tau$, $\rho$ & fidelity target; reliability level \\
$S_\tau(q)$ & configurations sufficient for $q$ \\
$W$ & workload \\
$A_{\mathrm{static}}$, $A^\ast(q)$ & static plan; per-query optimum \\
$\hat{A}(q)$ & selected configuration \\
$c_{\mathrm{est}}$ & per-query cost of selecting it \\
$L$, $K$ & landmarks; their number \\
$\ell(q)$ & $q$'s nearest landmark (its cell) \\
$Q$, $Q_{\mathrm{test}}$ & construction sample; held-out workload \\
$\mu_{ij}$, $\pi_{ij}(\tau)$ & mean fidelity and coverage of $A_j$ in cell $i$ \\
$\fidhat$, $\hat{\pi}_j(q)$ & estimated fidelity; estimated probability that $A_j$ suffices for $q$ \\
$\mathcal{P}$, $M$ & pool of candidate configurations; memory budget (advisor) \\
$\lambda$ & price of an expected miss (budget rule) \\
\bottomrule
\end{tabular}
\caption{Notation.}
\label{tab:notation}
\end{table}

\section{Plan Selection}
\label{sec:selection}
\label{sec:map}

\begin{figure*}[t]
    \centering
    \includegraphics[width=\textwidth]{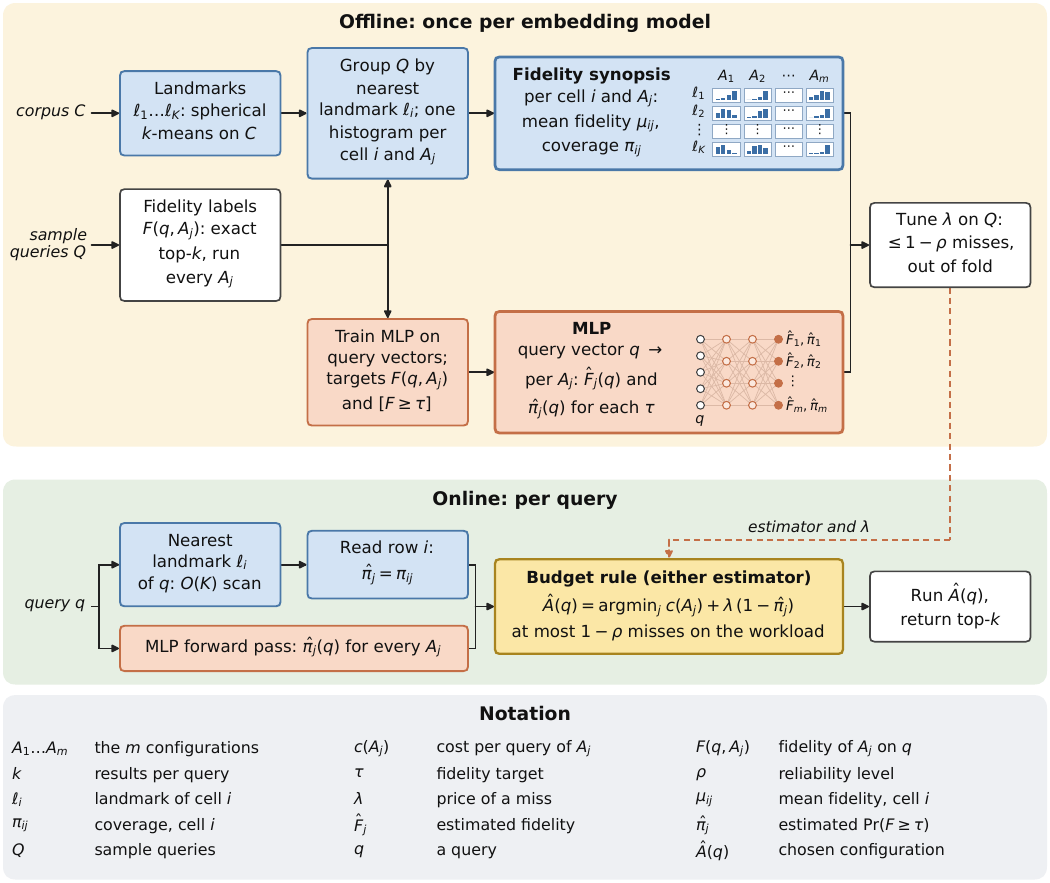}
    \caption{Plan selection with either estimator. Offline (top), a sample of queries is labeled with the fidelity of every configuration. The labels build a fidelity synopsis (blue: landmarks, then one fidelity histogram per cell and configuration) or an MLP over the query vector (orange: a predicted fidelity and a probability of meeting each target, per configuration), and $\lambda$ is tuned on their out-of-fold estimates. Online (bottom), the estimator gives each configuration's probability of meeting the target, and the budget rule chooses the configuration to run.}
    \label{fig:overview}
\end{figure*}

Plan selection has two parts, an estimator trained offline on a labeled sample and a selection rule applied online (Figure~\ref{fig:overview}). We first describe the fidelity synopsis, which motivates the approach (Section~\ref{sec:synopsis:offline}). The selection rule works with any estimator (Section~\ref{sec:synopsis:online}). We then describe a learned estimator (Section~\ref{sec:mlp}) and how to choose the catalog (Section~\ref{sec:advisor}).

\label{sec:map:overview}

The fidelity of a configuration on a query is determined by the geometry of the corpus around the query: how densely its true neighbors are packed, how far they lie from it, whether they fall within one IVF partition or straddle several, and how well connected the part of the HNSW graph that contains them is. Two queries that land close to each other in the embedding space largely share these surroundings, and often many of their true neighbors, so we expect them to fare similarly under a given configuration.

This suggests estimating fidelity by location. We divide the embedding space into regions and record, for each region and each configuration, how that configuration performed on a sample of queries that fell into the region. A new query is located and the record of its region is read off. This is the idea of a histogram in a relational optimizer, which records statistics per bucket of a column's domain so that the selectivity of a new predicate can be read off without scanning the table. Here the buckets are regions of the embedding space, and the statistic is fidelity.

Two properties make this more than a density estimate. First, the statistics are measured, not modeled: they capture whatever combination of density, partitioning, and graph structure determines fidelity in a region, without our having to know which. Second, they are kept separately for every configuration, because the same region can be easy for one index family and hard for another: a dense region whose neighbors straddle partition boundaries can be hard for IVFPQ and easy for HNSW, and a sparse region far from most data can be the reverse.

The approach rests on two premises, which Section~\ref{sec:rq2} tests directly:
\begin{itemize}
    \item \emph{Smoothness}: fidelity varies slowly enough across the space that the sample queries in a region represent the queries that will fall into it later. This is least plausible near the boundary between regions and for queries far from all data, such as cross-modal queries under models that separate the modalities strongly.
    \item \emph{Configuration specificity}: the same region can have very different fidelity under different configurations, so statistics must be kept per pair of region and configuration.
\end{itemize}

The synopsis is built once per model in an offline stage and consulted once per query in an online stage (Figure~\ref{fig:overview}). Only the offline stage uses exact search, and only on a sample of queries.

\begin{figure}[t]
    \centering
    \includegraphics[width=\linewidth]{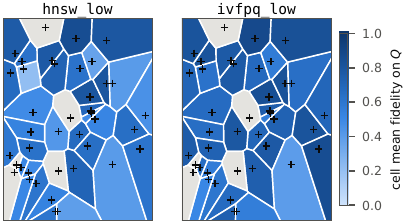}
    \caption{The offline stage on real data (Jina-v4, text-to-image, $K = 10{,}000$): 40 neighboring image landmarks (crosses), projected to two dimensions, with each cell shaded by the mean fidelity of \texttt{hnsw\_low} (left) and \texttt{ivfpq\_low} (right) on the construction sample. Grey: sparse cells. The cells are the same in both panels, but the fidelity in them is not.}
    \label{fig:offline-geometry}
\end{figure}

\subsection{The Fidelity Synopsis}
\label{sec:synopsis:offline}
\label{sec:synopsis}
\label{sec:map:offline}

The offline stage takes a model's corpus embeddings, its catalog $\mathcal{A} = \{A_1,\dots,A_m\}$, and a \emph{construction sample} $Q$ (a set of queries for which exact search is affordable), and builds the synopsis in four steps. Figure~\ref{fig:offline-geometry} shows the result on real data.

\paragraph{Step 1: Landmarks.} We choose $K \ll |C|$ \emph{landmarks} $L = \{\ell_1, \dots, \ell_K\}$, points in the embedding space obtained by spherical $k$-means on the corpus vectors. Each landmark defines a \emph{cell}, the set of points closer to it than to any other landmark:
\[
  V_i \;=\; \{\, x \;:\; \mathrm{dist}(x, \ell_i) \le \mathrm{dist}(x, \ell_{i'}) \text{ for all } i' \,\}.
\]
The cells partition the space, so every query belongs to exactly one cell (ties are broken arbitrarily). They play the role of histogram buckets. Clustering places more landmarks, and hence smaller cells, where the corpus is dense. Each model has one landmark set per modality, and a query is located among the landmarks of the modality it searches: in text-to-image search, a text query is located among image landmarks. All configurations share the same landmarks, so the cells are the same for every configuration and only the statistics differ.

\paragraph{Step 2: Fidelity labels.} For every query $q \in Q$, we compute its exact top-$k$ once, run every configuration $A_j$ on it, and record $y_{qj} = \fid(q, A_j)$. This gives a $|Q| \times m$ table of true fidelities. It is the only step that uses exact search, once per sample query, since all configurations are compared against the same exact top-$k$.

\paragraph{Step 3: Assignment.} Each sample query is assigned to the cell of its nearest landmark,
\[
  \ell(q) \;=\; \arg\min_{i} \; \mathrm{dist}(q, \ell_i), \qquad Q_i \;=\; \{\, q \in Q : \ell(q) = i \,\},
\]
and $n_i = |Q_i|$ is the number of sample queries in cell $i$.

\paragraph{Step 4: Aggregation.} For each cell $i$ and configuration $A_j$, we keep the mean fidelity of the cell's sample queries,
\[
  \mu_{ij} \;=\; \frac{1}{n_i} \sum_{q \in Q_i} y_{qj},
\]
and a histogram of their fidelities over $B$ bins, from which we read, for any target $\tau$, the fraction of the cell's queries on which $A_j$ met it:
\[
  \pi_{ij}(\tau) \;=\; \frac{1}{n_i}\, \big|\{\, q \in Q_i : y_{qj} \ge \tau \,\}\big| .
\]
We call $\pi_{ij}(\tau)$ the \emph{coverage} of $A_j$ in cell $i$: the estimated probability that $A_j$ suffices for a query in this region. The mean says how good a configuration is on average. The coverage says how often it is good enough, which is what selection needs. Two cells with the same mean can differ sharply in coverage, if fidelity is uniformly moderate in one and high for most queries but very low for a few in the other. We use $B = 20$ bins of width 0.05. For a target at a bin edge, such as 0.7, 0.8, or 0.9, the coverage read from the histogram is exact, and for a target inside a bin, the bin's count is divided evenly among the fidelity values it holds.

\paragraph{Sparse cells.} A cell with few or no sample queries has unreliable or undefined statistics. If fewer than $q_{\min}$ sample queries fall into cell $i$, its statistics are replaced by those of the whole sample, $\bar{\mu}_j$ and $\bar{\pi}_j(\tau)$. We use $q_{\min} = 10$.

\paragraph{Output.} The synopsis consists of the $K$ landmarks, the count $n_i$ of each cell, and, for each cell and configuration, the mean $\mu_{ij}$ and a $B$-bin histogram: a $K \times m$ table with one row per cell and one column per configuration. Its size is $O(K(d + mB))$ numbers for $d$-dimensional embeddings, dominated by the landmarks themselves, and it grows neither with the corpus nor with the sample. The label table of Step~2 is needed only to build the synopsis.

\subsection{Selection Rule}
\label{sec:synopsis:online}
\label{sec:map:online}

At query time, the synopsis chooses a configuration for a new query $q$ under the operator's target $\tau$ and reliability level $\rho$. Because each cell stores a histogram rather than a single number, the same synopsis serves any target.

\paragraph{Step 1: Locate the query.} We find the query's nearest landmark, $i = \ell(q)$, by comparing $q$ with all $K$ landmarks, at a cost of $O(K d)$.

\paragraph{Step 2: Read the estimates.} Row $i$ of the synopsis gives, for every configuration at once, an estimated fidelity and an estimated coverage:
\[
  \fidhat(q, A_j) = \mu_{ij}, \qquad \hat{\pi}(q, A_j, \tau) = \pi_{ij}(\tau), \qquad j = 1, \dots, m,
\]
with the whole-sample statistics in place of $\mu_{ij}$ and $\pi_{ij}(\tau)$ if the cell is sparse. No configuration has been run on $q$ at this point.

\paragraph{Step 3: Select.} The \emph{budget rule} prices each configuration by its cost plus a charge for its estimated chance of missing the target (Figure~\ref{fig:overview}, bottom). It uses only the estimated probabilities, so it works with any estimator:
\[
  \hat{A}(q) \;=\; \arg\min_{A_j \in \mathcal{A}} \; c(A_j) + \lambda\,\big(1 - \hat{\pi}(q, A_j, \tau)\big).
\]
The single parameter $\lambda$ is the price of an expected miss. A small $\lambda$ favors cheap configurations and lets misses rise. A large one favors reliable configurations. We set $\lambda$ once per target, offline, to the smallest value whose miss rate on the construction sample meets the requirement, at most $1 - \rho$ (Section~\ref{sec:setup:params}). Since the estimates depend only on the cell, the rule's choice is fixed per cell and can be precomputed, so Step~3 reduces to a table lookup.

The rule generalizes static deployment. With a single landmark ($K = 1$), every query falls into the same cell and receives the same configuration. This is a static plan, but not necessarily $A_{\mathrm{static}}$ of Section~\ref{sec:prelim:problem}. As $\lambda$ grows, a single cell's choice moves along the lower convex hull of the configurations' (miss rate, cost) points, so a configuration inside the hull is never chosen, and the grid of $\lambda$ values is discrete. The penalized rule can therefore pick a costlier configuration than the constrained definition would. With more landmarks, the choice varies across cells: cells where cheap configurations suffice are served cheaply, and cells where they do not are served by costlier ones. The saving over the static plan comes entirely from this variation.

\paragraph{Simpler rules.} Section~\ref{sec:rq3} compares the budget rule with two simpler ones. The \emph{mean rule} picks the cheapest configuration whose estimated fidelity $\fidhat$ reaches $\tau$, and the one with the highest estimate if none does. It ignores how fidelity is spread within a cell. The \emph{coverage rule} picks the cheapest configuration whose estimated coverage $\hat{\pi}$ reaches $\rho$ in the query's cell, and the one with the highest coverage if none does. It enforces the requirement in every cell rather than over the workload, which is stricter and therefore costlier.

\paragraph{Step 4: Execute.} We run $\hat{A}(q)$ and return its top-$k$ result. The per-query overhead is one nearest-landmark search and one table lookup, independent of the corpus size, with no exact search.

\paragraph{Tuning $\lambda$.} The budget rule's miss rate tends to fall as $\lambda$ grows, so we search a grid of values. For each value, the rule's choice is computed per cell, and its miss rate is measured on the construction sample. We measure it on held-out folds of the sample rather than on the queries the synopsis was built from (Section~\ref{sec:setup:params}). On its own training queries, an estimator is optimistic, and a $\lambda$ tuned there would let too many misses through. The search runs once per target, offline, and touches only the estimates and the labels.

\paragraph{Design choices.} Cells have three advantages over a learned estimator (Section~\ref{sec:mlp}). Their statistics are frequencies, so the coverage needs no separate calibration. An operator can inspect which regions are hard for which configuration. And a new configuration adds one column, leaving the rest unchanged. Landmarks come from the corpus, so the cells can serve any workload over it, and a text query that searches images is located among image landmarks, near the items it will retrieve.

\subsection{A Learned Estimator}
\label{sec:mlp}

The synopsis gives every query in a cell the same estimate. A learned estimator can also use where in the cell a query lies. We use a small multi-layer perceptron. Its input is the query vector. Its outputs are, for every configuration, the predicted fidelity and the probability of meeting each target we evaluate. It has two hidden layers of 512 ReLU units and is trained with a squared-error loss on fidelity plus a cross-entropy loss on meeting each target. We train one network per model and direction, on the same construction sample and labels as the synopsis, so it needs no further search or labeling. The selection rule is unchanged. The MLP's probability takes the place of the cell's coverage, and $\lambda$ is tuned on out-of-fold predictions with the same folds. The deployed MLP is retrained on all of $Q$. The five fold networks serve only to provide the out-of-fold predictions on which $\lambda$ is tuned. The lookup is one forward pass. It gives up the properties of cells listed in Section~\ref{sec:synopsis:online}. Its probabilities are not calibrated by construction, and it cannot be inspected region by region.

\subsection{Choosing the Catalog}
\label{sec:advisor}

Per-query selection can only choose among configurations that exist. Each one costs memory, and each one must be labeled on the construction sample. Which configurations are worth building? This is the physical-design problem of relational systems, and our answer follows the design of index advisors such as AutoAdmin~\cite{autoadmin}.

\paragraph{Relation to AutoAdmin.} AutoAdmin chooses indexes for a SQL workload in three steps. It derives candidate indexes from the individual queries of the workload. It costs each set of candidates by \emph{what-if} analysis, asking the optimizer to cost the workload as if those indexes existed, without building them. It then searches the sets of candidates within a storage bound, exhaustively for small sets and greedily beyond. Our advisor keeps this structure, with candidates derived from the sample queries, what-if evaluation, and a search under a memory budget. It differs in three ways. First, its what-if costs are measured, not modeled. No cost model predicts a configuration's fidelity, so every candidate must be built once and run on the sample to label it. Only after that is evaluating a set of candidates free of further searches. Second, it optimizes under a quality requirement. AutoAdmin minimizes the estimated cost of plans that all return exact answers. Our advisor minimizes cost subject to a miss rate. Third, the selector routes each query to one configuration of the set. The value of a set therefore depends on how its configurations complement each other, not only on the benefit of each.

\paragraph{Problem.} We are given a pool $\mathcal{P}$ of candidate configurations and a memory budget $M$. Each candidate $A_j$ has a cost $c(A_j)$ and uses one index. A sub-catalog $S \subseteq \mathcal{P}$ occupies the memory of the distinct indexes it uses. Search settings of one index, such as several values of \texttt{efSearch}, share that index's memory. We seek the sub-catalog whose selector has the lowest cost under the requirement, among those that fit in $M$.

\paragraph{What-if evaluation with either estimator.} The labels give every sample query the fidelity of every candidate. A selector for any sub-catalog $S$ can therefore be built and evaluated from them alone. Selection needs an estimate of each configuration's chance of meeting $\tau$. The synopsis reads it from the histogram of the query's cell, and the MLP predicts it from the query vector. Both are trained on the same labels. Restricting either to $S$ needs no retraining. The synopsis keeps its cells and drops the histograms of configurations outside $S$, and the MLP drops their outputs. We then choose the static plan of $S$, tune $\lambda$ by cross-fitting (Section~\ref{sec:setup}), and compute the selector's cost and miss rate on $Q$. With 11 candidates, all 2{,}047 sub-catalogs can be evaluated per setting.

\paragraph{The advisor.} The advisor proceeds in four steps.
\begin{enumerate}
  \item \emph{Candidates.} As in AutoAdmin, derive candidates from the queries. For each index family, start from a ladder of search settings that spans costs below and above the static plan, and keep the settings that are the cheapest sufficient configuration for many sample queries. Prefer search settings of one index, which add no memory.
  \item \emph{Screening.} Build the candidates, label them on a small part of the construction sample, and train a coarse estimator from it.
  \item \emph{Choice.} Evaluate the sub-catalogs that fit in $M$, or add candidates greedily by their saving per added gigabyte, and keep the cheapest that meets the requirement. Greedy choice starts from the static plan of the pool and considers only additions that keep the requirement on $Q$. A search setting of an index already chosen adds no memory, so it is taken first if it saves. Otherwise, greedy adds the candidate with the largest saving per added gigabyte. It stops when no addition lowers cost by at least 1\% of the static plan's cost.
  \item \emph{Final labeling.} Label only the chosen configurations on the full construction sample, and build the final estimator from them.
\end{enumerate}
The final step also lowers the labeling bill. Configurations that are never worth their memory are never labeled at full scale, and their indexes can be dropped.

\paragraph{Putting it together.} A deployment starts from a sample of the corpus itself, so no query log is needed. The advisor screens the candidate configurations on a tenth of this sample and chooses a catalog within the memory budget. Only the chosen configurations are built and labeled on the full sample. The estimator is trained on these labels and stored with the indexes, where a relational optimizer keeps its statistics. At query time, the operator looks up the estimate and sends the query to the configuration that the budget rule names. The interface of the operator does not change. This suits collections searched at high volume and in batches, as in retrieval-augmented generation and semantic query pipelines.

\paragraph{Changing targets, costs, and data.} A new target or reliability level needs no new labels. The synopsis's histograms serve any target, so only $\lambda$ is retuned. The MLP is retrained on the stored labels, which takes seconds. Cost enters only through $c(A_j)$ in the budget rule. Latency could be replaced by energy, or by a cost that charges memory or I/O, as for disk-resident indexes~\cite{diskann,spann}. A new configuration is labeled on the sample, whose exact neighbors are already known. Small changes to the corpus should leave the estimates roughly valid, much as histograms tolerate small updates. Large changes call for relabeling, which for the synopsis can be confined to the cells whose items changed. A shift in the query workload moves queries between cells, but it does not change what a cell's statistics mean, as long as new queries resemble the sample queries in the same cell. We have not evaluated these maintenance strategies.


\section{Evaluation Setup}
\label{sec:setup}

\subsection{Benchmark}
\label{sec:setup:benchmark}

We use image--caption pairs from LAION~\cite{laion}. Of the 6{,}999{,}488 pairs available to us, we remove 307 with a degenerate side (267 images with a side shorter than 16 pixels, mostly thin strips and banners, and 40 captions that are missing, unreadable, or one to two characters long), and then 172{,}894 duplicates found by grouping near-identical SigLIP2 embeddings: 147{,}566 pairs whose image belongs to a large group of identical images under mostly unrelated captions, which we take to be a placeholder for an image that failed to download, and 25{,}328 extra copies of pairs whose image and caption both repeat, at least one of them in a large group. Of each such pair we keep one copy. Duplicates make a query's exact top-100 arbitrary, since tied items can be returned in any order. We remove the large groups (the largest 5\% by size: eight or more identical images, or six or more identical captions), while smaller groups of repeated captions remain, so some queries still have ties at rank 100. The remaining 6{,}826{,}287 pairs are split once into three disjoint sets in proportions 85/10/5\%, stratified on the 616 non-empty combinations of two $k$-means clusterings with 25 clusters each, one of the image and one of the caption embeddings (seed 0):

\begin{itemize}
    \item the \emph{corpus} $C$ (5{,}802{,}772 pairs), which is indexed;
    \item the \emph{construction sample} $Q$ (682{,}444 pairs), used only offline, to build the synopses, choose the static plan, and tune $\lambda$;
    \item the \emph{held-out workload} $Q_{\mathrm{test}}$ (341{,}071 pairs), on which all results are reported.
\end{itemize}
The image and the caption of a pair always belong to the same set, so no query shares a pair with an item it searches. The split is computed in SigLIP2's embedding space and reused for the other models, so that all models are evaluated on the same pairs. Each model indexes the images and the captions of $C$ as two separate collections, and every pair in $Q$ and $Q_{\mathrm{test}}$ provides an image query and a text query, each searched against both collections. This gives the four query directions, with 341{,}071 held-out queries per setting. Unless stated otherwise, $k = 100$.

\subsection{Embedding Models}
\label{sec:setup:models}

We use four open-weight embedding models (Table~\ref{tab:models}), chosen to differ in properties that affect approximate search: their architecture (SigLIP2~\cite{siglip2} and OpenCLIP~\cite{openclip} are contrastively trained dual encoders, Jina-v4~\cite{jinav4} and Qwen3-VL~\cite{qwen3-vl} embedders built on vision--language models), their dimension, which affects the cost of every configuration and of the synopsis lookup, and how strongly they separate the two modalities, which determines how far cross-modal queries lie from the corpus they search~\cite{modality-gap}. Every structure (indexes, exact neighbors, landmarks, synopses) is built separately for each model. Using four models tests whether our conclusions hold across independently built embedding spaces.

\begin{table}[t]
\footnotesize\centering
\setlength{\tabcolsep}{3pt}
\resizebox{\columnwidth}{!}{%
\begin{tabular}{@{}l l l r r l@{}}
\toprule
Model & Checkpoint & Type & $d$ & Tokens & Precision \\
\midrule
SigLIP2  & \texttt{google/siglip2-so400m-patch16-512} & dual encoder & 1{,}152 & 64  & fp16 \\
OpenCLIP & \texttt{ViT-L-14}, \texttt{laion2b\_s32b\_b82k} & dual encoder & 768 & 77 & fp16 \\
Jina-v4  & \texttt{jinaai/jina-embeddings-v4} & VLM (Qwen2.5-VL) & 2{,}048 & 512 & bf16 \\
Qwen3-VL & \texttt{Qwen/Qwen3-VL-Embedding-2B} & VLM (Qwen3-VL) & 2{,}048 & 512 & bf16 \\
\bottomrule
\end{tabular}%
}
\caption{Embedding models. $d$: embedding dimension. Tokens: maximum caption length. Precision: inference precision.}
\label{tab:models}
\end{table}

\subsection{Configuration Catalog}
\label{sec:setup:catalog}

For each model, we build three configurations of each index family, \emph{low}, \emph{moderate}, and \emph{high}, giving nine configurations (Table~\ref{tab:catalog}), with FAISS~\cite{faiss} 1.15.1 (official cuVS build, with AVX2). The three tiers of a family differ in their build: HNSW in graph degree and \texttt{efConstruction}, the IVF families in code length. They were chosen by a parameter sweep on a 50{,}000-vector sample to reach mean fidelity bands of 0.65--0.75 (low), 0.75--0.85 (moderate), and at least 0.85 (high), averaged over the two directions an index serves. The search parameter was then set per model at full scale, on 5{,}000 construction queries. At full scale every IVF-family tier lies in its band, but \texttt{hnsw\_low} lies below its band for every model. Three tiers per family give a choice between cheap and expensive configurations both within and across families, while keeping the catalog small enough to run every configuration on every construction query.

Reaching the bands requires \texttt{nprobe} values of 2{,}048--4{,}096 for SigLIP2 and OpenCLIP but only 16--256 for Jina-v4 and Qwen3-VL. For SigLIP2 and OpenCLIP, every IVF-family configuration is therefore several times as expensive as \texttt{hnsw\_low}, and most of them are dominated. For Jina-v4 and Qwen3-VL, the low and moderate IVF tiers cost between \texttt{hnsw\_low} and \texttt{hnsw\_mod}. No configuration is removed, since a dominated configuration can still be the cheapest sufficient one for individual queries. The parameters are what the sweep selected to reach each band. They were not tuned per model for cost, and some look unusual. The high \texttt{nprobe} values for SigLIP2 and OpenCLIP are one example. The build of \texttt{hnsw\_high} also uses a smaller efConstruction than that of \texttt{hnsw\_mod}. A differently tuned configuration could change the static plan in some settings. We therefore do not rest the result on any single configuration. We also search the index of \texttt{hnsw\_high} with two lower \texttt{efSearch} values, its low and moderate \emph{search settings} (S-l and S-m), which add no memory. They are, per model, the first values of the parameter sweep at which mean fidelity over the four directions reaches 0.7 and 0.8 (\texttt{efSearch} of 192 and 384 for SigLIP2, 256 and 1{,}024 for OpenCLIP, 48 and 96 for Jina-v4, and 64 and 128 for Qwen3-VL). These eleven configurations form the \emph{pool}. Its static plan, the \emph{best static plan}, is the best static choice among everything we built and our main baseline. The static plan of the nine configurations, the \emph{full catalog}, is the secondary one. Table~\ref{tab:appendix-catalog} gives the cost and mean fidelity of every configuration in every setting.

\begin{table*}[t]
\small\centering
\setlength{\tabcolsep}{4pt}
\resizebox{\textwidth}{!}{%
\begin{tabular}{@{}l l l c r@{}}
\toprule
Configuration & Build parameters & Search parameter & Shared & Memory (GB) \\
\midrule
\texttt{hnsw\_low}     & $M{=}4$, efC${=}40$ & efSearch${=}$1024/1024/192/128 & no & 54.1/36.3/95.7/95.7 \\
\texttt{hnsw\_mod}     & $M{=}8$, efC${=}80$ & efSearch${=}$1536/1024/192/256 & no & 54.5/36.7/96.1/96.1 \\
\texttt{hnsw\_high}    & $M{=}16$, efC${=}40$ & efSearch${=}$6144/6144/512/1024 & no & 55.2/37.4/96.8/96.8 \\
\midrule
\texttt{ivfpq\_low}    & $n_{\mathrm{list}}{=}32768$, PQ288/192/512/512 & nprobe${=}$2048/2048/16/16 & no & 3.7/2.5/6.6/6.6 \\
\texttt{ivfpq\_mod}    & $n_{\mathrm{list}}{=}32768$, PQ384/256/1024/1024 & nprobe${=}$2048/2048/32/32 & no & 4.9/3.3/12.5/12.5 \\
\texttt{ivfpq\_high}   & $n_{\mathrm{list}}{=}32768$, PQ1152/768/2048/2048 & nprobe${=}$4096/4096/128/256 & no & 13.8/9.2/24.4/24.4 \\
\midrule
\texttt{oivfpq\_low}   & OPQ288/192/512/512, $n_{\mathrm{list}}{=}32768$, PQ288/192/512/512 & nprobe${=}$2048/2048/16/16 & no & 3.8/2.5/6.6/6.6 \\
\texttt{oivfpq\_mod}   & OPQ384/256/1024/1024, $n_{\mathrm{list}}{=}32768$, PQ384/256/1024/1024 & nprobe${=}$2048/2048/32/32 & no & 4.9/3.3/12.6/12.6 \\
\texttt{oivfpq\_high}  & OPQ1152/768/2048/2048, $n_{\mathrm{list}}{=}32768$, PQ1152/768/2048/2048 & nprobe${=}$4096/4096/128/256 & no & 13.8/9.2/24.4/24.4 \\
\bottomrule
\end{tabular}%
}
\caption{The nine configurations of the catalog, with per-model values in the order SigLIP2/OpenCLIP/Jina-v4/Qwen3-VL. Every configuration is a separate build. Memory is that of the image and the text index together. Table~\ref{tab:appendix-catalog} gives each configuration's cost and mean fidelity per setting.}
\label{tab:catalog}
\end{table*}

\begin{table*}[t]
\small\centering
\setlength{\tabcolsep}{3pt}
\begin{tabular}{@{}llrrrrrrrr@{}}
\toprule
 & & \multicolumn{2}{c}{SigLIP2} & \multicolumn{2}{c}{OpenCLIP} & \multicolumn{2}{c}{Jina-v4} & \multicolumn{2}{c}{Qwen3-VL} \\
\cmidrule(lr){3-4} \cmidrule(lr){5-6} \cmidrule(lr){7-8} \cmidrule(lr){9-10}
Dir. & Configuration & Cost & $\bar F$ & Cost & $\bar F$ & Cost & $\bar F$ & Cost & $\bar F$ \\
\midrule
i$\to$i & \texttt{hnsw\_low} & 651 & 0.683 & 484 & 0.674 & 208 & 0.728 & 160 & 0.656 \\
 & \texttt{hnsw\_mod} & 1{,}609 & 0.964 & 878 & 0.950 & 451 & 0.936 & 527 & 0.949 \\
 & \texttt{hnsw\_high} & 13{,}439 & 0.978 & 8{,}796 & 0.973 & 1{,}141 & 0.978 & 1{,}976 & 0.981 \\
 & \texttt{ivfpq\_low} & 3{,}582 & 0.890 & 3{,}050 & 0.874 & 297 & 0.817 & 313 & 0.815 \\
 & \texttt{ivfpq\_mod} & 18{,}365 & 0.935 & 12{,}263 & 0.923 & 420 & 0.899 & 399 & 0.895 \\
 & \texttt{ivfpq\_high} & 23{,}743 & 0.992 & 14{,}833 & 0.989 & 1{,}855 & 0.969 & 3{,}549 & 0.976 \\
 & \texttt{oivfpq\_low} & 3{,}455 & 0.889 & 2{,}341 & 0.874 & 323 & 0.819 & 353 & 0.816 \\
 & \texttt{oivfpq\_mod} & 18{,}251 & 0.935 & 12{,}194 & 0.924 & 428 & 0.899 & 407 & 0.895 \\
 & \texttt{oivfpq\_high} & 23{,}468 & 0.993 & 14{,}818 & 0.991 & 2{,}023 & 0.969 & 3{,}842 & 0.977 \\
\midrule
i$\to$t & \texttt{hnsw\_low} & 647 & 0.311 & 494 & 0.358 & 234 & 0.601 & 171 & 0.468 \\
 & \texttt{hnsw\_mod} & 1{,}579 & 0.748 & 961 & 0.714 & 525 & 0.868 & 581 & 0.869 \\
 & \texttt{hnsw\_high} & 13{,}145 & 0.831 & 8{,}560 & 0.744 & 1{,}384 & 0.941 & 1{,}870 & 0.940 \\
 & \texttt{ivfpq\_low} & 2{,}958 & 0.486 & 2{,}247 & 0.540 & 309 & 0.703 & 311 & 0.729 \\
 & \texttt{ivfpq\_mod} & 17{,}648 & 0.644 & 12{,}106 & 0.714 & 397 & 0.809 & 397 & 0.834 \\
 & \texttt{ivfpq\_high} & 19{,}339 & 0.862 & 13{,}854 & 0.926 & 1{,}813 & 0.925 & 3{,}848 & 0.958 \\
 & \texttt{oivfpq\_low} & 3{,}027 & 0.493 & 2{,}232 & 0.562 & 302 & 0.702 & 337 & 0.729 \\
 & \texttt{oivfpq\_mod} & 17{,}609 & 0.653 & 12{,}040 & 0.727 & 403 & 0.809 & 393 & 0.835 \\
 & \texttt{oivfpq\_high} & 19{,}233 & 0.865 & 13{,}786 & 0.929 & 2{,}008 & 0.925 & 3{,}483 & 0.959 \\
\midrule
t$\to$i & \texttt{hnsw\_low} & 673 & 0.490 & 531 & 0.525 & 228 & 0.562 & 146 & 0.386 \\
 & \texttt{hnsw\_mod} & 1{,}735 & 0.871 & 923 & 0.860 & 506 & 0.823 & 532 & 0.733 \\
 & \texttt{hnsw\_high} & 13{,}150 & 0.933 & 8{,}646 & 0.939 & 1{,}130 & 0.915 & 2{,}071 & 0.897 \\
 & \texttt{ivfpq\_low} & 3{,}116 & 0.468 & 2{,}018 & 0.493 & 320 & 0.670 & 337 & 0.530 \\
 & \texttt{ivfpq\_mod} & 17{,}755 & 0.626 & 11{,}725 & 0.645 & 384 & 0.782 & 384 & 0.648 \\
 & \texttt{ivfpq\_high} & 20{,}083 & 0.876 & 12{,}335 & 0.899 & 1{,}703 & 0.907 & 3{,}369 & 0.871 \\
 & \texttt{oivfpq\_low} & 2{,}991 & 0.477 & 1{,}972 & 0.512 & 302 & 0.672 & 322 & 0.531 \\
 & \texttt{oivfpq\_mod} & 17{,}693 & 0.634 & 11{,}741 & 0.654 & 418 & 0.782 & 395 & 0.648 \\
 & \texttt{oivfpq\_high} & 19{,}983 & 0.878 & 12{,}133 & 0.901 & 1{,}857 & 0.907 & 3{,}780 & 0.871 \\
\midrule
t$\to$t & \texttt{hnsw\_low} & 688 & 0.500 & 450 & 0.470 & 221 & 0.611 & 176 & 0.522 \\
 & \texttt{hnsw\_mod} & 1{,}687 & 0.912 & 971 & 0.842 & 501 & 0.886 & 602 & 0.905 \\
 & \texttt{hnsw\_high} & 14{,}220 & 0.929 & 9{,}146 & 0.809 & 1{,}226 & 0.937 & 2{,}192 & 0.953 \\
 & \texttt{ivfpq\_low} & 3{,}751 & 0.878 & 2{,}473 & 0.878 & 311 & 0.765 & 306 & 0.770 \\
 & \texttt{ivfpq\_mod} & 18{,}735 & 0.926 & 12{,}329 & 0.926 & 391 & 0.849 & 404 & 0.850 \\
 & \texttt{ivfpq\_high} & 27{,}356 & 0.988 & 16{,}032 & 0.988 & 1{,}779 & 0.931 & 4{,}010 & 0.950 \\
 & \texttt{oivfpq\_low} & 3{,}865 & 0.879 & 2{,}449 & 0.879 & 316 & 0.764 & 323 & 0.770 \\
 & \texttt{oivfpq\_mod} & 18{,}638 & 0.927 & 12{,}338 & 0.927 & 400 & 0.849 & 412 & 0.850 \\
 & \texttt{oivfpq\_high} & 27{,}188 & 0.989 & 15{,}812 & 0.989 & 1{,}874 & 0.931 & 3{,}737 & 0.950 \\
\bottomrule
\end{tabular}
\caption{Per-query cost ($\mu$s per query, controlled benchmark, batches of 4{,}096 on 64 threads) and mean fidelity $\bar F$ on $Q_{\mathrm{test}}$ of every configuration, per model and direction.}
\label{tab:appendix-catalog}
\end{table*}

\subsection{Cost Measurement}
\label{sec:setup:cost}

All costs are measured on dedicated nodes with an AMD EPYC 7763 processor (128 logical CPUs, 64 used) and 1\,TB of RAM, with all indexes in memory. Exact search and $k$-means run on NVIDIA A100-SXM4-80GB GPUs. We measure two regimes: \emph{single query}, one query at a time on one thread, and \emph{batched}, batches of 4{,}096 queries on 64 threads. Every measurement uses the same 5{,}000 queries drawn at random from $Q_{\mathrm{test}}$, after an untimed warm-up of 500 queries. In the batched regime, passes over the sample are repeated until at least 20\,s and three passes have been timed, and we report the median pass. In the single-query regime, we report the mean and the 99th percentile. The cost $c(A)$ of a configuration is its mean per-query latency for the given model and direction. The lookup cost $c_{\mathrm{est}}$ is timed with the same harness, queries, batch size, and number of threads, in a separate run on the same type of node. All headline results use the batched regime. Costs assume that every index receives full batches. Under per-query selection, a batch is split across the indexes the estimator chooses, so each index sees only part of it. A service under sustained load can accumulate the queries for each index into full batches, at some cost in waiting time. We do not model this effect.

\subsection{Parameters and Reference Points}
\label{sec:setup:params}

\paragraph{Estimator parameters.} Unless stated otherwise, each modality has $K = 10{,}000$ landmarks, and RQ5 varies $K$. Histograms have $B = 20$ bins, and $q_{\min} = 10$. The MLP (Section~\ref{sec:mlp}) is trained with Adam for 10 epochs, with outputs for all eleven configurations of the pool and for the three targets. The MLP is trained with a learning rate of $10^{-3}$, batches of 4{,}096 queries, and seed 0. Its loss is the mean squared error of the predicted fidelities plus the mean binary cross-entropy of the predicted probabilities of meeting each target, with equal weights. We evaluate fidelity targets $\tau \in \{0.7, 0.8, 0.9\}$, with $\tau = 0.7$ as the default, and a reliability level of $\rho = 0.9$, so the requirement allows at most 10\% of queries to miss the target.

\paragraph{Choice of target.} A target of $\tau = 0.7$ at $k = 100$ asks a configuration to return 70 of a query's 100 exact nearest neighbors. Fidelity is measured against exact search, not against relevance. The exact neighbors depend on the embedding model and are only a proxy for what a user would judge relevant. The catalog's tiers span mean fidelities from 0.31 to 0.99 across settings, and from 0.50 to 0.95 averaged over directions (Table~\ref{tab:appendix-catalog}). At $\tau = 0.7$, every model therefore has a real choice between cheap and costly configurations. At $\tau = 0.9$, many queries miss the target under every configuration, so the requirement can be met in only 9 of 16 settings, even with per-query choice. This is mostly a limit of the catalog, not of plan selection. At $k = 10$, mean fidelity differs from that at $k = 100$ by $-0.04$ to $+0.08$ for every configuration and setting, and the configurations rank almost identically (Table~\ref{tab:k10}). Section~\ref{sec:rq2} reports how well location predicts fidelity at $k = 10$. We have not evaluated plan selection at $k = 10$.

We fix $\rho = 0.9$ throughout. The trade-off curves of Section~\ref{sec:rq3} cover every other reliability level. Section~\ref{sec:rq3} reports plan selection at all three targets (Table~\ref{tab:targets}).

\begin{table*}[t]
\footnotesize\centering
\setlength{\tabcolsep}{2.5pt}
\resizebox{\textwidth}{!}{%
\begin{tabular}{@{}l rrrr rrrr@{}}
\toprule
 & \multicolumn{4}{c}{SigLIP2} & \multicolumn{4}{c}{OpenCLIP} \\
\cmidrule(lr){2-5}\cmidrule(lr){6-9}
Configuration & i$\to$i & i$\to$t & t$\to$i & t$\to$t & i$\to$i & i$\to$t & t$\to$i & t$\to$t \\
\midrule
\texttt{hnsw\_low} & 0.693 (0.683) & 0.324 (0.311) & 0.500 (0.490) & 0.508 (0.500) & 0.684 (0.674) & 0.391 (0.358) & 0.543 (0.525) & 0.488 (0.470) \\
\texttt{hnsw\_mod} & 0.968 (0.964) & 0.791 (0.748) & 0.892 (0.871) & 0.916 (0.912) & 0.958 (0.950) & 0.763 (0.714) & 0.889 (0.860) & 0.852 (0.842) \\
\texttt{hnsw\_high} & 0.981 (0.978) & 0.855 (0.831) & 0.942 (0.933) & 0.933 (0.929) & 0.977 (0.973) & 0.778 (0.744) & 0.950 (0.939) & 0.824 (0.809) \\
\midrule
\texttt{ivfpq\_low} & 0.886 (0.890) & 0.461 (0.486) & 0.459 (0.468) & 0.881 (0.878) & 0.866 (0.874) & 0.502 (0.540) & 0.491 (0.493) & 0.881 (0.878) \\
\texttt{ivfpq\_mod} & 0.928 (0.935) & 0.638 (0.644) & 0.629 (0.626) & 0.924 (0.926) & 0.913 (0.923) & 0.683 (0.714) & 0.649 (0.645) & 0.922 (0.926) \\
\texttt{ivfpq\_high} & 0.991 (0.992) & 0.894 (0.862) & 0.909 (0.876) & 0.986 (0.988) & 0.988 (0.989) & 0.927 (0.926) & 0.920 (0.899) & 0.985 (0.988) \\
\midrule
\texttt{oivfpq\_low} & 0.886 (0.889) & 0.464 (0.493) & 0.472 (0.477) & 0.883 (0.879) & 0.867 (0.874) & 0.526 (0.562) & 0.514 (0.512) & 0.882 (0.879) \\
\texttt{oivfpq\_mod} & 0.929 (0.935) & 0.645 (0.653) & 0.640 (0.634) & 0.924 (0.927) & 0.914 (0.924) & 0.698 (0.727) & 0.662 (0.654) & 0.923 (0.927) \\
\texttt{oivfpq\_high} & 0.992 (0.993) & 0.898 (0.865) & 0.915 (0.878) & 0.987 (0.989) & 0.990 (0.991) & 0.935 (0.929) & 0.925 (0.901) & 0.987 (0.989) \\
\midrule\midrule
 & \multicolumn{4}{c}{Jina-v4} & \multicolumn{4}{c}{Qwen3-VL} \\
\cmidrule(lr){2-5}\cmidrule(lr){6-9}
Configuration & i$\to$i & i$\to$t & t$\to$i & t$\to$t & i$\to$i & i$\to$t & t$\to$i & t$\to$t \\
\midrule
\texttt{hnsw\_low} & \hi{0.780 (0.728)} & \hi{0.682 (0.601)} & 0.600 (0.562) & \hi{0.664 (0.611)} & \hi{0.723 (0.656)} & \hi{0.530 (0.468)} & 0.413 (0.386) & \hi{0.582 (0.522)} \\
\texttt{hnsw\_mod} & 0.965 (0.936) & \hi{0.929 (0.868)} & 0.867 (0.823) & 0.923 (0.886) & 0.972 (0.949) & 0.917 (0.869) & 0.761 (0.733) & 0.936 (0.905) \\
\texttt{hnsw\_high} & 0.985 (0.978) & 0.962 (0.941) & 0.929 (0.915) & 0.951 (0.937) & 0.986 (0.981) & 0.955 (0.940) & 0.903 (0.897) & 0.963 (0.953) \\
\midrule
\texttt{ivfpq\_low} & 0.851 (0.817) & \hi{0.761 (0.703)} & 0.711 (0.670) & 0.814 (0.765) & 0.850 (0.815) & 0.766 (0.729) & \hi{0.586 (0.530)} & 0.819 (0.770) \\
\texttt{ivfpq\_mod} & 0.931 (0.899) & \hi{0.872 (0.809)} & 0.829 (0.782) & 0.895 (0.849) & 0.928 (0.895) & 0.882 (0.834) & \hi{0.720 (0.648)} & 0.896 (0.850) \\
\texttt{ivfpq\_high} & 0.981 (0.969) & 0.957 (0.925) & 0.932 (0.907) & 0.956 (0.931) & 0.984 (0.976) & 0.972 (0.958) & 0.906 (0.871) & 0.967 (0.950) \\
\midrule
\texttt{oivfpq\_low} & 0.853 (0.819) & \hi{0.760 (0.702)} & 0.713 (0.672) & 0.813 (0.764) & 0.851 (0.816) & 0.765 (0.729) & \hi{0.586 (0.531)} & 0.819 (0.770) \\
\texttt{oivfpq\_mod} & 0.932 (0.899) & \hi{0.872 (0.809)} & 0.830 (0.782) & 0.895 (0.849) & 0.928 (0.895) & 0.882 (0.835) & \hi{0.720 (0.648)} & 0.896 (0.850) \\
\texttt{oivfpq\_high} & 0.981 (0.969) & 0.958 (0.925) & 0.933 (0.907) & 0.956 (0.931) & 0.985 (0.977) & 0.972 (0.959) & 0.906 (0.871) & 0.967 (0.950) \\
\bottomrule
\end{tabular}%
}
\caption{Mean fidelity of every configuration at $k = 10$, with $k = 100$ in parentheses, on the held-out queries. The two differ by $-0.04$ to $+0.08$, and the configurations rank almost identically at both depths. Bold: fidelity at $k = 10$ higher by at least 0.05.}
\label{tab:k10}
\end{table*}

\paragraph{Tuning $\lambda$.} The budget rule's $\lambda$ is chosen on $Q$ by 5-fold cross-fitting: the estimator is built on four folds and its miss rate measured on the fifth, and we take the smallest $\lambda$ whose out-of-fold miss rate meets the chosen limit. The final estimator is then built on all of $Q$. We use two limits: the requirement, at most $1 - \rho$ (\emph{budget rule at the requirement}, our method), and the miss rate of the static plan on $Q$ (\emph{budget rule at the static rate}, the conservative comparison of Section~\ref{sec:prelim:problem}). If no $\lambda$ reaches the limit, we use the one with the fewest misses.

\paragraph{Reference points.} We compare the synopsis with location-blind estimates (the \emph{global average} fidelity and the \emph{global coverage} of each configuration on $Q$, which equal a synopsis with $K = 1$), with a \emph{density-only} estimate (RQ2), with a \emph{shared synopsis} that keeps one difficulty value per cell for all configurations (RQ2), with the static plan of the nine and the best static plan, and with three versions of the per-query optimum. The \emph{matched optimum} serves the same queries as the static plan and possibly more: it runs every query whose cheapest sufficient configuration costs no more than the static plan, and charges every other query the static plan's cost. The \emph{optimum at the requirement} is the cheapest assignment of configurations that misses at most $1 - \rho$ of the queries. No selector that meets the requirement can be cheaper. The \emph{unconstrained optimum} serves every query that some configuration can serve. Table~\ref{tab:reckoner} defines each precisely.

\subsection{Metrics}
\label{sec:setup:metrics}

All metrics are computed on $Q_{\mathrm{test}}$, for synopses, static plans, and $\lambda$ fitted on $Q$. Estimation metrics are computed per configuration and averaged over configurations, excluding \emph{flat} configurations, whose true fidelity has a standard deviation below 0.03 in a direction, since skill is not meaningful on a near-constant target. Table~\ref{tab:reckoner} summarizes all metrics and the abbreviations used in the tables. We define the main ones here.

\paragraph{Accuracy of estimates.} For the estimated fidelity, we report the mean absolute error and the \emph{MSE skill score}~\cite{murphy1988},
\[
  \mathrm{Skill} \;=\; 1 - \frac{\mathrm{MSE}(\fidhat)}{\mathrm{MSE}(\text{global average})},
\]
which is 0 for an estimate no better than the global average, 1 for a perfect one, and negative for a worse one. For the estimated coverage, a predicted probability of the event $\fid(q, A_j) \ge \tau$, we report the \emph{Brier skill score}~\cite{brier1950}, the same construction applied to the mean squared error of that probability against the global coverage. We also report the AUC for separating queries that meet the target from those that do not, and the expected calibration error (ECE)~\cite{naeini2015,guo2017}. The ECE is the mean gap between predicted and observed rates over 10 equal-width bins of $\hat{\pi}$, restricted to pairs of query and configuration with $0.05 < \hat{\pi} < 0.95$, since ECE over all pairs is dominated by configurations whose coverage is trivially 0 or 1.

\paragraph{Plan selection.} For a selector, we report its \emph{relative cost}, the mean of $c_{\mathrm{est}} + c(\hat{A}(q))$ divided by the cost of the static plan of the nine (or of the best static plan, where stated). We also report its \emph{miss rate} and the share of the available saving it captures, the \emph{savings recovered},
\[
  \frac{c(A_{\mathrm{static}}) - \bar{c}_{\mathrm{selector}}}{c(A_{\mathrm{static}}) - \bar{c}_{\mathrm{opt}}},
\]
which is 0 for no improvement over the static plan and 1 for a selector as cheap as the corresponding optimum, and is reported only where that optimum is at least 5\% cheaper than the static plan. A selector can exceed 1 only by missing more often than the optimum allows. In our results, only selectors that miss more often than the static plan do.

\begin{table*}[t]
\footnotesize\centering
\setlength{\tabcolsep}{4pt}
\renewcommand{\arraystretch}{1.1}
\begin{tabular}{@{}p{0.2\textwidth}p{0.76\textwidth}@{}}
\toprule
Term (abbreviation) & Meaning \\
\midrule
\multicolumn{2}{@{}l}{\emph{Workload and baselines}} \\
Setting & A pair of embedding model and query direction; 16 in total. \\
Requirement & At most a fraction $1 - \rho = 10\%$ of queries may miss the fidelity target $\tau$. \\
Static plan & The cheapest single configuration that meets $\tau$ on at least a fraction $\rho$ of $Q$; if none does, the one that meets it most often ($^*$, \S). \\
Serves & Share of held-out queries on which a configuration meets $\tau$. \\
Cheapest sufficient configuration & The cheapest configuration that meets $\tau$ on a given query; buckets give its cost relative to the static plan (none: no configuration meets $\tau$). \\
Optimum, matched & Runs every query whose cheapest sufficient configuration costs no more than the static plan on that configuration, and charges every other query the cost of the static plan; it serves at least every query that the static plan serves. \\
Optimum, unconstrained (all) & Runs every query that some configuration can serve on its cheapest sufficient configuration, and every other query on its highest-fidelity configuration. \\
Optimum, $\le 1{-}\rho$ & The cheapest assignment of configurations that misses $\tau$ on at most $1 - \rho$ of the held-out queries, a missed query running on the cheapest configuration; no selector that meets the requirement is cheaper (infeas.: more than $1 - \rho$ of queries cannot be served by any configuration). \\
\midrule
\multicolumn{2}{@{}l}{\emph{Selection rules} (all use the synopsis's estimates for the query's cell)} \\
Mean rule & The cheapest configuration whose estimated fidelity $\fidhat$ reaches $\tau$. \\
Coverage rule & The cheapest configuration whose estimated coverage $\hat\pi$ reaches $\rho$; if none does, the one with the highest coverage. \\
Budget rule & $\arg\min_j c(A_j) + \lambda (1 - \hat\pi_j)$, with $\lambda$ tuned on $Q$ to a miss limit: the requirement, $1 - \rho$ (\emph{at the requirement}, $\le 1{-}\rho$; our method), or the static plan's miss rate on $Q$ (\emph{at the static rate}, static miss); $\ddagger$: limit not reachable, fewest misses used. \\
\midrule
\multicolumn{2}{@{}l}{\emph{Selection metrics} (on $Q_{\mathrm{test}}$)} \\
Cost & Mean of $c_{\mathrm{est}} + c(\hat{A}(q))$, relative to the static plan ($1\times$). \\
Miss rate (Miss) & Share of queries whose chosen configuration falls short of $\tau$. \\
Shortfall (Short.) & Mean of $\tau - \fid(q, \hat{A}(q))$ over missed queries. \\
Savings recovered (Rec.) & Share of the optimum's saving over the static plan that a selector captures: against the matched optimum, except for the budget rule at the requirement, measured against the optimum at $\le 1{-}\rho$; shown where that optimum saves at least 5\%. \\
$\dagger$ & The coverage rule's missed queries fall less far below $\tau$ than the mean rule's (mean shortfall over missed queries). \\

\midrule
\multicolumn{2}{@{}l}{\emph{Estimation metrics} (on $Q_{\mathrm{test}}$, averaged over non-flat configurations)} \\
Non-flat configuration & True fidelity has a standard deviation of at least 0.03 in the direction. \\
Global average (GA), global coverage & Location-blind estimates: a configuration's mean fidelity, or share of queries meeting $\tau$, on all of $Q$; equivalent to a synopsis with $K = 1$. \\
Density-only (DO) & Global average plus a linear term in the local density of the query's cell (mean distance from its landmark to its 100 nearest corpus vectors), with one slope per configuration (per-cfg) or one for all (shared). \\
Skill & MSE skill score of the estimated fidelity against the global average: 0 no better, 1 perfect. \\
$\rho_s$ & Spearman rank correlation between estimated and true fidelity. \\
ICC & Share of fidelity variance that lies between cells on $Q_{\mathrm{test}}$; approximately the highest skill any estimate with one value per cell can reach. \\
Interior (Int.), boundary (Bnd.) & Skill on the third of queries with the lowest, or highest, ratio of distances to the nearest and second-nearest landmark. \\
Brier skill (BSS) & Skill of the estimated coverage, as a probability of meeting $\tau$, against the global coverage. \\
AUC, ECE & How well the estimated coverage separates queries that meet $\tau$ from those that do not; its expected calibration error over 10 bins, for $0.05 < \hat\pi < 0.95$. \\
Keyed, shared synopsis & The synopsis, with statistics per configuration; a variant with one difficulty offset per cell for all configurations. \\
Thin & Share of held-out queries in cells with fewer than $q_{\min}$ sample queries. \\
\midrule
\multicolumn{2}{@{}l}{\emph{Configuration abbreviations}} \\
H, I, O; l, m, h & HNSW, IVFPQ, O-IVFPQ; low, moderate, high tier. In the single index, l, m, h are its three search settings. S-l and S-m (Sl, Sm): the low and moderate search settings of the \texttt{hnsw\_high} index. The union is the pool. \\
\bottomrule
\end{tabular}
\caption{Ready reckoner: the terms, metrics, and abbreviations used in the tables of Section~\ref{sec:experiments}.}
\label{tab:reckoner}
\end{table*}

\subsection{Statistical Analysis}
\label{sec:setup:stats}

With 341{,}071 held-out queries per setting, almost any difference is statistically significant, so we lead with effect sizes and confidence intervals and use hypothesis tests as supporting evidence. We compute 95\% bootstrap confidence intervals with 1{,}000 resamples. Since queries in the same cell share their estimate, we resample cells rather than individual queries. For the MLP, whose estimates are per query, we resample the same cells, so that correlated queries stay together. Each comparison uses a paired test suited to its outcome: the Wilcoxon signed-rank test on per-query squared error for estimation error (synopsis against global average, and keyed against shared synopsis, in RQ2), the same test with Pratt's handling of zero differences for shortfall (RQ3), and McNemar's test on per-query outcomes for misses (RQ3). Within each family of tests, $p$-values are corrected across settings with the Benjamini--Hochberg procedure at a false discovery rate of $q < 0.05$.

\paragraph{Robustness.} Over three seeds of the landmark clustering, at $K = 10{,}000$ and $\tau = 0.7$, the standard deviation of skill, Brier skill, relative cost, and miss rate is small in every setting (at most 0.008 for skill and 0.010 for Brier skill, 0.03 for relative cost, and a quarter of a percentage point for miss rates), so no conclusion depends on the seed.

\section{Experiments}
\label{sec:experiments}

We answer five research questions. RQ1 measures the opportunity. RQ2 tests the two premises and compares the accuracy of the two estimators. RQ3 evaluates selection end to end. RQ4 evaluates the catalog advisor. RQ5 measures what the estimators cost.

\subsection{RQ1: What could per-query selection save?}
\label{sec:rq1}

\begin{table*}[t]
\small\centering
\setlength{\tabcolsep}{4pt}
\begin{tabular}{@{}ll l r rrrrr rr@{}}
\toprule
 & & & Static & \multicolumn{5}{c}{Queries whose cheapest sufficient plan costs} & \multicolumn{2}{c}{Optimum cost} \\
\cmidrule(lr){5-9}\cmidrule(l){10-11}
Model & Dir. & Static plan & serves & $\le 0.33\times$ & $0.33$--$1\times$ & $1\times$ & $>1\times$ & none & all & matched \\
\midrule
SigLIP2 & i$\to$i & \texttt{hnsw\_mod} & 99.2\% & 0.0\% & 53.8\% & 45.3\% & 0.8\% & 0.0\% & 0.72$\times$ & 0.68$\times$  \\
 & i$\to$t & \texttt{oivfpq\_high}$^*$ & 89.9\% & 68.1\% & 21.6\% & 6.8\% & 0.1\% & 3.4\% & 0.30$\times$ & \hi{0.31$\times$}  \\
 & t$\to$i & \texttt{hnsw\_high} & 97.4\% & 89.7\% & 0.0\% & 7.7\% & 1.6\% & 1.0\% & 0.21$\times$ & \hi{0.20$\times$}  \\
 & t$\to$t & \texttt{hnsw\_mod} & 96.6\% & 0.0\% & 21.2\% & 75.4\% & 3.1\% & 0.2\% & 0.94$\times$ & \lo{0.87$\times$}  \\
\midrule
OpenCLIP & i$\to$i & \texttt{hnsw\_mod} & 97.5\% & 0.0\% & 55.5\% & 42.0\% & 2.5\% & 0.0\% & 0.92$\times$ & \lo{0.75$\times$}  \\
 & i$\to$t & \texttt{oivfpq\_high} & 97.2\% & 64.7\% & 22.7\% & 10.5\% & 0.1\% & 2.1\% & 0.33$\times$ & \hi{0.33$\times$}  \\
 & t$\to$i & \texttt{hnsw\_high} & 98.4\% & 88.8\% & 0.0\% & 9.6\% & 1.5\% & 0.1\% & 0.20$\times$ & \hi{0.20$\times$}  \\
 & t$\to$t & \texttt{oivfpq\_low} & 99.1\% & 19.3\% & 69.0\% & 11.2\% & 0.3\% & 0.2\% & 0.44$\times$ & 0.43$\times$  \\
\midrule
Jina-v4 & i$\to$i & \texttt{ivfpq\_mod} & 90.9\% & 0.0\% & 82.4\% & 8.6\% & 8.7\% & 0.2\% & 0.68$\times$ & 0.62$\times$  \\
 & i$\to$t & \texttt{hnsw\_high} & 97.2\% & 79.0\% & 12.0\% & 6.7\% & 0.9\% & 1.4\% & 0.30$\times$ & \hi{0.30$\times$}  \\
 & t$\to$i & \texttt{hnsw\_high} & 93.7\% & 58.3\% & 25.7\% & 10.1\% & 1.2\% & 4.7\% & 0.41$\times$ & \hi{0.39$\times$}  \\
 & t$\to$t & \texttt{hnsw\_mod} & 91.6\% & 0.0\% & 85.7\% & 8.3\% & 4.4\% & 1.6\% & 0.71$\times$ & 0.61$\times$  \\
\midrule
Qwen3-VL & i$\to$i & \texttt{ivfpq\_mod} & 90.1\% & 0.0\% & 81.6\% & 8.6\% & 9.8\% & 0.0\% & 0.73$\times$ & 0.61$\times$  \\
 & i$\to$t & \texttt{hnsw\_mod} & 90.6\% & 16.5\% & 67.1\% & 9.1\% & 6.9\% & 0.4\% & 0.79$\times$ & 0.60$\times$  \\
 & t$\to$i & \texttt{hnsw\_high} & 92.6\% & 76.8\% & 0.0\% & 18.2\% & 3.6\% & 1.4\% & 0.39$\times$ & \hi{0.37$\times$}  \\
 & t$\to$t & \texttt{hnsw\_mod} & 93.6\% & 28.6\% & 56.2\% & 10.1\% & 4.4\% & 0.7\% & 0.67$\times$ & 0.54$\times$  \\
\bottomrule
\end{tabular}
\caption{Opportunity for per-query selection at $\tau = 0.7$: the static plan and the share of held-out queries it serves, the held-out queries by the cost of their cheapest sufficient configuration relative to the static plan, and the mean cost of the unconstrained and the matched optimum. Bold: large opportunity (matched optimum below $0.4\times$). Grey: small (above $0.7\times$). Terms: Table~\ref{tab:reckoner}.}

\label{tab:rq1-headroom}
\end{table*}

Before evaluating plan selection, we measure the opportunity it targets: how much cheaper a workload would be if every query ran on its cheapest sufficient configuration. Table~\ref{tab:rq1-headroom} reports, for each setting at $\tau = 0.7$, the static plan, how the held-out queries are distributed by the cost of their cheapest sufficient configuration, and the cost of the per-query optimum. We use the matched optimum, which serves at least every query the static plan serves and is therefore the like-for-like comparison. The unconstrained optimum also pays for queries the static plan misses, and can cost more.

\paragraph{The opportunity is large in most settings.} At matched service, per-query selection would lower cost in every setting, and to about half of the static plan in a typical one. By the matched optimum, the settings fall into three groups (bold and grey in Table~\ref{tab:rq1-headroom}). In seven cross-modal settings, the opportunity is large (below 0.4$\times$): the static plan is the most accurate tier of a family, yet most queries would meet the target on a configuration costing at most a third as much. In seven settings, the opportunity is moderate (0.43--0.68$\times$): most queries could use a cheaper configuration, mostly one costing between a third and the whole of the static plan. In two settings, SigLIP2 t$\to$t and OpenCLIP i$\to$i, it is small (0.87$\times$ and 0.75$\times$): the static plan is \texttt{hnsw\_mod}, and \texttt{hnsw\_low} is its only cheaper alternative.

\paragraph{The hardest queries set the price for all of them.} The size of the opportunity follows from two quantities: how expensive the static plan is, and how many queries actually need it. The static plan must serve a fraction $\rho$ of the workload, so it is sized for the hardest queries within that fraction. Where these force \texttt{hnsw\_high}, which costs several times as much as \texttt{hnsw\_mod}, every query pays for them, although in the large-opportunity settings fewer than a fifth of queries need the static plan's configuration. Conversely, where most queries do need it, as in SigLIP2 t$\to$t, the opportunity is small. Every setting with a hard tail of this kind is cross-modal, consistent with cross-modal queries lying away from the corpus distribution. The only cross-modal setting without one is Qwen3-VL i$\to$t.

\begin{figure*}[t]
\centering
\includegraphics[width=\textwidth]{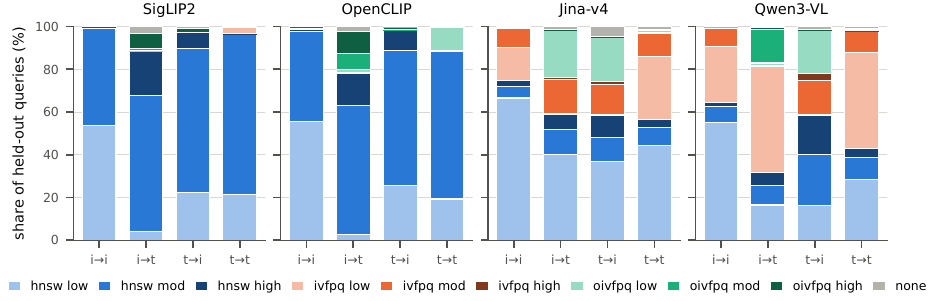}
\caption{Held-out queries by their cheapest sufficient configuration ($\tau = 0.7$). None: no configuration meets the target. Shades: low, moderate, and high tier of each family.}
\label{fig:rq1-cheapest}
\end{figure*}

\paragraph{Which configurations provide the saving.} Figure~\ref{fig:rq1-cheapest} breaks the held-out queries down by their cheapest sufficient configuration. For SigLIP2 and OpenCLIP, it is usually an HNSW configuration: their IVF-family configurations probe thousands of partitions and are the cheapest choice mainly where HNSW falls short, as in image-to-text search. For Jina-v4 and Qwen3-VL, whose low and moderate IVF tiers are cheaper than \texttt{hnsw\_mod}, an IVF-family configuration is the cheapest sufficient one for a quarter to two thirds of the queries. Since every configuration in this catalog is a separate index, capturing the saving requires keeping several indexes built. Section~\ref{sec:small-catalog} and RQ4 (Section~\ref{sec:rq4}) examine how much of it a smaller catalog retains.

\paragraph{Reliability as well as cost.} Per-query selection can also serve queries that the static plan misses: up to a tenth of queries (column ``$>1\times$'') miss the target under the static plan but meet it under a costlier configuration. The share is largest for Jina-v4 and Qwen3-VL i$\to$i, whose static plan only just meets the requirement. Queries that no configuration can serve are rare, at most 5\% in any setting.

\paragraph{The static plan generalizes.} The static plan, chosen on $Q$, meets the target on at least 90\% of the held-out queries in every setting where some configuration meets it on 90\% of $Q$. The baseline is therefore stable across samples, and the opportunity is not an artifact of evaluating a plan on the queries it was tuned on.

\paragraph{Other targets.} At $\tau = 0.8$ and $0.9$, the matched optimum is again at least 5\% cheaper than the static plan in every setting, with a median close to half of it.

\paragraph{Answer to RQ1.} At the same level of service, per-query configuration choice would roughly halve the cost of a static deployment in a typical setting, and save most where the static plan is sized for a small hard tail of cross-modal queries. Whether this opportunity can be captured without knowing each query's fidelity in advance is the subject of RQ2--RQ4.

\subsection{RQ2: Do the premises hold, and how accurate are the estimators?}
\label{sec:rq2}

\begin{table*}[t]
\small\centering
\setlength{\tabcolsep}{3pt}
\resizebox{\textwidth}{!}{%
\begin{tabular}{@{}ll rr l rr r r rr l r r@{}}
\toprule
 & & \multicolumn{2}{c}{MAE} & & \multicolumn{2}{c}{DO skill} & & & \multicolumn{2}{c}{Skill} & \multicolumn{3}{c}{Coverage $\hat\pi$, $\tau{=}0.7$} \\
Model & Dir. & GA & Synopsis & Skill [95\% CI] & per-cfg & shared & $\rho_s$ & ICC & Int. & Bnd. & BSS [95\% CI] & AUC & ECE \\
\midrule
SigLIP2 & i$\to$i & 0.053 & 0.031 & 0.581 [0.574, 0.588] & 0.348 & 0.294 & 0.728 & 0.591 & \hi{0.529} & 0.512 & 0.199 [0.181, 0.216] & 0.889 & 0.0028  \\
 & i$\to$t & 0.118 & 0.090 & 0.349 [0.320, 0.378] & 0.033 & 0.030 & 0.609 & 0.357 & \hi{0.390} & 0.274 & 0.219 [0.200, 0.237] & 0.832 & 0.0021  \\
 & t$\to$i & 0.136 & 0.104 & 0.320 [0.310, 0.330] & 0.017 & 0.004 & 0.583 & 0.331 & \hi{0.285} & 0.251 & 0.204 [0.196, 0.212] & 0.814 & 0.0018  \\
 & t$\to$t & 0.054 & 0.034 & 0.587 [0.466, 0.663] & 0.059 & 0.032 & 0.632 & 0.595 & \hi{0.758} & 0.284 & 0.493 [0.360, 0.574] & 0.921 & 0.0053  \\
\midrule
OpenCLIP & i$\to$i & 0.062 & 0.034 & 0.594 [0.587, 0.601] & 0.389 & 0.339 & 0.734 & 0.604 & 0.516 & \hi{0.529} & 0.237 [0.226, 0.250] & 0.907 & 0.0030  \\
 & i$\to$t & 0.090 & 0.067 & 0.386 [0.365, 0.407] & 0.016 & 0.008 & 0.599 & 0.398 & \hi{0.374} & 0.329 & 0.230 [0.218, 0.241] & 0.840 & 0.0025  \\
 & t$\to$i & 0.120 & 0.090 & 0.364 [0.330, 0.392] & 0.075 & 0.053 & 0.592 & 0.374 & \hi{0.382} & 0.310 & 0.214 [0.195, 0.229] & 0.830 & 0.0035  \\
 & t$\to$t & 0.067 & 0.043 & 0.578 [0.471, 0.641] & 0.038 & $-$0.029 & 0.597 & 0.584 & \hi{0.730} & 0.261 & 0.503 [0.377, 0.564] & 0.924 & 0.0027  \\
\midrule
Jina-v4 & i$\to$i & 0.087 & 0.052 & 0.527 [0.518, 0.535] & 0.342 & 0.275 & 0.725 & 0.535 & 0.281 & \hi{0.458} & 0.273 [0.263, 0.281] & 0.897 & 0.0050  \\
 & i$\to$t & 0.116 & 0.078 & 0.461 [0.440, 0.481] & 0.204 & 0.196 & 0.692 & 0.489 & \hi{0.413} & 0.380 & 0.286 [0.264, 0.309] & 0.866 & 0.0049  \\
 & t$\to$i & 0.151 & 0.098 & 0.475 [0.456, 0.494] & 0.300 & 0.293 & 0.702 & 0.486 & \hi{0.438} & 0.375 & 0.331 [0.314, 0.349] & 0.867 & 0.0035  \\
 & t$\to$t & 0.109 & 0.073 & 0.461 [0.445, 0.477] & 0.221 & 0.218 & 0.700 & 0.470 & \hi{0.500} & 0.336 & 0.274 [0.261, 0.287] & 0.865 & 0.0045  \\
\midrule
Qwen3-VL & i$\to$i & 0.085 & 0.054 & 0.481 [0.472, 0.490] & 0.296 & 0.145 & 0.693 & 0.491 & 0.235 & \hi{0.408} & 0.233 [0.225, 0.240] & 0.858 & 0.0042  \\
 & i$\to$t & 0.102 & 0.076 & 0.364 [0.339, 0.393] & 0.104 & 0.072 & 0.621 & 0.393 & \hi{0.295} & 0.270 & 0.203 [0.181, 0.226] & 0.832 & 0.0056  \\
 & t$\to$i & 0.189 & 0.146 & 0.317 [0.279, 0.351] & 0.109 & 0.083 & 0.565 & 0.325 & \hi{0.278} & 0.238 & 0.236 [0.216, 0.253] & 0.812 & 0.0038  \\
 & t$\to$t & 0.106 & 0.069 & 0.505 [0.481, 0.534] & 0.157 & 0.146 & 0.708 & 0.513 & \hi{0.564} & 0.377 & 0.314 [0.289, 0.341] & 0.880 & 0.0034  \\
\bottomrule
\end{tabular}%
}
\caption{Accuracy of the synopsis's two estimates on the held-out queries ($K = 10{,}000$), averaged over non-flat configurations: estimated fidelity against the global average (GA) and the density-only estimate (DO), and estimated coverage. ECE: pooled over the pairs of query and configuration of all nine configurations, as in Figure~\ref{fig:rq2-reliability}. Bold: the better-predicted of interior and boundary queries. Terms: Table~\ref{tab:reckoner}.}

\label{tab:rq2-smoothness}
\end{table*}

We test the two premises of Section~\ref{sec:selection} in turn and then compare the accuracy of the two estimators.

\subsubsection{Smoothness: location predicts fidelity}
\label{sec:rq2:smooth}

The smoothness premise says that queries in the same cell have similar fidelity under a given configuration. The synopsis provides two estimates per configuration, the cell mean $\fidhat$ and the coverage $\hat{\pi}$. Selection uses the coverage, while the mean is the most direct test of whether location carries information about fidelity at all. Table~\ref{tab:rq2-smoothness} evaluates both on every held-out query, against the location-blind global average and global coverage.

\paragraph{Location carries information in every setting.} The synopsis is more accurate than the global average in every setting, with skill well above zero everywhere (all confidence intervals exclude zero, and every per-configuration Wilcoxon test is significant). Location alone explains roughly half of the variance in fidelity in a typical setting, and the synopsis orders queries by fidelity moderately well. Local density does not explain it: the density-only estimate has much lower skill than the synopsis in every setting, so the synopsis captures more than how crowded a region is.

The information is partial, however: the other half of the variance lies among queries within the same cell, where the synopsis cannot see it. Whether the synopsis nevertheless chooses configurations well is the question of RQ3.

\paragraph{The synopsis reaches almost all that its partition allows.} The ICC column (the \emph{ceiling}), the share of fidelity variance that lies between cells on the held-out queries themselves, approximates the highest skill any estimate with one value per cell could reach. The synopsis's skill, with cell means estimated on the construction sample, comes within a few hundredths of it in every setting. The cell statistics therefore carry over to new queries almost unchanged: the synopsis is limited not by its sample or by overfitting, but by how finely the space is partitioned and how much fidelity varies within a cell. Consistent with this, skill rises with the number of landmarks until cells become too thin (Section~\ref{sec:rq5}).

\begin{figure}[t]
\centering
\includegraphics[width=\linewidth]{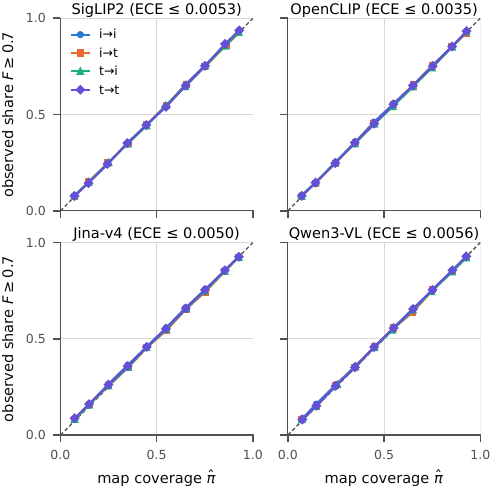}
\caption{Calibration of the estimated coverage ($\tau = 0.7$, $K = 10{,}000$, nine configurations pooled): mean predicted coverage against the share of held-out (query, configuration) pairs that meet the target, in 10 bins over pairs with $0.05 < \hat\pi < 0.95$. Dashed: perfect calibration.}
\label{fig:rq2-reliability}
\end{figure}

\paragraph{The coverage estimate is calibrated.} The coverage that selection relies on is also informative: its Brier skill is positive in every setting, and it separates queries that meet the target from those that do not well (AUC above 0.8 everywhere). It is also well calibrated (Figure~\ref{fig:rq2-reliability}): among queries for which the synopsis predicts a coverage of, say, 0.8, close to 80\% meet the target. Calibration is what makes the reliability level meaningful: if the coverage is calibrated, a selector that controls the expected number of misses controls the actual number.

\paragraph{Location predicts better for uni-modal search.} For SigLIP2, OpenCLIP, and Qwen3-VL, skill is higher for both uni-modal directions than for both cross-modal ones, as the smoothness premise predicts (for Jina-v4, cross-modal skill, 0.461 and 0.475, is as high as its text-to-text skill, 0.461): a cross-modal query is located among the landmarks of the other modality, which represent its surroundings less well. Measured by the mean distance from a query to its nearest landmark, cross-modal queries lie about twice as far as uni-modal ones for SigLIP2, and less far for OpenCLIP, Qwen3-VL, and Jina-v4. Jina-v4, with the smallest gap, on average, has the highest cross-modal skill, but Qwen3-VL, with a smaller gap than SigLIP2 and OpenCLIP, has cross-modal skill as low as theirs, so the gap explains the difference only in part.

\paragraph{Queries near cell boundaries.} The premise is least plausible near the boundary between two cells. We compare \emph{interior} queries, the third whose nearest landmark is much closer than the second-nearest, with \emph{boundary} queries, the third for which the two are almost equally close (bold in Table~\ref{tab:rq2-smoothness}: the better-predicted group). For text-to-text search, interior queries are predicted much better in all four models. For cross-modal search, better but by less. For image-to-image search, skill is higher for boundary queries under OpenCLIP, Jina-v4 and Qwen3-VL. This is an artifact of how skill is normalized: skill in each group is measured against that group's own variance, and interior image queries vary much less. In absolute error, interior queries are predicted more accurately in all 16 settings, as the premise expects.

\begin{figure}[t]
\centering
\includegraphics[width=\linewidth]{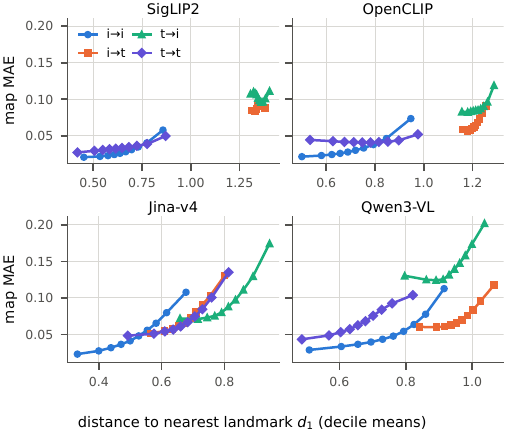}
\caption{Error of the estimated fidelity (mean absolute error over non-flat configurations) against the distance $d_1$ from a held-out query to its nearest landmark, in ten equal-count bins per direction ($K = 10{,}000$).}
\label{fig:rq2-error-d1}
\end{figure}

Figure~\ref{fig:rq2-error-d1} shows the same continuously: in every setting, the synopsis's error is higher for the queries farthest from their nearest landmark than for the nearest ones. The cross-modal queries of SigLIP2 and OpenCLIP lie far from all landmarks, and their error is high throughout.

\begin{table}[t]
\footnotesize\centering
\setlength{\tabcolsep}{3pt}
\begin{tabular}{@{}l rr l@{}}
\toprule
 & \multicolumn{2}{c}{Median} & \\
\cmidrule(lr){2-3}
Metric ($\tau = 0.7$, $K = 10{,}000$) & $k = 10$ & $k = 100$ & $k = 10$ better \\
\midrule
Skill & 0.226 & 0.468 & 0 of 16 \\
ICC (ceiling) & 0.236 & 0.487 & 0 of 16 \\
Spearman $\rho_s$ & 0.392 & 0.662 & 0 of 16 \\
Brier skill (coverage) & 0.131 & 0.236 & 0 of 16 \\
AUC & 0.812 & 0.866 & 0 of 16 \\
ECE & 0.005 & 0.004 & 2 of 16 \\
Unservable queries & 0.3\% & 0.5\% & fewer in 11, more in 5 \\
\midrule
Gap to ceiling (ICC $-$ skill), median & \hi{0.012} & \hi{0.010} & \\
\bottomrule
\end{tabular}
\caption{Accuracy of the synopsis when fidelity is measured at $k = 10$ instead of $k = 100$, with medians over the 16 settings. Every metric but ECE is lower at $k = 10$ in every setting, but so is the ceiling, and the synopsis stays about as close to it (bold: the median over the settings of each setting's gap). Unservable: held-out queries that no configuration serves at $\tau = 0.7$. Terms: Table~\ref{tab:reckoner}.}
\label{tab:k10-accuracy}
\end{table}

\paragraph{At $k = 10$.} Location predicts fidelity less well at $k = 10$ (Table~\ref{tab:k10-accuracy}). Skill, the ceiling, $\rho_s$, Brier skill, and AUC are lower in every setting, with a median skill of 0.23 against 0.47. The ceiling falls just as much, from 0.49 to 0.24, so the synopsis still captures nearly all that its cells allow. The drop is likely due to the coarser measure. Fidelity at $k = 10$ takes only 11 values and so varies more between queries in the same cell. Calibration stays good (ECE below 0.008 in every setting, against below 0.006 at $k = 100$), and fewer queries are unservable in 11 of the 16 settings.

\subsubsection{Configuration specificity: statistics must be kept per configuration}
\label{sec:rq2:specific}

\begin{table}[t]
\small\centering
\setlength{\tabcolsep}{3pt}
\resizebox{\columnwidth}{!}{%
\begin{tabular}{@{}l rl rl rl@{}}
\toprule
 & \multicolumn{2}{c}{HNSW} & \multicolumn{2}{c}{IVFPQ} & \multicolumn{2}{c}{O-IVFPQ} \\
Model & Shared & Keyed [CI of diff.] & Shared & Keyed [CI] & Shared & Keyed [CI] \\
\midrule
SigLIP2 & 0.256 & \hi{0.466 [0.205, 0.216]} & 0.283 & \hi{0.445 [0.128, 0.244]} & 0.278 & \hi{0.442 [0.129, 0.247]}  \\
OpenCLIP & 0.261 & \hi{0.516 [0.244, 0.268]} & 0.167 & \hi{0.451 [0.211, 0.471]} & 0.162 & \hi{0.450 [0.215, 0.474]}  \\
Jina-v4 & 0.252 & \hi{0.430 [0.173, 0.184]} & 0.434 & \hi{0.506 [0.070, 0.074]} & 0.435 & \hi{0.506 [0.069, 0.074]}  \\
Qwen3-VL & $-$0.021 & \hi{0.312 [0.317, 0.352]} & 0.324 & \hi{0.469 [0.141, 0.149]} & 0.324 & \hi{0.469 [0.141, 0.150]}  \\
\bottomrule
\end{tabular}%
}
\caption{Skill of the keyed synopsis (statistics per configuration, bold) and the shared synopsis (one offset per cell), per index family, averaged over directions and non-flat configurations ($K = 10{,}000$). Brackets: 95\% confidence interval of the difference.}

\label{tab:rq3-keyed}
\end{table}

The configuration-specificity premise says that the same region can be easy for one configuration and hard for another, so statistics must be kept for every pair of cell and configuration. We compare the synopsis with a \emph{shared} variant that keeps one difficulty value per cell: the average, over configurations, of how far the cell's mean fidelity lies from the configuration's global mean. The shared synopsis knows which regions are harder or easier than average, but assumes they are equally so for every configuration. Table~\ref{tab:rq3-keyed} reports both per index family.

\paragraph{Keeping the synopsis per configuration is always better.} The keyed synopsis has higher skill than the shared synopsis for every model and index family, with confidence intervals for the difference that exclude zero and every per-configuration test significant. The shared synopsis is not useless, since its skill is positive in all but one combination: some regions are harder than others for every configuration alike. But one value per cell captures only part of what the keyed synopsis captures.

\begin{figure}[t]
\centering
\includegraphics[width=\linewidth]{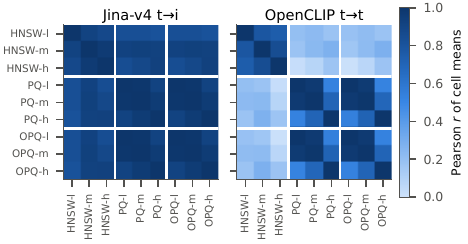}
\caption{Correlation, across cells, of the mean fidelity of every pair of configurations ($K = 10{,}000$, l/m/h: low/moderate/high tier), in the setting where index families agree most (left, Jina-v4 t$\to$i) and least (right, OpenCLIP t$\to$t).}
\label{fig:rq3-similarity}
\end{figure}

\paragraph{The gap follows the index families.} The shared offset averages over all nine configurations, six of which are IVF-family, so it mostly describes how IVF-family configurations behave. For Jina-v4 and Qwen3-VL, the two synopses are therefore close for the IVF families but far apart for HNSW, whose pattern of easy and hard regions differs. For SigLIP2 and OpenCLIP, the gap is large for the IVF families as well: there, HNSW and IVF-family configurations agree much less on which regions are hard (Figure~\ref{fig:rq3-similarity}), so no single offset fits both. Configurations of the same family largely agree with each other. Giving each configuration its own scale on the shared offset narrows the gap but does not close it in any model or family: the regional pattern differs across configurations, not only its size.

\begin{table}[t]
\small\centering
\footnotesize
\setlength{\tabcolsep}{2.5pt}
\begin{tabular}{@{}ll l rr r@{}}
\toprule
Model & Dir. & Pair $A$ / $B$ (Similar cost) & $A$ wins & $B$ wins & \#pairs \\
\midrule
SigLIP2 & i$\to$i & \texttt{hnsw\_high} / \texttt{ivfpq\_mod} & 2.5\% & 0.3\% & 0  \\
 & i$\to$t & \texttt{hnsw\_high} / \texttt{oivfpq\_high} & 3.7\% & 17.6\% & 0  \\
 & t$\to$i & \texttt{hnsw\_high} / \texttt{oivfpq\_high} & 13.7\% & 1.2\% & \hi{2}  \\
 & t$\to$t & \texttt{hnsw\_high} / \texttt{oivfpq\_mod} & 0.0\% & 4.9\% & 0  \\
\midrule
OpenCLIP & i$\to$i & \texttt{hnsw\_high} / \texttt{ivfpq\_mod} & 1.3\% & 0.2\% & 0  \\
 & i$\to$t & \texttt{hnsw\_high} / \texttt{oivfpq\_mod} & 13.1\% & 9.1\% & \hi{2}  \\
 & t$\to$i & \texttt{hnsw\_high} / \texttt{oivfpq\_high} & 9.0\% & 2.0\% & \hi{4}  \\
 & t$\to$t & \texttt{hnsw\_high} / \texttt{oivfpq\_high} & 0.0\% & 64.2\% & 0  \\
\midrule
Jina-v4 & i$\to$i & \texttt{hnsw\_high} / \texttt{oivfpq\_high} & 0.3\% & 0.2\% & 0  \\
 & i$\to$t & \texttt{oivfpq\_mod} / \texttt{hnsw\_mod} & 0.6\% & 20.4\% & 0  \\
 & t$\to$i & \texttt{ivfpq\_mod} / \texttt{hnsw\_mod} & 0.7\% & 15.7\% & 0  \\
 & t$\to$t & \texttt{oivfpq\_mod} / \texttt{hnsw\_mod} & 1.0\% & 10.8\% & 0  \\
\midrule
Qwen3-VL & i$\to$i & \texttt{hnsw\_high} / \texttt{ivfpq\_high} & 0.1\% & 0.3\% & 0  \\
 & i$\to$t & \texttt{oivfpq\_mod} / \texttt{hnsw\_mod} & 0.8\% & 11.4\% & 0  \\
 & t$\to$i & \texttt{hnsw\_high} / \texttt{oivfpq\_high} & 16.3\% & 12.4\% & \hi{4}  \\
 & t$\to$t & \texttt{oivfpq\_mod} / \texttt{hnsw\_mod} & 0.6\% & 21.6\% & 0  \\
\bottomrule
\end{tabular}
\caption{Rank reversals between configurations of different families with similar cost (within $2\times$, $A$ the cheaper), over cells with at least $q_{\min}$ sample queries ($K = 10{,}000$). For each setting: the pair that reverses most often, the share of cells in which each wins by more than 0.1 in mean fidelity, and how many of the 36 pairs reverse in at least 5\% of cells each way (bold: any).}

\label{tab:rq3-flips}
\end{table}

\paragraph{Configurations differ in how much fidelity they lose, not mainly in their order.} One might expect configuration specificity to show up as reversals: regions where configuration $A$ beats $B$ and others where $B$ beats $A$. Table~\ref{tab:rq3-flips} shows that reversals are the exception: in most settings, no pair of the 36 formed by the nine configurations reverses in a substantial share of cells, and the pair that reverses most often is usually lopsided. Two-way reversals occur only in four cross-modal settings (bold). In a given region, configurations largely agree on which of them is better but differ in how much fidelity they lose, and the latter is what selection depends on: whether a configuration meets the target in a region changes with the size of its loss even when the order of configurations does not.

\subsubsection{Accuracy of the MLP}
\label{sec:rq2:mlp}

\begin{figure}[t]
\centering
\includegraphics[width=\linewidth]{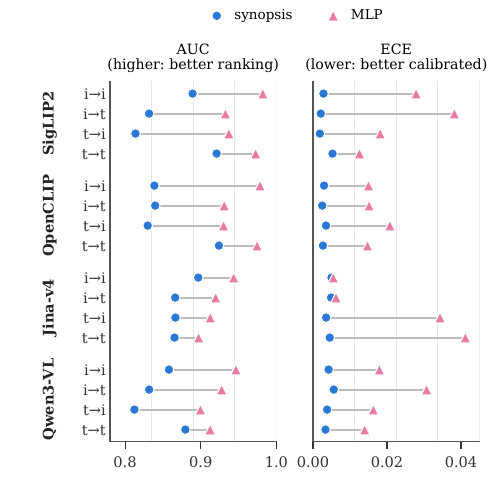}
\caption{Accuracy of the two estimators per setting ($\tau = 0.7$, nine configurations). \keys{(i)~Left: AUC of the estimated probability of meeting the target, averaged over the nine configurations (higher: better ranking). (ii)~Right: expected calibration error over ten bins, for pairs of query and configuration with a predicted probability between 0.05 and 0.95, configurations pooled (lower: better calibrated).}}
\label{fig:rq2-accuracy}
\end{figure}

The MLP is not limited to one value per cell. Per setting, the MLP's skill ranges from 0.49 (Jina-v4 t$\to$t) to 0.72 (OpenCLIP i$\to$i), with a median of 0.62, against 0.47 for the synopsis. It exceeds both the synopsis's skill and its ceiling in all 16 settings. Its AUC is 0.93 at the median (0.90--0.98), against 0.86 for the synopsis, both averaged over all nine configurations. Table~\ref{tab:rq2-smoothness} averages the synopsis's AUC only over the configurations whose fidelity varies across queries, which gives 0.87. Its probabilities are less well calibrated (Figure~\ref{fig:rq2-accuracy}). At $\tau = 0.7$, with the ECE of Section~\ref{sec:setup:metrics} pooled over the nine configurations, the MLP's error is 0.017 at the median (0.005--0.041), against 0.004 (0.002--0.006) for the synopsis, and it is higher in all 16 settings.  Part of the variation within cells is therefore predictable from where in the cell a query lies, but the MLP's probabilities are less reliable as probabilities, which matters for holding the workload to the requirement (Section~\ref{sec:rq3}). A paired Wilcoxon signed-rank test of per-query squared error, with Benjamini--Hochberg correction across the 16 settings, favors the MLP in 15 settings (Table~\ref{tab:mlp-wilcoxon}). In Jina-v4 t$\to$t, the synopsis has the smaller error for slightly more queries (51\%), although the MLP's mean error is lower. Per configuration, 137 of 140 tests are significant, and the MLP's mean error is lower in 136.

\begin{table}[t]
\footnotesize\centering
\setlength{\tabcolsep}{3pt}
\begin{tabular}{@{}ll rr r l@{}}
\toprule
 & & \multicolumn{2}{c}{MSE ($\times 10^{3}$)} & MLP & \\
\cmidrule(lr){3-4}
Model & Dir. & Synopsis & MLP & lower & Favors \\
\midrule
SigLIP2 & i$\to$i & 2.76 & 2.26 & 53.5\% & MLP \\
 & i$\to$t & 15.02 & 7.85 & 69.1\% & MLP \\
 & t$\to$i & 21.39 & 10.55 & 71.2\% & MLP \\
 & t$\to$t & 3.95 & 3.38 & 53.0\% & MLP \\
\midrule
OpenCLIP & i$\to$i & 3.21 & 2.44 & 53.8\% & MLP \\
 & i$\to$t & 8.99 & 5.33 & 64.8\% & MLP \\
 & t$\to$i & 17.29 & 10.27 & 66.2\% & MLP \\
 & t$\to$t & 5.94 & 4.44 & 57.5\% & MLP \\
\midrule
Jina-v4 & i$\to$i & 7.75 & 6.81 & 53.4\% & MLP \\
 & i$\to$t & 13.42 & 10.59 & 55.6\% & MLP \\
 & t$\to$i & 20.16 & 16.34 & 54.1\% & MLP \\
 & t$\to$t & 12.68 & 12.13 & 49.1\% & synopsis \\
\midrule
Qwen3-VL & i$\to$i & 8.96 & 7.18 & 56.8\% & MLP \\
 & i$\to$t & 13.51 & 9.79 & 59.0\% & MLP \\
 & t$\to$i & 39.51 & 27.12 & 63.9\% & MLP \\
 & t$\to$t & 11.92 & 11.43 & 50.5\% & MLP \\
\bottomrule
\end{tabular}
\caption{Paired Wilcoxon signed-rank test of the per-query squared error of the estimated fidelity, MLP against synopsis, on the held-out queries. \keys{(i)~Per query, the squared error is averaged over the configurations that skill uses. (ii)~MSE: mean over the queries. (iii)~MLP lower: share of queries for which the MLP's error is lower. (iv)~Favors: the estimator that the one-sided test finds significantly better, with Benjamini--Hochberg correction across the 16 settings ($q < 0.05$).}}
\label{tab:mlp-wilcoxon}
\end{table}

\paragraph{Answer to RQ2.} Both premises hold. A query's cell predicts its fidelity in every setting, more strongly for uni-modal search and less near cell boundaries, and the synopsis captures almost all of the variation between cells available at its resolution. The synopsis's coverage is calibrated. Keeping estimates per configuration beats one value per cell for every model and index family, most where index families behave differently. Configurations largely agree on which of them is better in a region and disagree on how much fidelity the region costs them, which is what selection depends on. About half of the variance in fidelity lies within cells, invisible to the synopsis. The MLP recovers part of it but is less well calibrated.

\subsection{RQ3: Does plan selection lower cost under the requirement?}
\label{sec:rq3}

\begin{table}[t]
\footnotesize\centering
\setlength{\tabcolsep}{3pt}
\begin{tabular}{@{}l cc cc@{}}
\toprule
 & \multicolumn{2}{c}{Best static plan (11)} & \multicolumn{2}{c}{Static plan of nine} \\
\cmidrule(lr){2-3}\cmidrule(l){4-5}
 & MLP & Synopsis & MLP & Synopsis \\
\midrule
Median cost & \hi{0.59$\times$} & 0.70$\times$ & 0.62$\times$ & 0.72$\times$ \\
Cheaper in & 14 of 16 & 13 of 16 & 16 of 16 & 16 of 16 \\
Above 10\% misses & 0 of 16 & 0 of 16 & 1 of 16 & 0 of 16 \\
\bottomrule
\end{tabular}
\caption{Plan selection at $\tau = 0.7$ against two baselines, with all eleven configurations of the pool (left) or the nine of the full catalog (right). \keys{(i)~Cost: on the held-out queries, relative to the baseline, lookup included, median over the 16 settings. (ii)~Above 10\%: settings whose held-out miss rate exceeds the requirement. (iii)~Bold: our main result. (iv)~One MLP is trained for all eleven configurations. Against the nine, its outputs for the two other search settings of \texttt{hnsw\_high} are unused. Per setting: Table~\ref{tab:small-catalog} (synopsis, union column) and Table~\ref{tab:mlp} (MLP against the nine).}}
\label{tab:headline}
\end{table}

\begin{figure}[t]
\centering
\includegraphics[width=0.85\linewidth]{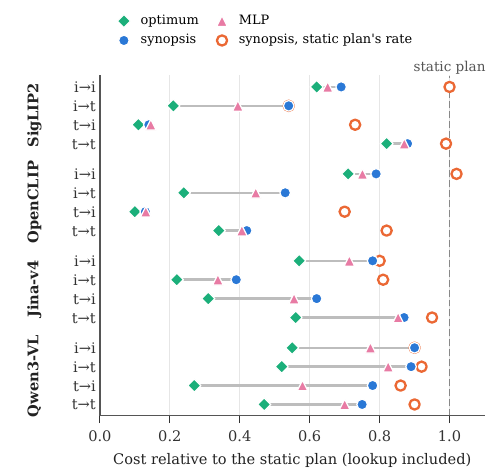}
\caption{Plan selection at $\tau = 0.7$, with costs relative to the static plan of the nine configurations and the lookup included. \keys{(i)~Dashed: the static plan. (ii)~Blue circles and magenta triangles: the budget rule with the synopsis and with the MLP, held to the requirement (at most 10\% misses). (iii)~Green: the optimum under the same requirement, which needs no lookup. (iv)~Grey line: the gap between the synopsis and the optimum. (v)~Orange: the synopsis held to the static plan's own, lower, miss rate, omitted where that rate is unreachable on $Q$.}}
\label{fig:selection}
\end{figure}

\paragraph{Against the best static plan.} Table~\ref{tab:headline} gives the main result. Against the best static choice among all eleven configurations, selection costs a median of 0.59$\times$ with the MLP and 0.70$\times$ with the synopsis. The synopsis is cheaper in 13 settings and equal in one. In the other two, the best static plan is already the cheapest search setting of the \texttt{hnsw\_high} index, and the synopsis costs about its lookup more (Section~\ref{sec:small-catalog}). The MLP is cheaper than the best static plan in 14 settings. Against the best static plan, neither estimator exceeds the requirement in any setting. Table~\ref{tab:mlp} (last two columns) gives the MLP against the best static plan per setting. It is not cheaper only for SigLIP2 and OpenCLIP i$\to$i (1.01$\times$), where the best static plan is the cheapest search setting of the \texttt{hnsw\_high} index. Its held-out miss rate is 6.9--9.9\%.

\paragraph{Against the static plan of the nine.} The rest of this section compares against the static plan of the nine configurations, for which we have per-setting optima and the simpler rules. Figure~\ref{fig:selection} places both estimators between the static plan and the optimum. The MLP costs a median of 0.62$\times$ and is cheaper in all 16 settings. It exceeds the requirement in one, SigLIP2 i$\to$i, with 10.9\% misses (95\% CI [10.7, 11.1]). Its probabilities are not calibrated by construction, so the $\lambda$ tuned on $Q$ carries over less exactly. With $\lambda$ tuned to at most 9.5\% misses on $Q$, the MLP stays within the requirement in all 16 settings, at the same median cost of 0.62$\times$. The MLP is cheaper than the synopsis in 15 of 16 settings (magenta).

\begin{table*}[t]
\footnotesize\centering
\setlength{\tabcolsep}{2.4pt}
\resizebox{\textwidth}{!}{%
\begin{tabular}{@{}ll rr rrr rrr rrr rrr rr@{}}
\toprule
 & & \multicolumn{2}{c}{Static plan} & \multicolumn{3}{c}{Synopsis, mean rule} & \multicolumn{3}{c}{Synopsis, coverage rule ($\rho{=}0.9$)} & \multicolumn{3}{c}{Synopsis, budget (static miss)} & \multicolumn{3}{c}{Synopsis, budget ($\le 1{-}\rho$)} & \multicolumn{2}{c}{Opt. cost} \\
\cmidrule(lr){3-4}\cmidrule(lr){5-7}\cmidrule(lr){8-10}\cmidrule(lr){11-13}\cmidrule(lr){14-16}\cmidrule(l){17-18}
Model & Dir. & Miss & Short. & Cost & Miss & Rec. & Cost & Miss & Rec. & Cost & Miss & Rec. & Cost & Miss [95\% CI] & Rec. & matched & $\le 1{-}\rho$ \\
\midrule
SigLIP2 & i$\to$i & 0.8\% & 0.11 & 0.72$\times$ [0.71, 0.72] & 8.0\% & 89\% & \lo{1.03$\times$ [1.00, 1.05]} & 1.5\% & $-$8\% & \lo{1.00$\times$ [0.99, 1.02]} & 0.8\% & $-$2\% & \hi{0.69$\times$ [0.68, 0.69]} & 9.6\% [9.3, 9.9] & 82\% & 0.68$\times$ & 0.62$\times$  \\
 & i$\to$t & \ex{10.1\%$^\S$} & 0.14 & 0.30$\times$ [0.24, 0.35] & \ex{19.5\%} & 102\% & 0.62$\times$ [0.56, 0.67] & 8.6\% & 55\% & 0.54$\times$ [0.47, 0.60] & 9.9\% & 67\% & \hi{0.54$\times$ [0.47, 0.60]} & 9.9\% [8.5, 11.3] & 59\% & 0.31$\times$ & 0.21$\times$  \\
 & t$\to$i & 2.6\% & 0.11 & 0.18$\times$ [0.17, 0.19] & 9.9\% & 103\% & 0.42$\times$ [0.38, 0.46] & 4.0\%$^\dagger$ & 73\% & 0.73$\times$ [0.71, 0.75] & 2.6\% & 34\% & \hi{0.14$\times$ [0.14, 0.14]} & 9.9\% [9.4, 10.4] & 96\% & 0.20$\times$ & 0.11$\times$  \\
 & t$\to$t & 3.4\% & 0.15 & 0.96$\times$ [0.95, 0.99] & 5.7\% & 29\% & \lo{1.17$\times$ [1.15, 1.20]} & 1.8\% & $-$138\% & 0.99$\times$ [0.99, 1.00] & 3.1\% & 6\% & \hi{0.88$\times$ [0.87, 0.89]} & 9.7\% [9.3, 10.1] & 65\% & 0.87$\times$ & 0.82$\times$  \\
\midrule
OpenCLIP & i$\to$i & 2.5\% & 0.12 & 0.85$\times$ [0.84, 0.87] & 9.0\% & 59\% & \lo{1.54$\times$ [1.50, 1.58]} & 2.1\% & $-$218\% & \lo{1.02$\times$ [1.01, 1.03]} & 2.5\% & $-$9\% & \hi{0.79$\times$ [0.78, 0.79]} & 9.0\% [8.7, 9.2] & 73\% & 0.75$\times$ & 0.71$\times$  \\
 & i$\to$t & 2.8\% & 0.13 & 0.35$\times$ [0.33, 0.36] & \ex{19.6\%} & 98\% & 0.74$\times$ [0.73, 0.75] & 5.2\% & 39\% & 0.96$\times$ [0.95, 0.96]$^\ddagger$ & 2.9\% & 7\% & \hi{0.53$\times$ [0.52, 0.55]} & 9.2\% [8.8, 9.6] & 62\% & 0.33$\times$ & 0.24$\times$  \\
 & t$\to$i & 1.6\% & 0.09 & 0.17$\times$ [0.16, 0.18] & \ex{11.5\%} & 104\% & 0.40$\times$ [0.36, 0.45] & 3.4\% & 75\% & 0.70$\times$ [0.67, 0.74] & 1.6\% & 37\% & \hi{0.13$\times$ [0.13, 0.14]} & 9.9\% [9.4, 10.5] & 96\% & 0.20$\times$ & 0.10$\times$  \\
 & t$\to$t & 0.9\% & 0.11 & 0.44$\times$ [0.43, 0.45] & 9.2\% & 98\% & 0.62$\times$ [0.61, 0.63] & 2.2\% & 66\% & 0.82$\times$ [0.81, 0.83] & 0.9\% & 31\% & \hi{0.42$\times$ [0.41, 0.42]} & 9.3\% [9.0, 9.7] & 89\% & 0.43$\times$ & 0.34$\times$  \\
\midrule
Jina-v4 & i$\to$i & 9.1\% & 0.13 & 0.72$\times$ [0.72, 0.73] & \ex{16.8\%} & 72\% & \lo{1.03$\times$ [1.02, 1.04]} & 4.9\%$^\dagger$ & $-$8\% & 0.80$\times$ [0.79, 0.80] & 9.1\% & 53\% & \hi{0.78$\times$ [0.78, 0.79]} & 9.8\% [9.6, 10.1] & 50\% & 0.62$\times$ & 0.57$\times$  \\
 & i$\to$t & 2.8\% & 0.13 & 0.31$\times$ [0.30, 0.33] & \ex{25.0\%} & 98\% & 0.52$\times$ [0.50, 0.53] & 6.6\%$^\dagger$ & 69\% & 0.81$\times$ [0.80, 0.83] & 2.8\% & 27\% & \hi{0.39$\times$ [0.39, 0.40]} & 9.1\% [8.5, 9.9] & 78\% & 0.30$\times$ & 0.22$\times$  \\
 & t$\to$i & 6.3\% & 0.16 & 0.43$\times$ [0.42, 0.44] & \ex{26.0\%} & 94\% & 0.67$\times$ [0.66, 0.68] & 9.8\%$^\dagger$ & 54\% & \lo{1.04$\times$ [1.03, 1.05]$^\ddagger$} & 6.5\% & $-$6\% & \hi{0.62$\times$ [0.61, 0.63]} & 9.7\% [9.1, 10.2] & 55\% & 0.39$\times$ & 0.31$\times$  \\
 & t$\to$t & 8.4\% & 0.20 & 0.76$\times$ [0.75, 0.76] & \ex{22.3\%} & 63\% & \lo{1.20$\times$ [1.18, 1.22]} & 7.0\%$^\dagger$ & $-$52\% & 0.95$\times$ [0.94, 0.95] & 8.2\% & 14\% & \hi{0.87$\times$ [0.87, 0.88]} & 9.8\% [9.5, 10.1] & 29\% & 0.61$\times$ & 0.56$\times$  \\
\midrule
Qwen3-VL & i$\to$i & 9.9\% & 0.13 & 0.77$\times$ [0.76, 0.78] & \ex{17.7\%} & 59\% & \lo{1.24$\times$ [1.22, 1.26]} & 4.7\%$^\dagger$ & $-$62\% & 0.90$\times$ [0.89, 0.90] & 9.1\% & 27\% & \hi{0.90$\times$ [0.89, 0.90]} & 9.1\% [8.9, 9.3] & 23\% & 0.61$\times$ & 0.55$\times$  \\
 & i$\to$t & 9.4\% & 0.15 & 0.76$\times$ [0.71, 0.84] & \ex{21.5\%} & 59\% & \lo{1.48$\times$ [1.41, 1.56]} & 5.8\%$^\dagger$ & $-$118\% & 0.92$\times$ [0.90, 0.93] & 9.2\% & 21\% & \hi{0.89$\times$ [0.88, 0.91]} & 9.6\% [8.9, 10.6] & 22\% & 0.60$\times$ & 0.52$\times$  \\
 & t$\to$i & 7.4\% & 0.30 & 0.48$\times$ [0.43, 0.53] & \ex{22.6\%} & 83\% & \lo{1.03$\times$ [0.99, 1.10]} & 5.0\% & $-$5\% & 0.86$\times$ [0.84, 0.88] & 7.2\% & 22\% & \hi{0.78$\times$ [0.75, 0.81]} & 8.7\% [8.2, 9.3] & 30\% & 0.37$\times$ & 0.27$\times$  \\
 & t$\to$t & 6.4\% & 0.17 & 0.69$\times$ [0.68, 0.71] & \ex{18.7\%} & 67\% & \lo{1.21$\times$ [1.19, 1.24]} & 5.3\%$^\dagger$ & $-$46\% & 0.90$\times$ [0.89, 0.92] & 6.1\% & 21\% & \hi{0.75$\times$ [0.74, 0.75]} & 9.6\% [9.4, 9.9] & 48\% & 0.54$\times$ & 0.47$\times$  \\
\bottomrule
\end{tabular}%
}
\caption{Plan selection at $\tau = 0.7$ ($K = 10{,}000$). Costs are relative to the static plan, include the lookup, and carry 95\% confidence intervals. The optimum needs no lookup. Bold: our method, the budget rule at the requirement. Grey: not cheaper than the static plan. Italic: miss rate above the requirement of 10\%. Terms and markers: Table~\ref{tab:reckoner}.}
\label{tab:rq4-selection}
\end{table*}

\begin{figure*}[t]
\centering
\includegraphics[width=\textwidth]{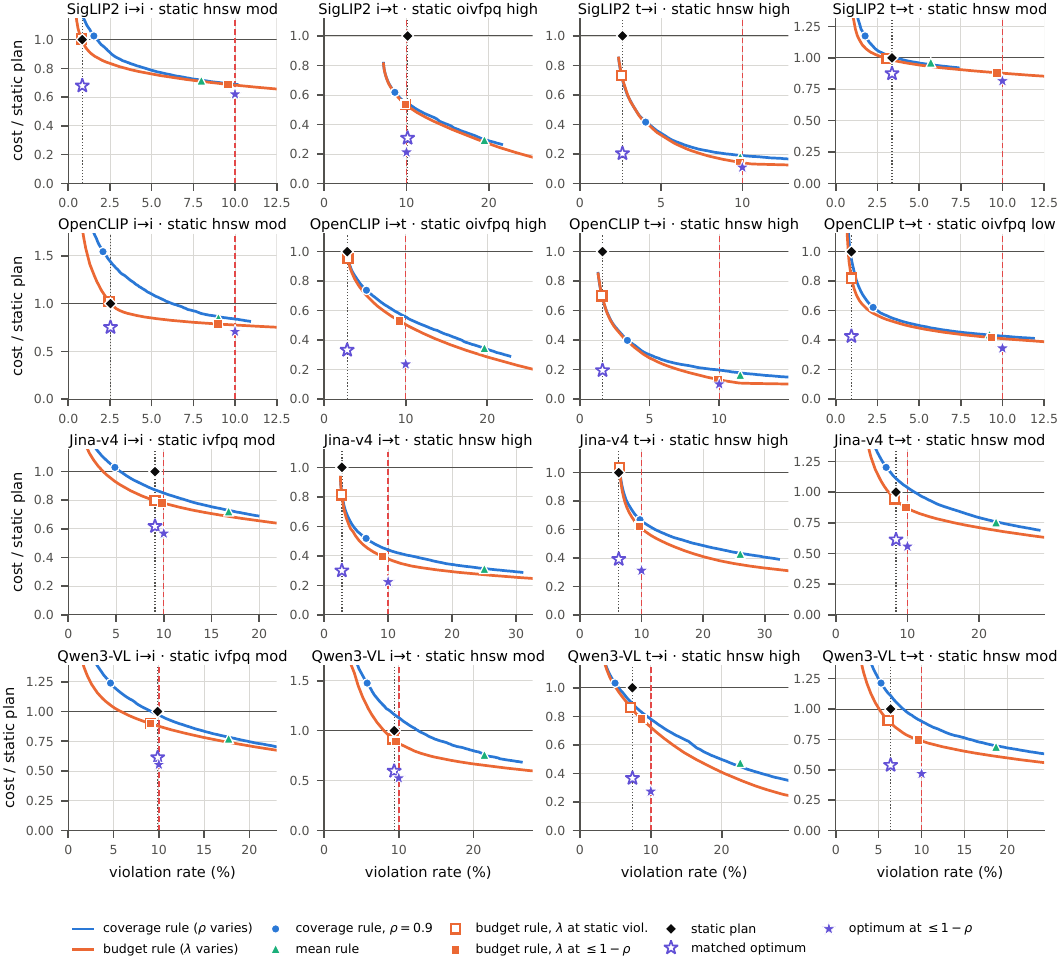}
\caption{Cost against miss rate on the held-out queries ($\tau = 0.7$, $K = 10{,}000$) as $\rho$ (coverage rule) or $\lambda$ (budget rule) varies. Markers: static plan, mean rule, coverage rule at $\rho = 0.9$, budget rule at the static rate (open square) and at the requirement (filled square), matched optimum (open star), and optimum at the requirement (filled star). Dotted: the static plan's miss rate. Dashed: the requirement.}
\label{fig:rq4-tradeoff}
\end{figure*}

\paragraph{The synopsis in detail.} Table~\ref{tab:rq4-selection} compares the static plan of the nine with four ways of using the synopsis at $\tau = 0.7$: the mean rule, the coverage rule, and the budget rule at the static rate and at the requirement. Figure~\ref{fig:rq4-tradeoff} shows, for each setting, the full trade-off between cost and miss rate as the coverage rule's $\rho$ or the budget rule's $\lambda$ varies.

\begin{figure}[t]
\centering
\includegraphics[width=\linewidth]{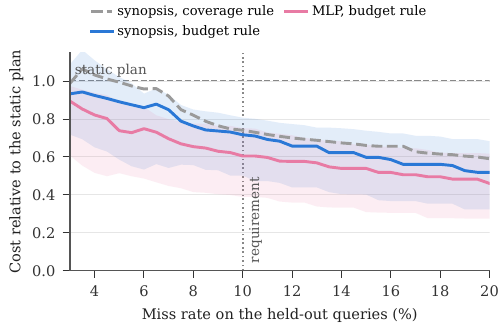}
\caption{Cost against miss rate on the held-out queries ($\tau = 0.7$), relative to the static plan of the nine, lookup included: the median over the 16 settings. \keys{(i)~Bands: the middle half of the settings. (ii)~Each point uses the cheapest $\lambda$ (or $\rho$) that reaches that miss rate. (iii)~At a given miss rate, the median is over the settings that can reach it, at least 12 of 16 from 3\%. (iv)~Vertical line: the requirement. Horizontal line: the static plan.}}
\label{fig:tradeoff-median}
\end{figure}

\paragraph{Under the requirement, the synopsis lowers cost in every setting.} Held to the requirement, at most 10\% of queries missing the target, the budget rule is cheaper than the static plan in all 16 settings, at a median of 0.72$\times$ its cost, and recovers more than half of the saving that the optimum at the requirement achieves in a typical setting. In one setting, SigLIP2 i$\to$t, no single configuration meets the requirement on $Q$, and the static plan misses it on 10.1\% of held-out queries. There, the budget rule is both cheaper and within the requirement, with 9.9\% misses (95\% CI [8.5, 11.3]), and in the other 15 settings, both meet it. Its miss rate on the held-out queries stays within the requirement in every setting. In half of them, the upper end of its confidence interval exceeds 10\%, by at most a little over one percentage point. The comparison is fair because both selectors face the same requirement: the static plan misses less often than it is allowed to only because a catalog offers few configurations, and the synopsis spends that unused allowance on cheaper configurations where its estimates say they are safe. However, this saving is not merely the allowance spent blindly, as the comparison at the static plan's own miss rate below shows. Figure~\ref{fig:rq4-tradeoff} places each selector on the full trade-off curve, so that the comparison can also be read at any other miss rate. Every point on a curve corresponds to a different reliability level. Figure~\ref{fig:tradeoff-median} summarizes the curves over the 16 settings. At every miss rate from 3\% to 20\%, the median MLP is cheaper than the synopsis, and the synopsis's budget rule is cheaper than its coverage rule. An operator who needs $\rho = 0.95$, for instance, reads the cost at a 5\% miss rate. Our results are therefore not tied to $\rho = 0.9$.

\paragraph{At the static plan's own miss rate, it still saves in most settings.} Held to the lower miss rate the static plan actually incurs, the budget rule has less room. Where that rate is reachable on $Q$ (14 settings), it meets it on the held-out queries and is cheaper than the static plan in 12, by at least 8\% in 10. By McNemar's test with Benjamini--Hochberg correction, it misses significantly fewer queries than the static plan in 10 of these settings, and the difference is not significant in the other 4. It recovers a smaller share of the matched optimum's saving, about a quarter in a typical setting. This comparison also separates the synopsis's information from the allowance. At the static plan's own miss rate, a deployment that ignores where a query lies has no allowance left to spend. Every configuration cheaper than the static plan misses the target on more than 10\% of queries. Sending a share of queries to one of them at random would therefore raise the miss rate above the static plan's. The saving at this rate comes from the synopsis's estimates, not from the allowance. A random blend of a cheaper and a costlier configuration could in principle match the static plan's miss rate at lower cost, if the costlier configuration is accurate enough. Such a blend uses no information about the query either. Chosen on $Q$ to match the static plan's miss rate, it is cheaper than the static plan in only one setting (Jina-v4 i$\to$i, 0.95$\times$).

\paragraph{The simpler rules fall short.} The mean rule is cheap only because it misses often: a cell mean at the target leaves many of the cell's queries below it, and the rule misses the target on more than 10\% of queries in most settings (italic). The coverage rule controls misses, as its calibration predicts, but not cost: requiring coverage $\rho$ in every cell rather than over the workload moves cells where the static plan's configuration falls short to costlier ones, and it costs more than the static plan in nine settings (grey). Against the mean rule, its per-query shortfall is significantly lower in every setting (Wilcoxon--Pratt, Benjamini--Hochberg $q < 0.05$), mostly because it misses far fewer queries. The queries it does miss fall less far below the target in half of the settings ($\dagger$). The budget rule's curve in Figure~\ref{fig:rq4-tradeoff} lies on or below the coverage rule's in most settings, so for a given miss budget it is the better way to use the synopsis. All rules differ significantly from the static plan in which queries they miss (McNemar), in all but three of the 64 comparisons.

\paragraph{Where the synopsis gains little.} At the static rate, the synopsis saves nothing for SigLIP2 and OpenCLIP i$\to$i, although the optimum would save a quarter to a third: the only cheaper configuration, \texttt{hnsw\_low}, serves part of the queries, and the synopsis cannot single out which ones reliably enough. More generally, half of the variation in fidelity lies within cells (RQ2), where the synopsis cannot see it, which is why the budget rule captures only part of the optimum's saving.

\paragraph{Other targets.} Table~\ref{tab:targets} repeats the evaluation at $\tau = 0.8$ and $0.9$. A higher target leaves fewer feasible settings. Even the optimum can meet the requirement in only 15 settings at $\tau = 0.8$ and 9 at $\tau = 0.9$ (infeas.). In the others, more than $1 - \rho$ of the queries miss the target under every configuration. The budget rule reaches the requirement on $Q$ in 14 and 6 settings. It cannot reach it where the optimum cannot, and in four more settings: Jina-v4 t$\to$i at $\tau = 0.8$, and OpenCLIP t$\to$i and Qwen3-VL i$\to$t and t$\to$t at $\tau = 0.9$. There, the optimum is feasible, but the synopsis's estimates are not precise enough to reach $1 - \rho$ ($\ddagger$).

Where the budget rule reaches the requirement, its miss rate on the held-out queries ranges from 9.3 to 9.9\% at $\tau = 0.8$ and from 8.5 to 9.7\% at $\tau = 0.9$. At $\tau = 0.8$, it is cheaper than the static plan in 13 of 14 settings. In the remaining one, SigLIP2 t$\to$t, it matches the static plan (0.999$\times$) and recovers almost none of the optimum's saving. At $\tau = 0.9$, it is cheaper in all 6 settings. Its median cost is 0.73$\times$ at $\tau = 0.8$, close to the 0.72$\times$ at $\tau = 0.7$, and 0.45$\times$ at $\tau = 0.9$. It recovers a median of 56\% and 72\% of the saving of the optimum at the requirement, which itself costs a median of 0.44$\times$ and 0.21$\times$ the static plan. The six feasible settings at $\tau = 0.9$ are all uni-modal, so medians are not directly comparable across targets.

A higher target moves the static plan to costlier configurations. \texttt{hnsw\_high} is the static plan in 5 settings at $\tau = 0.7$, 8 at $\tau = 0.8$, and 10 at $\tau = 0.9$, and a high IVF-family tier in 2, 2, and 5. Most queries do not need these configurations: at $\tau = 0.9$, the optimum costs as little as 0.12$\times$ the static plan for SigLIP2 t$\to$t and OpenCLIP i$\to$i. The saving therefore does not depend on our default target.

\begin{table*}[t]
\footnotesize\centering
\setlength{\tabcolsep}{3pt}
\resizebox{\textwidth}{!}{%
\begin{tabular}{@{}ll lrrrrr lrrrrr@{}}
\toprule
 & & \multicolumn{6}{c}{$\tau = 0.8$} & \multicolumn{6}{c}{$\tau = 0.9$} \\
\cmidrule(lr){3-8}\cmidrule(l){9-14}
 & & \multicolumn{2}{c}{Static plan} & \multicolumn{3}{c}{Budget, $\le 1{-}\rho$} & Opt. & \multicolumn{2}{c}{Static plan} & \multicolumn{3}{c}{Budget, $\le 1{-}\rho$} & Opt. \\
\cmidrule(lr){3-4}\cmidrule(lr){5-7}\cmidrule(lr){8-8}\cmidrule(lr){9-10}\cmidrule(lr){11-13}\cmidrule(l){14-14}
Model & Dir. & Config. & Miss & Cost [95\% CI] & Miss [95\% CI] & Rec. & $\le 1{-}\rho$ & Config. & Miss & Cost [95\% CI] & Miss [95\% CI] & Rec. & $\le 1{-}\rho$ \\
\midrule
SigLIP2 & i$\to$i & \texttt{hnsw\_mod} & 2.4\% & \hi{0.83$\times$ [0.82, 0.84]} & 9.5\% [9.2, 9.8] & 68\% & 0.75$\times$ & \texttt{hnsw\_mod} & 8.5\% & \hi{0.98$\times$ [0.98, 0.99]} & 9.6\% [9.2, 10.0] & 14\% & 0.88$\times$ \\
 & i$\to$t & \texttt{oivfpq\_high}$^*$ & \ex{21.5\%} & 0.88$\times$ [0.86, 0.90]$^\ddagger$ & \ex{17.8\% [15.1, 20.4]} & -- & infeas. & \texttt{oivfpq\_high}$^*$ & \ex{45.4\%} & 0.93$\times$ [0.92, 0.94]$^\ddagger$ & \ex{42.4\% [38.7, 46.1]} & -- & infeas. \\
 & t$\to$i & \texttt{hnsw\_high} & 6.6\% & \hi{0.44$\times$ [0.41, 0.48]} & 9.6\% [9.1, 10.0] & 69\% & 0.20$\times$ & \texttt{hnsw\_high}$^*$ & \ex{19.8\%} & \lo{1.04$\times$ [1.04, 1.05]$^\ddagger$} & \ex{17.9\% [17.0, 18.8]} & -- & infeas. \\
 & t$\to$t & \texttt{hnsw\_mod} & 9.5\% & \lo{1.00$\times$ [1.00, 1.00]} & 9.6\% [9.2, 10.0] & 1\% & 0.88$\times$ & \texttt{oivfpq\_high} & 0.6\% & \hi{0.32$\times$ [0.31, 0.32]} & 9.7\% [9.4, 9.9] & 78\% & 0.12$\times$ \\
\midrule
OpenCLIP & i$\to$i & \texttt{hnsw\_mod} & 5.7\% & \hi{0.91$\times$ [0.91, 0.91]} & 9.5\% [9.2, 9.8] & 45\% & 0.80$\times$ & \texttt{hnsw\_high} & 5.7\% & \hi{0.18$\times$ [0.18, 0.19]} & 8.9\% [8.6, 9.2] & 92\% & 0.12$\times$ \\
 & i$\to$t & \texttt{oivfpq\_high} & 6.7\% & \hi{0.85$\times$ [0.84, 0.86]} & 9.9\% [9.2, 10.6] & 32\% & 0.52$\times$ & \texttt{oivfpq\_high}$^*$ & \ex{17.9\%} & \lo{1.00$\times$ [1.00, 1.00]$^\ddagger$} & \ex{18.0\% [16.8, 19.2]} & -- & infeas. \\
 & t$\to$i & \texttt{hnsw\_high} & 5.3\% & \hi{0.40$\times$ [0.36, 0.44]} & 9.5\% [8.4, 11.0] & 76\% & 0.20$\times$ & \texttt{hnsw\_high}$^*$ & \ex{17.7\%} & \lo{1.09$\times$ [1.08, 1.10]$^\ddagger$} & \ex{13.3\% [12.6, 14.0]} & $-$15\% & 0.44$\times$ \\
 & t$\to$t & \texttt{oivfpq\_low} & 8.4\% & \hi{0.62$\times$ [0.61, 0.63]} & 9.6\% [9.3, 9.9] & 72\% & 0.47$\times$ & \texttt{oivfpq\_high} & 0.4\% & \hi{0.57$\times$ [0.56, 0.58]} & 8.5\% [8.1, 8.7] & 55\% & 0.22$\times$ \\
\midrule
Jina-v4 & i$\to$i & \texttt{hnsw\_mod} & 8.3\% & \hi{0.94$\times$ [0.94, 0.95]} & 9.3\% [9.1, 9.6] & 14\% & 0.60$\times$ & \texttt{hnsw\_high} & 5.2\% & \hi{0.56$\times$ [0.55, 0.56]} & 9.7\% [9.4, 10.0] & 66\% & 0.33$\times$ \\
 & i$\to$t & \texttt{hnsw\_high} & 6.7\% & \hi{0.64$\times$ [0.62, 0.65]} & 9.4\% [8.9, 10.1] & 53\% & 0.32$\times$ & \texttt{hnsw\_high}$^*$ & \ex{17.2\%} & \lo{1.07$\times$ [1.07, 1.08]$^\ddagger$} & \ex{16.2\% [15.2, 17.3]} & -- & infeas. \\
 & t$\to$i & \texttt{hnsw\_high}$^*$ & \ex{12.1\%} & \lo{1.12$\times$ [1.11, 1.13]$^\ddagger$} & \ex{11.9\% [11.0, 12.8]} & $-$21\% & 0.42$\times$ & \texttt{hnsw\_high}$^*$ & \ex{25.3\%} & \lo{1.21$\times$ [1.21, 1.22]$^\ddagger$} & \ex{23.5\% [22.4, 24.7]} & -- & infeas. \\
 & t$\to$t & \texttt{hnsw\_high} & 6.8\% & \hi{0.58$\times$ [0.57, 0.59]} & 9.7\% [9.5, 10.0] & 59\% & 0.29$\times$ & \texttt{hnsw\_high}$^*$ & \ex{17.7\%} & \lo{1.12$\times$ [1.12, 1.13]$^\ddagger$} & \ex{16.3\% [15.8, 16.8]} & -- & infeas. \\
\midrule
Qwen3-VL & i$\to$i & \texttt{hnsw\_mod} & 6.0\% & \hi{0.83$\times$ [0.83, 0.83]} & 9.9\% [9.6, 10.2] & 35\% & 0.52$\times$ & \texttt{hnsw\_high} & 3.0\% & \hi{0.34$\times$ [0.34, 0.35]} & 9.6\% [9.3, 9.8] & 83\% & 0.21$\times$ \\
 & i$\to$t & \texttt{hnsw\_high} & 4.9\% & \hi{0.54$\times$ [0.52, 0.56]} & 9.6\% [9.3, 10.0] & 61\% & 0.24$\times$ & \texttt{ivfpq\_high}$^*$ & \ex{11.5\%} & 0.82$\times$ [0.81, 0.83]$^\ddagger$ & \ex{11.1\% [10.3, 12.3]} & 24\% & 0.23$\times$ \\
 & t$\to$i & \texttt{hnsw\_high}$^*$ & \ex{11.4\%} & \hi{0.97$\times$ [0.96, 0.98]} & 9.5\% [8.9, 10.1] & 5\% & 0.40$\times$ & \texttt{hnsw\_high}$^*$ & \ex{26.2\%} & \lo{1.26$\times$ [1.22, 1.30]$^\ddagger$} & \ex{20.0\% [19.2, 20.8]} & -- & infeas. \\
 & t$\to$t & \texttt{hnsw\_high} & 3.3\% & \hi{0.33$\times$ [0.32, 0.34]} & 9.3\% [9.1, 9.6] & 80\% & 0.17$\times$ & \texttt{hnsw\_high}$^*$ & \ex{12.0\%} & \lo{1.18$\times$ [1.17, 1.19]$^\ddagger$} & \ex{10.8\% [10.4, 11.1]} & $-$27\% & 0.31$\times$ \\
\midrule
\multicolumn{2}{@{}l}{Median (feasible)} & & & \hi{0.73$\times$} & & 56\% & 0.44$\times$ & & & \hi{0.45$\times$} & & 72\% & 0.21$\times$ \\
\bottomrule
\end{tabular}%
}
\caption{Plan selection at $\tau = 0.8$ and $0.9$ ($K = 10{,}000$, held-out queries), in the format of Table~\ref{tab:rq4-selection}. Costs are relative to the static plan at the same target and include the lookup. Bold: our method. Grey: not cheaper than the static plan. Italic: miss rate above the requirement of 10\%. $^*$: no configuration meets the target on a fraction $\rho$ of $Q$, so the static plan is the one that meets it most often. $^\ddagger$: the budget rule cannot reach $1 - \rho$ on $Q$, so the $\lambda$ with the fewest misses is used. Rec.: the share of the saving of the optimum at the requirement that the budget rule achieves. infeas.: more than $1 - \rho$ of the queries cannot be served by any configuration. Medians are over the feasible settings. Terms: Table~\ref{tab:reckoner}.}
\label{tab:targets}
\end{table*}

\paragraph{A learned estimator.} Table~\ref{tab:mlp} repeats the selection with the MLP of Section~\ref{sec:mlp} in place of the synopsis, with $\lambda$ tuned to the requirement or to a 9.5\% limit. With $\lambda$ tuned to 10\%, the MLP costs a median of 0.62$\times$ the static plan of the nine, against 0.72$\times$ for the synopsis, and is cheaper than the static plan in all 16 settings. It exceeds the requirement on the held-out queries in one setting (SigLIP2 i$\to$i, 10.9\%, 95\% CI [10.7, 11.1]). Tuned to 9.5\%, it costs the same median of 0.62$\times$ and stays within the requirement in all 16 settings. The synopsis stays within the requirement in both cases. Against the best static choice among all eleven configurations, the MLP costs a median of 0.59$\times$, is cheaper in 14 of 16 settings, and stays within the requirement in all 16. At $\tau = 0.8$, it costs 0.65$\times$ the static plan of the nine (synopsis 0.73$\times$), and at $\tau = 0.9$, 0.40$\times$ (synopsis 0.45$\times$), feasible in the same 14 and 6 settings. The MLP reaches the requirement on $Q$ in the same settings as the synopsis. With $\lambda$ tuned to 10\%, its held-out miss rates at $\tau = 0.8$ are 9.6\%, 9.9\% (SigLIP2 t$\to$i, t$\to$t), 9.2\%, 8.8\%, 9.99\%, 9.5\% (OpenCLIP), 8.7\%, 9.6\%, 9.0\% (Jina-v4 i$\to$i, i$\to$t, t$\to$t), and 9.1\%, 8.5\%, 9.0\%, 9.0\% (Qwen3-VL), and 11.3\% for SigLIP2 i$\to$i. At $\tau = 0.9$, they are 9.9\% and 9.4\% (SigLIP2 i$\to$i, t$\to$t), 7.1\% and 8.5\% (OpenCLIP i$\to$i, t$\to$t), 9.2\% (Jina-v4 i$\to$i), and 8.1\% (Qwen3-VL i$\to$i). They stay within 10\% except for SigLIP2 i$\to$i at $\tau = 0.8$, the setting that also exceeds it at $\tau = 0.7$.

\begin{table}[t]
\footnotesize\centering
\setlength{\tabcolsep}{2.5pt}
\resizebox{\columnwidth}{!}{%
\begin{tabular}{@{}ll rr rr rr rr rr@{}}
\toprule
 & & \multicolumn{4}{c}{$\lambda$ to 10\%} & \multicolumn{4}{c}{$\lambda$ to 9.5\%} & \multicolumn{2}{c}{Best static of eleven} \\
\cmidrule(lr){3-6}\cmidrule(lr){7-10}\cmidrule(l){11-12}
 & & \multicolumn{2}{c}{Synopsis} & \multicolumn{2}{c}{MLP} & \multicolumn{2}{c}{Synopsis} & \multicolumn{2}{c}{MLP} & \multicolumn{2}{c}{MLP, 10\%} \\
Model & Dir. & Cost & Miss & Cost & Miss & Cost & Miss & Cost & Miss & Cost & Miss \\
\midrule
SigLIP2 & i$\to$i & 0.69$\times$ & 9.6\% & \hi{0.65$\times$} & \textit{10.9\%} & 0.71$\times$ & 8.2\% & \hi{0.67$\times$} & 9.5\% & \hi{1.01$\times$} & 6.9\% \\
 & i$\to$t & 0.54$\times$ & 9.9\% & \hi{0.39$\times$} & 8.5\% & 0.55$\times$ & 9.5\% & \hi{0.39$\times$} & 8.5\% & \hi{0.39$\times$} & 8.6\% \\
 & t$\to$i & 0.14$\times$ & 9.9\% & \hi{0.15$\times$} & 9.2\% & 0.15$\times$ & 9.3\% & \hi{0.16$\times$} & 8.1\% & \hi{0.12$\times$} & 9.3\% \\
 & t$\to$t & 0.88$\times$ & 9.7\% & \hi{0.87$\times$} & 8.7\% & 0.90$\times$ & 8.1\% & \hi{0.87$\times$} & 8.7\% & \hi{0.32$\times$} & 9.4\% \\
\midrule
OpenCLIP & i$\to$i & 0.79$\times$ & 9.0\% & \hi{0.75$\times$} & 9.7\% & 0.79$\times$ & 9.0\% & \hi{0.75$\times$} & 9.7\% & \hi{1.01$\times$} & 7.9\% \\
 & i$\to$t & 0.53$\times$ & 9.2\% & \hi{0.45$\times$} & 8.0\% & 0.53$\times$ & 9.2\% & \hi{0.47$\times$} & 7.3\% & \hi{0.44$\times$} & 8.1\% \\
 & t$\to$i & 0.13$\times$ & 9.9\% & \hi{0.13$\times$} & 9.4\% & 0.15$\times$ & 9.1\% & \hi{0.16$\times$} & 7.5\% & \hi{0.79$\times$} & 9.9\% \\
 & t$\to$t & 0.42$\times$ & 9.3\% & \hi{0.41$\times$} & 8.9\% & 0.42$\times$ & 9.3\% & \hi{0.41$\times$} & 8.9\% & \hi{0.29$\times$} & 9.3\% \\
\midrule
Jina-v4 & i$\to$i & 0.78$\times$ & 9.8\% & \hi{0.71$\times$} & 8.6\% & 0.80$\times$ & 9.1\% & \hi{0.71$\times$} & 8.6\% & \hi{0.70$\times$} & 9.1\% \\
 & i$\to$t & 0.39$\times$ & 9.1\% & \hi{0.34$\times$} & 9.8\% & 0.39$\times$ & 9.1\% & \hi{0.35$\times$} & 8.8\% & \hi{0.32$\times$} & 9.4\% \\
 & t$\to$i & 0.62$\times$ & 9.7\% & \hi{0.56$\times$} & 9.4\% & 0.64$\times$ & 9.2\% & \hi{0.57$\times$} & 8.9\% & \hi{0.55$\times$} & 9.1\% \\
 & t$\to$t & 0.87$\times$ & 9.8\% & \hi{0.85$\times$} & 8.5\% & 0.90$\times$ & 9.0\% & \hi{0.85$\times$} & 8.5\% & \hi{0.77$\times$} & 8.3\% \\
\midrule
Qwen3-VL & i$\to$i & 0.90$\times$ & 9.1\% & \hi{0.77$\times$} & 8.7\% & 0.90$\times$ & 9.1\% & \hi{0.79$\times$} & 7.8\% & \hi{0.60$\times$} & 9.2\% \\
 & i$\to$t & 0.89$\times$ & 9.6\% & \hi{0.82$\times$} & 8.3\% & 0.92$\times$ & 9.2\% & \hi{0.82$\times$} & 8.3\% & \hi{0.78$\times$} & 8.7\% \\
 & t$\to$i & 0.78$\times$ & 8.7\% & \hi{0.58$\times$} & 8.6\% & 0.78$\times$ & 8.7\% & \hi{0.58$\times$} & 8.6\% & \hi{0.58$\times$} & 8.7\% \\
 & t$\to$t & 0.75$\times$ & 9.6\% & \hi{0.70$\times$} & 9.2\% & 0.76$\times$ & 9.2\% & \hi{0.71$\times$} & 8.8\% & \hi{0.62$\times$} & 9.2\% \\
\midrule
\multicolumn{2}{@{}l}{Median} & 0.72$\times$ & 9.6\% & \hi{0.62$\times$} & 8.8\% & 0.73$\times$ & 9.1\% & \hi{0.62$\times$} & 8.6\% & \hi{0.59$\times$} & 9.1\% \\
\bottomrule
\end{tabular}%
}
\caption{Plan selection with the synopsis and with the MLP ($\tau = 0.7$, held-out queries), with $\lambda$ tuned to the requirement or to 9.5\%. Costs are relative to the static plan of the nine configurations and include the lookup. Bold: the MLP's cost. Italic: miss rate above 10\%. Last two columns: the MLP choosing among all eleven configurations, with $\lambda$ tuned to 10\%, relative to the best static plan of the eleven.}
\label{tab:mlp}
\end{table}

\subsubsection{Smaller catalogs and the best static plan}
\label{sec:small-catalog}

\begin{table*}[t]
\footnotesize\centering
\setlength{\tabcolsep}{2.4pt}
\resizebox{\textwidth}{!}{%
\begin{tabular}{@{}ll lrrrr lrrrr lrrrr lrrrr@{}}
\toprule
 & & \multicolumn{5}{c}{Full catalog (nine indexes)} & \multicolumn{5}{c}{HNSW-only (three indexes)} & \multicolumn{5}{c}{Single index (three \texttt{efSearch})} & \multicolumn{5}{c}{Union (eleven configurations)} \\
\cmidrule(lr){3-7}\cmidrule(lr){8-12}\cmidrule(lr){13-17}\cmidrule(l){18-22}
 & & & & \multicolumn{2}{c}{Budget, $\le 1{-}\rho$} & Budget & & & \multicolumn{2}{c}{Budget, $\le 1{-}\rho$} & Budget & & & \multicolumn{2}{c}{Budget, $\le 1{-}\rho$} & Budget & & & \multicolumn{2}{c}{Budget, $\le 1{-}\rho$} & Budget \\
\cmidrule(lr){5-6}\cmidrule(lr){10-11}\cmidrule(lr){15-16}\cmidrule(lr){20-21}
Model & Dir. & Static & Opt. & Cost & Miss & static v. & Static & Opt. & Cost & Miss & static v. & Static & Opt. & Cost & Miss & static v. & Static & Opt. & Cost & Miss & / static \\
\midrule
SigLIP2 & i$\to$i & H-m & 0.62$\times$ & 0.69$\times$ & 9.6\% & 1.00$\times$ & H-m & 0.62$\times$ & 0.69$\times$ & 9.6\% & 1.01$\times$ & l & 0.21$\times$ & 0.22$\times$ & 6.9\% & 0.22$\times$ & S-l & 0.21$\times$ & 0.22$\times$ & 6.9\% & \lo{1.07$\times$}  \\
 & i$\to$t & O-h$^*$ & 0.21$\times$ & 0.54$\times$ & 9.9\% & 0.54$\times$ & H-h$^*$ & infeas. & 0.65$^\ddagger$$\times$ & 12.1\% & 0.65$^\ddagger$$\times$ & h$^*$ & infeas. & 0.68$^\ddagger$$\times$ & 12.1\% & 0.68$^\ddagger$$\times$ & O-h$^*$ & 0.18$\times$ & 0.54$\times$ & 9.9\% & \hi{0.54$\times$}  \\
 & t$\to$i & H-h & 0.11$\times$ & 0.14$\times$ & 9.9\% & 0.73$\times$ & H-h & 0.11$\times$ & 0.15$\times$ & 9.5\% & 0.81$^\ddagger$$\times$ & h & 0.16$\times$ & 0.38$\times$ & 9.9\% & 0.98$^\ddagger$$\times$ & H-h & 0.04$\times$ & 0.15$\times$ & 9.4\% & \hi{0.15$\times$}  \\
 & t$\to$t & H-m & 0.82$\times$ & 0.88$\times$ & 9.7\% & 0.99$\times$ & H-m & 0.82$\times$ & 0.88$\times$ & 9.7\% & 1.01$\times$ & h & 0.28$\times$ & 0.62$\times$ & 9.5\% & 6.55$^\ddagger$$\times$ & H-m & 0.22$\times$ & 0.35$\times$ & 9.8\% & \hi{0.35$\times$}  \\
\midrule
OpenCLIP & i$\to$i & H-m & 0.71$\times$ & 0.79$\times$ & 9.0\% & 1.02$\times$ & H-m & 0.71$\times$ & 0.79$\times$ & 9.0\% & 1.04$\times$ & l & 0.32$\times$ & 0.34$\times$ & 7.9\% & 0.34$\times$ & S-l & 0.32$\times$ & 0.34$\times$ & 7.9\% & \lo{1.08$\times$}  \\
 & i$\to$t & O-h & 0.24$\times$ & 0.53$\times$ & 9.2\% & 0.96$^\ddagger$$\times$ & H-h$^*$ & infeas. & 0.46$^\ddagger$$\times$ & 25.3\% & 0.19$\times$ & h$^*$ & infeas. & 0.61$^\ddagger$$\times$ & 30.8\% & 0.61$^\ddagger$$\times$ & O-h & 0.20$\times$ & 0.52$\times$ & 9.4\% & \hi{0.52$\times$}  \\
 & t$\to$i & H-h & 0.10$\times$ & 0.13$\times$ & 9.9\% & 0.70$\times$ & H-h & 0.10$\times$ & 0.15$\times$ & 9.3\% & 0.80$^\ddagger$$\times$ & m & 0.05$\times$ & 0.13$\times$ & 9.6\% & 0.14$\times$ & S-m & 0.05$\times$ & 0.13$\times$ & 9.5\% & \lo{1.00$\times$}  \\
 & t$\to$t & O-l & 0.34$\times$ & 0.42$\times$ & 9.3\% & 0.82$\times$ & H-m$^*$ & 0.39$\times$ & 1.08$^\ddagger$$\times$ & 10.7\% & 0.40$\times$ & h$^*$ & infeas. & 2.95$^\ddagger$$\times$ & 20.3\% & 2.95$^\ddagger$$\times$ & O-l & 0.18$\times$ & 0.32$\times$ & 9.7\% & \hi{0.32$\times$}  \\
\midrule
Jina-v4 & i$\to$i & I-m & 0.57$\times$ & 0.78$\times$ & 9.8\% & 0.80$\times$ & H-m & 0.63$\times$ & 0.81$\times$ & 9.9\% & 1.07$\times$ & m & 0.44$\times$ & 0.60$\times$ & 9.6\% & 0.71$\times$ & S-m & 0.44$\times$ & 0.59$\times$ & 9.7\% & \hi{0.80$\times$}  \\
 & i$\to$t & H-h & 0.22$\times$ & 0.39$\times$ & 9.1\% & 0.81$\times$ & H-h & 0.28$\times$ & 0.41$\times$ & 9.9\% & 0.90$^\ddagger$$\times$ & h & 0.25$\times$ & 0.46$\times$ & 9.1\% & 0.98$^\ddagger$$\times$ & H-h & 0.19$\times$ & 0.39$\times$ & 8.9\% & \hi{0.39$\times$}  \\
 & t$\to$i & H-h & 0.31$\times$ & 0.62$\times$ & 9.7\% & 1.04$^\ddagger$$\times$ & H-h & 0.38$\times$ & 0.64$\times$ & 9.8\% & 0.96$^\ddagger$$\times$ & h & 0.34$\times$ & 0.71$\times$ & 9.9\% & 1.01$^\ddagger$$\times$ & H-h & 0.26$\times$ & 0.63$\times$ & 9.5\% & \hi{0.63$\times$}  \\
 & t$\to$t & H-m & 0.56$\times$ & 0.87$\times$ & 9.8\% & 0.95$\times$ & H-m & 0.70$\times$ & 1.00$\times$ & 9.6\% & 1.05$\times$ & h & 0.51$\times$ & 0.96$\times$ & 9.5\% & 2.33$^\ddagger$$\times$ & H-m & 0.42$\times$ & 0.83$\times$ & 9.4\% & \hi{0.83$\times$}  \\
\midrule
Qwen3-VL & i$\to$i & I-m & 0.55$\times$ & 0.90$\times$ & 9.1\% & 0.90$\times$ & H-m & 0.72$\times$ & 1.03$\times$ & 9.0\% & 1.39$\times$ & m & 0.58$\times$ & 0.70$\times$ & 9.6\% & 0.99$\times$ & I-m & 0.47$\times$ & 0.70$\times$ & 9.6\% & \hi{0.70$\times$}  \\
 & i$\to$t & H-m & 0.52$\times$ & 0.89$\times$ & 9.6\% & 0.92$\times$ & H-m & 0.81$\times$ & 1.04$\times$ & 9.7\% & 1.06$\times$ & h & 0.71$\times$ & 1.48$\times$ & 9.0\% & 3.23$^\ddagger$$\times$ & H-m & 0.42$\times$ & 0.89$\times$ & 9.6\% & \hi{0.89$\times$}  \\
 & t$\to$i & H-h & 0.27$\times$ & 0.78$\times$ & 8.7\% & 0.86$\times$ & H-h & 0.35$\times$ & 0.82$\times$ & 9.9\% & 1.00$^\ddagger$$\times$ & h & 0.41$\times$ & 0.86$\times$ & 9.7\% & 1.00$^\ddagger$$\times$ & H-h & 0.24$\times$ & 0.78$\times$ & 8.9\% & \hi{0.78$\times$}  \\
 & t$\to$t & H-m & 0.47$\times$ & 0.75$\times$ & 9.6\% & 0.90$\times$ & H-m & 0.73$\times$ & 0.95$\times$ & 9.7\% & 1.05$\times$ & h & 0.48$\times$ & 0.83$\times$ & 9.5\% & 3.25$^\ddagger$$\times$ & H-m & 0.38$\times$ & 0.70$\times$ & 9.5\% & \hi{0.70$\times$}  \\
\bottomrule
\end{tabular}%
}\caption{Plan selection with smaller catalogs and their union at $\tau = 0.7$. All costs are relative to the static plan of the nine-configuration catalog, except the last column, which is relative to the union's own static plan (bold: cheaper, grey: not cheaper). Abbreviations and markers: Table~\ref{tab:reckoner}.}
\label{tab:small-catalog}
\end{table*}

\begin{table}[t]
\footnotesize\centering
\setlength{\tabcolsep}{2.5pt}
\resizebox{\columnwidth}{!}{%
\begin{tabular}{@{}l rrrr@{}}
\toprule
 & Full & HNSW & Single & Union \\
\midrule
Indexes per collection & 9 & 3 & 1 & 9 \\
Memory (GB), SigLIP2 & 208.7 & 163.9 & 55.2 & 208.7 \\
Memory (GB), OpenCLIP & 140.4 & 110.4 & 37.4 & 140.4 \\
Memory (GB), Jina-v4 & 375.8 & 288.7 & 96.8 & 375.8 \\
Memory (GB), Qwen3-VL & 375.8 & 288.7 & 96.8 & 375.8 \\
Labeling $Q$ (CPU-h) & 344--5{,}036 & 94--767 & 84--705 & 369--5{,}086 \\
Opt., median & 0.41$\times$ & 0.63$\times$$^{(14)}$ & 0.34$\times$$^{(13)}$ & 0.23$\times$ \\
Budget, median & 0.72$\times$ & 0.81$\times$$^{(13)}$ & 0.62$\times$$^{(13)}$ & 0.53$\times$ \\
Budget / own static, med. & 0.72$\times$ & 0.78$\times$$^{(13)}$ & 0.68$\times$$^{(13)}$ & 0.70$\times$ \\
\bottomrule
\end{tabular}%
}
\caption{The catalogs of Section~\ref{sec:small-catalog}. Memory: image and text indexes together. The union adds no index. Costs: medians over the settings where the catalog can meet the requirement (their number in parentheses where fewer than 16), relative to the nine-configuration static plan, except the last row, which is relative to each catalog's own static plan.}
\label{tab:small-catalog-summary}
\end{table}

Per-query selection requires every configuration it may choose to be built and kept in memory. Our nine-configuration catalog occupies 140--376\,GB per model, most of it in the three HNSW indexes (Table~\ref{tab:small-catalog-summary}). We ask how much of the saving survives with less, using two smaller catalogs and one larger one:
\begin{itemize}
    \item \emph{HNSW-only}: the three HNSW configurations, still three separately built indexes.
    \item \emph{Single index}: the index of \texttt{hnsw\_high}, searched with three values of \texttt{efSearch}. The highest is \texttt{hnsw\_high}'s own. The two lower ones are, per model, the first values of the parameter sweep at which mean fidelity over the four directions reaches 0.7 and 0.8 (\texttt{efSearch} of 192 and 384 for SigLIP2, 256 and 1{,}024 for OpenCLIP, 48 and 96 for Jina-v4, and 64 and 128 for Qwen3-VL). All three share one index, so per-query selection adds no memory. Their costs were measured on one node together with \texttt{hnsw\_high} and put on the scale of the main benchmark through it.
    \item \emph{Union}: the nine configurations plus the single index's two lower settings, eleven configurations with no additional index. Its static plan is the best static choice among everything we built, the strongest static baseline available without new indexes.
\end{itemize}
For each catalog, the static plan, the optimum, and the synopsis are computed within that catalog alone, so that a static and a per-query deployment choose among the same configurations. The synopsis uses the same landmarks. Only the configurations it records change. Table~\ref{tab:small-catalog} reports every cost relative to the static plan of the nine-configuration catalog, so that all catalogs share one scale.

\paragraph{Against the strongest static baseline, the synopsis still saves.} The union's static plan is the nine-configuration static plan in 12 of 16 settings: the cheaper single-index settings do not help a static deployment there, because its hardest queries still need the same configuration. In the other four, the union's static plan is a cheap single-index setting, which shows that \texttt{hnsw\_low} is a poor operating point for these models. Measured against the union's own static plan, the budget rule at the requirement costs a median of 0.70$\times$, about the same as with the nine-configuration catalog. It is cheaper in 13 settings, equal in one, and slightly costlier, by about its lookup cost, in the two where the union's static plan is already its cheapest setting (grey in Table~\ref{tab:small-catalog}).

\paragraph{A smaller catalog keeps most of the opportunity.} Without the IVF-family indexes, the optimum keeps most of the full catalog's saving in a typical setting. The loss lies in Jina-v4 and Qwen3-VL, whose low and moderate IVF tiers are cheaper than \texttt{hnsw\_mod}, and in SigLIP2 and OpenCLIP i$\to$t, where only IVF-family configurations meet the target often enough to stay within the requirement. The single index differs from the full catalog in both directions: its lower settings are cheaper and more accurate than \texttt{hnsw\_low} for SigLIP2, OpenCLIP, and Jina-v4, but its single grid of \texttt{efSearch} values per model leaves the lower settings short of the target in some directions, where the only remaining choice is \texttt{hnsw\_high}'s expensive setting, and in three settings no assignment meets the requirement.

\paragraph{The synopsis captures it within each catalog.} Within each smaller catalog, the budget rule at the requirement is cheaper than that catalog's own static plan in most settings where the requirement can be met, by about a fifth with three HNSW indexes and about a third with a single index (Table~\ref{tab:small-catalog-summary}). With a single index, this saving needs no additional memory, and labeling the construction sample costs far less, since only three configurations are run. Where the single index's static plan is already its cheapest setting (SigLIP2 and OpenCLIP i$\to$i), and for OpenCLIP t$\to$i, the synopsis saves nothing and costs 3--8\% more, mostly its lookup. Where the static plan is the middle setting for Jina-v4 and Qwen3-VL i$\to$i, it still saves 19\% and 32\%.

\paragraph{Summary.} The saving does not depend on a large catalog or a weak static baseline. Against the best static choice among all indexes we built, the synopsis still costs a median of 0.70$\times$. A single HNSW index searched at three settings, a quarter of the full catalog's memory, lets the synopsis save within that index alone. The single index is not a uniform substitute for the full catalog, however: with one grid of \texttt{efSearch} values per model it cannot meet the requirement in three settings, and per setting it is much cheaper or much costlier than the full catalog.

\paragraph{Labeling cost and payback.} The smallest catalog keeps one HNSW index searched at several search settings, which vector databases already support. At no extra memory, the synopsis costs a median of 0.68$\times$ the index's own static plan in the 13 settings where the index can meet the requirement, and labeling the sample costs 84--705 CPU-hours rather than 344--5{,}036. The construction sample needs about 70 queries per cell (Section~\ref{sec:rq5}). At a fixed $K$ this does not grow with the corpus, although a larger corpus may need more cells, so for a corpus of hundreds of millions of items it is likely a small fraction of the collection. Labeling is a one-time cost. From the per-configuration costs of Table~\ref{tab:appendix-catalog} and the savings of Table~\ref{tab:rq4-selection}, we estimate that labeling the full catalog is repaid after a median of about 40 million queries per setting, from about 6 million for cross-modal settings with an expensive static plan to more than 100 million in five settings (SigLIP2 i$\to$i and t$\to$t, OpenCLIP i$\to$i, Qwen3-VL i$\to$i and i$\to$t, at most 394 million). For SigLIP2 and OpenCLIP, labeling is dominated by IVF-family configurations that are rarely chosen. The single index, in the 10 settings where it saves, is repaid after a median of about 2 million queries. This estimate counts only the CPU time of running the configurations on the sample. Exact search on the sample adds about an hour of GPU time per model.

\paragraph{Answer to RQ3.} Yes. Against the best static choice among all our configurations, selection costs a median of 0.59$\times$ with the MLP and 0.70$\times$ with the synopsis. Against the static plan of the nine configurations, the synopsis is cheaper in all 16 settings, at a median of 0.72$\times$, and in 12 of 14 at the static plan's lower miss rate. The MLP saves more, but exceeds the requirement slightly in one setting against the static plan of the nine (at $\tau = 0.7$ and $0.8$), and in none against the best static plan. The simpler rules show why the budget rule is needed: the cell mean ignores how fidelity is spread within a cell, and per-cell coverage overpays in hard cells. The saving does not depend on a large catalog or a weak static baseline. Which estimator to use depends on the deployment. Where cost matters most, the MLP is the better choice. It is cheaper than the synopsis in 15 of 16 settings, and with $\lambda$ tuned to a slightly stricter limit (9.5\%), it stays within the requirement in every setting at $\tau = 0.7$, at a median of 0.62$\times$. Where the requirement must hold without such a margin, where operators need to see which regions are hard, or where configurations are added often, the synopsis is the safer choice. Its coverage is calibrated by construction, and a new configuration adds one column.

\subsection{RQ4: How should the catalog be chosen?}
\label{sec:rq4}
\label{sec:advisor-eval}

We evaluate the advisor on the configurations we built, using the stored labels and costs only. We do not derive candidates (step~1 in Section~\ref{sec:advisor}). We use the pool of Section~\ref{sec:setup} instead: the nine configurations of the full catalog and the two cheaper search settings of the \texttt{hnsw\_high} index. For every setting, we evaluate all 2{,}047 sub-catalogs of the pool with the protocol of Section~\ref{sec:setup} ($\tau = 0.7$, $\rho = 0.9$, $K = 10{,}000$), with the synopsis and with the MLP. A sub-catalog is ranked by its cross-fitted cost on $Q$ and reported on $Q_{\mathrm{test}}$, relative to the static plan of the nine and with the lookup included. Its memory is that of the distinct indexes it uses in the searched collection. For the full catalog and the pool, this evaluation reproduces the results of Section~\ref{sec:rq3} exactly.

\paragraph{What we find.} A catalog of three configurations captures the full catalog's saving in every setting (Table~\ref{tab:advisor-size}). Chosen from the nine configurations alone, the best three cost a median of 0.71$\times$ the static plan, against 0.72$\times$ for all nine, and cost at most 0.4\% more than all nine in any setting. More configurations add little, because the budget rule sends most queries to two or three of them. Chosen from the whole pool, the best three cost a median of 0.53$\times$ at a median of 56\,GB per collection, against 70--188\,GB for all nine. Their held-out miss rate stays within 10\% in 15 settings and is 10.05\% in the other (Jina-v4 i$\to$t). The difference comes from the cheaper search settings of \texttt{hnsw\_high}. For SigLIP2 i$\to$i, one of them alone meets the requirement at about a fifth of the cost of \texttt{hnsw\_mod}, the static plan of the nine. In four settings, such a search setting is also a cheaper static plan (0.13--0.74$\times$ that of the nine). Relative to the pool's own static plan, the best three cost a median of 0.69$\times$ and are cheaper in 14 of 16 settings.

\begin{table}[t]
\footnotesize\centering
\setlength{\tabcolsep}{3pt}
\resizebox{\columnwidth}{!}{%
\begin{tabular}{@{}l ccc r@{}}
\toprule
 & \multicolumn{3}{c}{Median cost} & Memory \\
\cmidrule(lr){2-4}
Catalog & Nine, synopsis & Pool, synopsis & Pool, MLP & (GB) \\
\midrule
Full catalog (nine) & 0.72$\times$ & 0.72$\times$ & 0.62$\times$ & 70--188 \\
Best of size 1 & 1.02$\times$ & 1.01$\times$ & 1.00$\times$ & 28 \\
Best of size 2 & 0.75$\times$ & 0.57$\times$ & 0.49$\times$ & 54 \\
Best of size 3 & \hi{0.71$\times$} & \hi{0.53$\times$} & 0.42$\times$ & 56 \\
Best of size 4 & 0.71$\times$ & 0.53$\times$ & 0.41$\times$ & 73 \\
\bottomrule
\end{tabular}%
}
\caption{The best sub-catalog of each size, chosen from the nine configurations or from the whole pool of eleven. \keys{(i)~Cost: on the held-out queries, relative to the static plan of the nine, lookup included, median over the 16 settings. Size 1: median over 15 settings, since no single configuration meets the requirement for SigLIP2 i$\to$t. (ii)~Memory: per collection, median over the settings, for the synopsis chosen from the pool. The MLP's best catalogs of three and four need 79 and 93\,GB. (iii)~A sub-catalog of one configuration needs no lookup, but its cost here includes one.}}
\label{tab:advisor-size}
\end{table}

\begin{table}[t]
\footnotesize\centering
\setlength{\tabcolsep}{3pt}
\resizebox{\columnwidth}{!}{%
\begin{tabular}{@{}ll l r l r r@{}}
\toprule
Model & Dir. & Best 2 & Cost & Best 3 & Cost & Full \\
\midrule
SigLIP2 & i$\to$i & Sl+Hl & 0.22 & Sl+Hl+Sm & 0.22 & 0.69 \\
 & i$\to$t & Hh+Oh & 0.73$^\ast$ & Hm+Hh+Oh & 0.54$^\ast$ & 0.54 \\
 & t$\to$i & Hm+Hh & 0.14$^\ast$ & Hm+Ol+Hh & 0.14$^\ast$ & 0.14 \\
 & t$\to$t & Sl+Hm & 0.39$^\ast$ & Sl+Sm+Hm & 0.36$^\ast$ & 0.88 \\
\midrule
OpenCLIP & i$\to$i & Sl+Hl & 0.34 & Sl+Hl+Hm & 0.34$^\ast$ & 0.79 \\
 & i$\to$t & Hm+Oh & 0.53$^\ast$ & Hm+Sm+Oh & 0.52$^\ast$ & 0.53 \\
 & t$\to$i & Hm+Sm & 0.12 & Sl+Hm+Sm & 0.12 & 0.13 \\
 & t$\to$t & Hm+Il & 0.42 & Sl+Hm+Il & 0.32 & 0.42 \\
\midrule
Jina-v4 & i$\to$i & Sl+Sm & 0.60 & Sl+Sm+Ol & 0.59 & 0.78 \\
 & i$\to$t & Hm+Hh & 0.41$^\ast$ & Sm+Hm+Ih & 0.38 & 0.39 \\
 & t$\to$i & Hm+Hh & 0.63$^\ast$ & Om+Hm+Hh & 0.62$^\ast$ & 0.62 \\
 & t$\to$t & Sm+Hm & 0.86$^\ast$ & Sm+Im+Hm & 0.82$^\ast$ & 0.87 \\
\midrule
Qwen3-VL & i$\to$i & Sl+Sm & 0.70 & Sl+Om+Sm & 0.69 & 0.90 \\
 & i$\to$t & Im+Hm & 0.89$^\ast$ & Om+Hm+Hh & 0.88$^\ast$ & 0.89 \\
 & t$\to$i & Im+Hh & 0.81$^\ast$ & Im+Hm+Hh & 0.77$^\ast$ & 0.78 \\
 & t$\to$t & Sl+Hm & 0.73$^\ast$ & Sl+Im+Hm & 0.70$^\ast$ & 0.75 \\
\bottomrule
\end{tabular}%
}
\caption{The best sub-catalogs of two and three configurations, chosen from the pool with the synopsis. \keys{(i)~H, I, O: HNSW, IVFPQ, O-IVFPQ. l, m, h: low, moderate, high tier. Sl, Sm: \texttt{hnsw\_high} at its low and moderate search setting. (ii)~Cost: relative to the static plan of the nine, held-out queries, lookup included. (iii)~$^\ast$: includes the static plan of the nine. (iv)~For SigLIP2 and OpenCLIP i$\to$i, every query runs on Sl. The other configurations listed there serve no query, and any other choice alongside Sl ties (with Sm and Hh instead, the best three need only the memory of \texttt{hnsw\_high}).}}
\label{tab:advisor-pattern}
\end{table}

\begin{figure}[t]
\centering
\includegraphics[width=\linewidth]{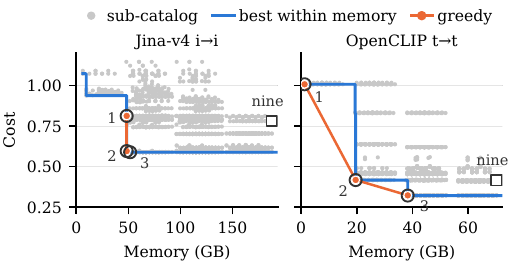}
\caption{Cost against memory for every sub-catalog of the pool, with the synopsis. \keys{(i)~Cost: on the held-out queries, relative to the static plan of the nine (dashed), lookup included. (ii)~Grey: sub-catalogs that meet the requirement on $Q$ and cost at most 1.2$\times$. Not shown: the 15 (Jina-v4) and 31 (OpenCLIP) that do not meet it, and the 32 and 240 that cost more. (iii)~Blue: the cheapest sub-catalog within each memory budget. (iv)~Orange: the greedy path. (v)~Circles 1--3: the best sub-catalog of that size. Screening on 10\% of $Q$ picks it, or one within 0.01\% of its cost. (vi)~Square: the full catalog of nine.}}
\label{fig:advisor}
\end{figure}

\paragraph{A few configurations keep the saving.} Figure~\ref{fig:advisor} shows every sub-catalog in two settings, by its memory and its cost. We chose Jina-v4 i$\to$i, where a cheap IVF-family tier joins the best catalog of three, and OpenCLIP t$\to$t, where an accurate one does. In Jina-v4 i$\to$i, two search settings of the \texttt{hnsw\_high} index cost 0.60$\times$ the static plan, with no memory beyond that one index. A low tier of O-IVFPQ lowers this only to 0.59$\times$. The full catalog of nine costs 0.78$\times$ at 188\,GB, because it lacks the cheaper settings. In OpenCLIP t$\to$t, \texttt{hnsw\_mod} with a low tier of IVFPQ already costs 0.42$\times$, as much as the full catalog at 70\,GB. With the low search setting of \texttt{hnsw\_high} added, three configurations cost 0.32$\times$ at 38\,GB. In both settings, cost falls in a few large steps and then stays flat as memory grows. The greedy path stays close to the cheapest sub-catalog at every step, and screening on 10\% of $Q$ finds the best catalog of each size.

\paragraph{The best small catalogs share a pattern.} Every best catalog of three configurations contains a search setting of the \texttt{hnsw\_high} index, 13 of 16 contain \texttt{hnsw\_mod} (in one of them, OpenCLIP i$\to$i, it serves no query), and 12 of 16 add one IVF-family configuration (Table~\ref{tab:advisor-pattern}). Eight of these 12 settings are those of Jina-v4 and Qwen3-VL. In seven of the eight, the configuration is a low or moderate tier, cheaper there than \texttt{hnsw\_mod}. For SigLIP2 and OpenCLIP, it always costs more than \texttt{hnsw\_mod}. In their i$\to$t settings, it is \texttt{oivfpq\_high}, the static plan, which serves 27\% and 48\% of queries. Only five of the 16 catalogs use two search settings of the \texttt{hnsw\_high} index. A small catalog is therefore usually \texttt{hnsw\_mod}, one or two search settings of the \texttt{hnsw\_high} index, and one IVF-family configuration, which is cheap for Jina-v4 and Qwen3-VL and accurate for SigLIP2 and OpenCLIP i$\to$t. The MLP chooses catalogs of the same kind.

\paragraph{The advisor finds them cheaply.} Many sub-catalogs cost almost the same. In each setting, between 60 and 1{,}024 of them lie within 1\% of the cheapest, so we report the best catalog of each size rather than one overall winner. Greedy choice ends at one to four configurations and comes within 0.008$\times$ of the best sub-catalog with no more memory, with no gap in 9 of 16 settings. Holding memory fixed hides one failure. For Qwen3-VL i$\to$i, no single addition lowers cost by 1\%, so greedy stops at the static plan alone (1.08$\times$ with the lookup), while the best two configurations cost 0.70$\times$. Choosing from a random 10\% of $Q$, with a coarser synopsis of 1{,}250 cells, yields the full-sample choice or one within 1\% of it in 14, 13, and 11 of 16 settings for two, three, and four configurations. A wrong choice costs at most 0.046$\times$. The synopsis and the MLP choose the same best catalog in 11, 8, and 3 of 16 settings for two, three, and four configurations. Their catalogs within 1\% of the cheapest overlap in 13, 13, and 11. In a given setting, the MLP's best catalog of two to four configurations costs a median of 0.05--0.07$\times$ less than the synopsis's, and between 0.01$\times$ more and 0.24$\times$ less. Its best catalogs of two to four configurations stay within 10\% misses on the held-out queries in every setting, but with all nine configurations it misses 10.9\% for SigLIP2 i$\to$i.

\paragraph{What it saves.} Labeling the best three configurations takes a median of 30 CPU-hours per setting, against 83--1{,}409 for all nine, and is repaid after a median of 2.4 million queries. Screening all eleven candidates on 10\% of $Q$ first adds a median of 46 CPU-hours. With it, the total is about 78 CPU-hours, repaid after about 6 million queries at the median. The advisor does not avoid building the candidates once. It saves the labeling of configurations that are never worth their memory, and the memory of the indexes it drops. Search settings of one index are the cheapest candidates, since they need no build and no memory. Memory is counted per searched collection. A deployment that serves both directions of a collection needs the union of their catalogs.

\paragraph{Answer to RQ4.} A few configurations suffice. Three keep the full catalog's saving, at a median of 56\,GB per collection, and the cheaper search settings of one index provide much of it at no extra memory. With eleven candidates, every sub-catalog can be evaluated from the stored labels. Greedy choice comes close to the best, and screening on a tenth of the sample usually finds the same catalog. The advisor still builds every candidate once, but it labels only the chosen ones at full scale.

\subsection{RQ5: What do the estimators cost?}
\label{sec:rq5}

\begin{table}[t]
\footnotesize\centering
\setlength{\tabcolsep}{2pt}
\resizebox{\columnwidth}{!}{%
\begin{tabular}{@{}l rrrr@{}}
\toprule
 & SigLIP2 & OpenCLIP & Jina-v4 & Qwen3-VL \\
\midrule
Dimension $d$ & 1152 & 768 & 2048 & 2048 \\
$c_{\mathrm{est}}$, 1 query, mean ($\mu$s) & 1640 & 822 & \ex{2859} & \ex{2842}  \\
$c_{\mathrm{est}}$, 1 query, p99 ($\mu$s) & 1659 & 848 & 2874 & 2873 \\
Cheapest plan, 1 query ($\mu$s) & 1678 & 1482 & 445 & 349 \\
$c_{\mathrm{est}}$, batched ($\mu$s/q) & \hi{23.3} & \hi{20.6} & \hi{31.4} & \hi{32.3}  \\
Cheapest plan, batched ($\mu$s/q) & 665 & 490 & 223 & 163 \\
MLP lookup, batched ($\mu$s/q) & 3.2 & 3.2 & 3.0 & 4.5 \\
MLP lookup, 1 query ($\mu$s) & 82 & 88 & 119 & 117 \\
Synopsis size (MiB) & 119.5 & 90.2 & 187.8 & 187.8 \\
MLP size (MiB) & 13.4 & 10.4 & 20.4 & 20.4 \\
$k$-means (GPU-h) & 0.12 & 0.08 & 0.21 & 0.20 \\
Exact $k$-NN on $Q$ (GPU-h) & 0.57 & 0.40 & 0.98 & 0.98 \\
All configs on $Q$ (h, 64 thr.) & 78.7 & 51.2 & 5.4 & 8.8 \\
All configs on $Q$ (CPU-h) & 5{,}036 & 3{,}275 & 344 & 563 \\
\bottomrule
\end{tabular}%
}
\caption{Cost of the two estimators (synopsis at $K = 10{,}000$). Online rows: averaged over directions. Offline rows: per model, over its four directions. Rows: the lookup, one query at a time and batched, against the cheapest configuration (\texttt{hnsw\_low}), the estimators' size, and the one-time offline cost. Bold: the batched regime used throughout. Italic: single-query lookup costlier than the cheapest configuration. MLP lookup: batches of 4{,}096 on 64 threads, or one query on one thread. MLP size: the four networks of a model.}
\label{tab:rq5-overhead}
\end{table}

\begin{table}[t]
\small\centering
\setlength{\tabcolsep}{4pt}
\begin{tabular}{@{}r rrr@{}}
\toprule
$K$ & Skill & Thin & Size (MiB) \\
\midrule
1 & 0.000 & -- & $<$0.1 \\
500 & 0.280 & 0.01\% & 4.5--9.4 \\
1{,}250 & 0.335 & 0.02\% & 11.3--23.5 \\
2{,}500 & 0.374 & 0.04\% & 22.5--47.0 \\
5{,}000 & 0.416 & 0.14\% & 45.1--93.9 \\
 \hi{10{,}000} & \hi{0.451} & \hi{0.51\%} & \hi{90.2--187.8} \\
20{,}000 & 0.469 & 2.17\% & 180.4--375.7 \\
25{,}000 & 0.468 & 3.61\% & 225.4--469.6 \\
40{,}000 & 0.440 & 10.46\% & 360.7--751.3 \\
\bottomrule
\end{tabular}
\caption{Sensitivity to the number of landmarks $K$ over the 16 settings ($K = 1$ is a single cell, the global average). Skill: mean over settings of the mean over all nine configurations. Bold: the default. Thin: held-out queries in cells with fewer than $q_{\min}$ sample queries. Terms: Table~\ref{tab:reckoner}.}
\label{tab:rq5-K}
\end{table}

\paragraph{Lookup cost.} In the batched regime used throughout, the synopsis's lookup takes 21--32\,$\mu$s per query (Table~\ref{tab:rq5-overhead}), at most a fifth of the cheapest configuration's cost. The MLP's forward pass takes 3.0--4.5\,$\mu$s per query in batches of 4{,}096 on 64 threads, and 82--119\,$\mu$s for one query on one thread. Lookups are included in every cost of Section~\ref{sec:rq3}. For the synopsis, one query at a time, it is dominated by the exact scan of the $K$ landmarks: it costs less than \texttt{hnsw\_low} for OpenCLIP and SigLIP2, but several times as much for the 2{,}048-dimensional Jina-v4 and Qwen3-VL (italic). An HNSW index over the landmarks makes the single-query lookup 3.1$\times$ (SigLIP2 t$\to$i) to 11.0$\times$ (Jina-v4 i$\to$i) faster. The budget rule then chooses the same configuration for at least 98.5\% of queries, and its relative cost and miss rate change by at most 0.003 and 0.05 percentage points. Only for Qwen3-VL does the lookup still cost more than \texttt{hnsw\_low} (1.1--1.5$\times$). Table~\ref{tab:landmark-hnsw} gives the timings in every setting.

\begin{table*}[t]
\footnotesize\centering
\setlength{\tabcolsep}{3pt}
\resizebox{\textwidth}{!}{%
\begin{tabular}{@{}ll r rr r r rrrr@{}}
\toprule
 & & Exact scan & \multicolumn{2}{c}{HNSW over landmarks ($\mu$s)} & & Same & \multicolumn{4}{c}{Same configuration chosen} \\
\cmidrule(lr){4-5}\cmidrule(l){8-11}
Model (HNSW parameters) & Dir. & ($\mu$s) & mean & p99 & Speed-up & cell & Mean & Coverage & Budget (static) & Budget ($\le 1{-}\rho$) \\
\midrule
SigLIP2 ($M{=}16$, ef${=}256$) & i$\to$i & 1{,}591 & \hi{483} & 605 & 3.3$\times$ & 99.96\% & 100.00\% & 99.99\% & 99.99\% & \hi{99.99\%} \\
 & i$\to$t & 1{,}579 & \hi{437} & 494 & 3.6$\times$ & 97.49\% & 98.84\% & 98.61\% & 98.54\% & \hi{98.54\%} \\
 & t$\to$i & 1{,}587 & \hi{508} & 598 & 3.1$\times$ & 99.35\% & 99.80\% & 99.66\% & 99.76\% & \hi{99.97\%} \\
 & t$\to$t & 1{,}584 & \hi{493} & 596 & 3.2$\times$ & 99.88\% & 99.98\% & 99.95\% & 99.98\% & \hi{99.98\%} \\
\midrule
OpenCLIP ($M{=}16$, ef${=}128$) & i$\to$i & 947 & \hi{229} & 296 & 4.1$\times$ & 99.86\% & 99.95\% & 99.92\% & 99.97\% & \hi{99.96\%} \\
 & i$\to$t & 963 & \hi{238} & 282 & 4.0$\times$ & 98.92\% & 99.31\% & 99.53\% & 99.58\% & \hi{99.40\%} \\
 & t$\to$i & 959 & \hi{242} & 291 & 4.0$\times$ & 98.11\% & 99.49\% & 99.03\% & 99.03\% & \hi{99.85\%} \\
 & t$\to$t & 951 & \hi{236} & 284 & 4.0$\times$ & 99.89\% & 99.95\% & 99.94\% & 99.95\% & \hi{99.97\%} \\
\midrule
Jina-v4 ($M{=}32$, ef${=}64$) & i$\to$i & 2{,}814 & \hi{255} & 370 & 11.0$\times$ & 99.94\% & 99.96\% & 99.95\% & 99.96\% & \hi{99.96\%} \\
 & i$\to$t & 2{,}813 & \hi{290} & 403 & 9.7$\times$ & 99.85\% & 99.88\% & 99.91\% & 99.94\% & \hi{99.91\%} \\
 & t$\to$i & 2{,}810 & \hi{264} & 377 & 10.6$\times$ & 99.52\% & 99.67\% & 99.80\% & 99.86\% & \hi{99.76\%} \\
 & t$\to$t & 2{,}814 & \hi{292} & 433 & 9.6$\times$ & 99.68\% & 99.75\% & 99.79\% & 99.79\% & \hi{99.80\%} \\
\midrule
Qwen3-VL ($M{=}32$, ef${=}128$) & i$\to$i & 2{,}832 & \hi{500} & 668 & 5.7$\times$ & 99.98\% & 99.99\% & 99.99\% & 99.99\% & \hi{99.99\%} \\
 & i$\to$t & 2{,}871 & \hi{396} & 527 & 7.3$\times$ & 99.79\% & 99.85\% & 99.86\% & 99.87\% & \hi{99.88\%} \\
 & t$\to$i & 2{,}834 & \hi{533} & 685 & 5.3$\times$ & 99.88\% & 99.92\% & 99.97\% & 99.98\% & \hi{99.96\%} \\
 & t$\to$t & 2{,}855 & \hi{465} & 695 & 6.1$\times$ & 99.89\% & 99.92\% & 99.93\% & 99.94\% & \hi{99.93\%} \\
\bottomrule
\end{tabular}%
}
\caption{Single-query lookup (Step~1, one thread) with an HNSW index over the $K = 10{,}000$ landmarks of each modality, one setting per model for all four directions: the fastest whose choice under the coverage rule agrees with the exact scan's on at least 99.9\% of a 5{,}000-query held-out sample in at least two directions. Exact scan and HNSW are timed alternately, query by query. Speed-up: the exact scan divided by the HNSW mean. Same cell: held-out queries located in the same cell as by the exact scan. Same configuration chosen: held-out queries for which each selection rule picks the same configuration with either lookup, at $\tau = 0.7$. Bold: the HNSW lookup time and the agreement for the budget rule at the requirement, our method.}
\label{tab:landmark-hnsw}
\end{table*}

\paragraph{Size and construction.} The synopsis occupies 90--188\,MiB, mostly the landmarks themselves, well under 1\% of the HNSW indexes of a single tier. Building it requires a few minutes of GPU time for clustering and an hour or less for exact search on $Q$, shared by all configurations. Running every configuration on $Q$ dominates the offline cost, most of all for SigLIP2 and OpenCLIP, whose IVF-family configurations probe thousands of partitions. With a single index (Section~\ref{sec:small-catalog}) or the advisor's small catalogs (Section~\ref{sec:rq4}), it is several times cheaper. This cost is paid once per model and catalog. The MLP occupies 10.4--20.4\,MiB for the four networks of a model. Its forward pass takes 3.0--4.5\,$\mu$s per query in batches of 4{,}096 on 64 threads, and 82--119\,$\mu$s for one query on one thread. Training one network takes 1.7--3.8\,s on one NVIDIA A100 (median 2.7\,s, excluding the GPU's start-up). The six networks of a setting, five for tuning $\lambda$ and the deployed one, take a median of 17\,s in total.

\paragraph{Number of landmarks.} Skill rises with $K$ up to about 20{,}000 landmarks and then falls, as more held-out queries land in cells with fewer than $q_{\min}$ sample queries (Table~\ref{tab:rq5-K}). Lookup cost and synopsis size grow linearly with $K$. Our default of 10{,}000 is close to the best skill at half the cost of 20{,}000.

\paragraph{Construction sample.} The synopsis needs a large construction sample at this $K$: built from 10\% of $Q$, most held-out queries fall into sparse cells and skill drops by more than half, and with 1\% the synopsis is barely better than the global average. Roughly 70 sample queries per cell are needed. The minimum support matters less: $q_{\min} = 1$ leaves skill unchanged, while $q_{\min} = 50$ lowers it.

\paragraph{Answer to RQ5.} Both estimators are small and their batched lookups are cheap, the MLP's the cheaper. The synopsis's batched lookup costs at most a fifth of the cheapest configuration. Their one-time cost is dominated by running the catalog on the construction sample, which needs several hundred thousand queries per direction at $K = 10{,}000$. Accuracy grows with $K$ as long as cells keep enough sample queries. A single-query lookup by exact landmark scan is too slow for 2{,}048-dimensional models. An HNSW index over the landmarks makes it 3--11$\times$ faster.

\section{Related Work}
\label{sec:related}

\paragraph{Approximate nearest-neighbor indexes and their tuning.} Graph-based indexes such as HNSW~\cite{hnsw} and DiskANN~\cite{diskann}, and partition- and quantization-based ones such as IVF with product quantization~\cite{pq} and its optimized variant~\cite{opq}, SPANN~\cite{spann}, and ScaNN~\cite{scann}, cover a wide range of trade-offs between cost and fidelity. Benchmarks and surveys compare them by their average recall over a workload~\cite{annbench,graphann-survey,vectordb}, and tuners such as VDTuner~\cite{vdtuner} search index types and parameters for a workload-level target of throughput and recall. All of these choose one configuration for a workload. They produce a better static plan. Plan selection takes a catalog as given and chooses among its configurations per query. The two are complementary: a better-tuned catalog gives plan selection better options, as the pool of Section~\ref{sec:setup:catalog} shows.

\paragraph{Adaptive search within an index.} Learned adaptive early termination~\cite{laet} predicts, per query, when a search has found its neighbors, and DARTH~\cite{darth} terminates a search once a declared target recall is predicted to be met. Ada-ef~\cite{adaef} sets HNSW's \texttt{efSearch} per query from offline statistics of the distance distribution, and the adaptive partition scanning of Quake~\cite{quake} sets the number of partitions an IVF-style index scans per query from a running estimate of recall. All of these adapt how far a search goes inside an index that has already been chosen, and all need that index to be able to reach the target. Closest to our setting, MINT~\cite{mint} plans multi-vector queries across several indexes, choosing which indexes to use and how many items to take from each under a recall threshold. Its recall estimates come from models fitted on a sample of the data rather than from the query's own surroundings, and it does not hold a workload to a miss rate. Plan selection makes the earlier decision, which index and which settings to use, before any search runs, and it can move a query to a different index family when the default one cannot reach the target. The approaches compose: an early-termination method can run inside whichever configuration is selected.

\paragraph{Query hardness.} Local intrinsic dimensionality~\cite{lid} and Steiner-hardness~\cite{steiner} characterize how hard a query is for nearest-neighbor search, the latter for graph indexes specifically, and work on cross-modal search observes that queries of one modality lie outside the distribution of the other~\cite{roargraph,modality-gap}. These measures explain why some queries are hard, but they are not calibrated estimates of whether a particular configuration will meet a particular target, which is what plan selection needs. The synopsis measures that quantity directly, per region and per configuration. In our experiments, local density alone predicts fidelity much less well (Section~\ref{sec:rq2}).

\paragraph{Cardinality estimation and synopses.} Cost-based optimizers choose among access paths using estimates of each plan's output size~\cite{selinger}, obtained from synopses such as histograms built offline and consulted at plan time~\cite{histograms}, and more recently from learned models~\cite{kipf-mscn}. The quality of these estimates largely determines the quality of the chosen plans~\cite{leis-qo}. The fidelity synopsis applies this design to vector search, with cells of the embedding space as buckets and fidelity as the statistic. Unlike selectivity, which is a property of the data and the predicate, fidelity also depends on the physical plan, so the synopsis is keyed by configuration as well as by region.

\paragraph{Index advisors.} Physical-design tools such as AutoAdmin~\cite{autoadmin} choose indexes for a SQL workload by what-if analysis under a storage bound, for plans that all return exact answers. Our advisor (Section~\ref{sec:advisor}) adapts this design to plan selection. Its what-if costs are measured rather than modeled, it optimizes under a miss rate, and the value of a set of configurations depends on how they complement each other.

\paragraph{Accuracy-aware query processing.} A broad line of work trades accuracy for cost under an explicit target: approximate query processing with error bounds~\cite{blinkdb}, including over multimodal data~\cite{thalamusdb}. Others use cheap filters and model cascades for machine-learning inference~\cite{noscope,probpred}, cascades and routing among language models~\cite{frugalgpt}, and optimizers for semantic operators that choose models and plans for cost and quality~\cite{lotus,palimpzest,docetl}. Our work brings this view to the vector-search operator itself, with a per-query choice among physical configurations and a workload-level requirement on how often the fidelity target may be missed.

\paragraph{Vector search in database systems.} Vector databases and extensions such as Milvus~\cite{milvus}, AnalyticDB-V~\cite{analyticdbv}, and pgvector~\cite{pgvector} expose index types and search parameters, typically as settings for a collection or session~\cite{vectordb}. Work on hybrid queries, such as VBase~\cite{vbase} and ACORN~\cite{acorn}, chooses how to combine vector search with relational predicates, based on the predicates' selectivity. To our knowledge, no system estimates, before execution, how each candidate configuration will perform on a single query from where that query lies, and chooses among index families under a requirement on the whole workload.

\section{Limitations and Conclusion}
\label{sec:conclusion}

\paragraph{Limitations.} Our costs are estimated from per-configuration latencies measured with full batches, not measured end to end for a deployed selector that splits each batch across indexes. Half of the variation in fidelity lies within cells, so the synopsis captures only part of the per-query optimum's saving, and less when held to the static plan's realized miss rate than to the requirement. The comparison at the requirement lets the synopsis spend an allowance the static plan leaves unused. We report the stricter comparison as well. The MLP captures more of it, but its probabilities are not calibrated by construction, and it exceeded the requirement slightly in one setting against the static plan of the nine, at two targets. Both estimators require labeling a large construction sample with every candidate configuration, a one-time cost that is substantial for a large catalog and much smaller for a single index. The advisor confines the full labeling to a small catalog, but every candidate must still be built and labeled on a screening sample. With an exact scan of the landmarks, the single-query lookup is too slow for high-dimensional models. An HNSW index over the landmarks removes most of this cost. Finally, the size of the opportunity depends on the catalog: part of it, for some models, comes from how the catalog's HNSW tiers were built, although the synopsis's advantage persists against the best static choice among all the indexes we built.

\paragraph{Scope of the evaluation.} Our results come from one corpus, LAION image--caption pairs, with plan selection at one result depth, $k = 100$ (estimation accuracy also at $k = 10$). They cover three index families in one library, FAISS, with all indexes in memory on CPUs. The four models and four directions give 16 settings that differ widely in dimension, modality gap, and cost order, and the conclusions hold across all of them. Other corpora, smaller result depths, disk-resident indexes, and GPU search may change the size of the opportunity. They do not change how the synopsis is built or used. The construction and held-out queries are images and captions drawn from the same distribution as the corpus. Real workloads, such as short typed text queries, may differ. Cross-modal search is a partial test of such a shift, since its queries lie away from the items they search. The estimates are less accurate there, and for queries far from all landmarks (Section~\ref{sec:rq2}). A deployment should therefore track the share of queries that fall far from their nearest landmark, and extend the construction sample when that share grows. Our costs also assume that every index receives full batches, which per-query selection may not provide under light load.

\paragraph{Future work.} Three directions follow directly. Reducing the landmarks' dimension before indexing them would make single-query lookups cheaper still for high-dimensional models. Search-parameter grids chosen per query direction, rather than per model, would remove the coarse steps that limited the single-index catalog. And reducing the labeling cost, by sampling cells adaptively or by transferring a synopsis between catalogs that share an index, would make the synopsis cheaper to build and to keep current as the corpus changes.

\paragraph{Conclusion.} A static ANN deployment is sized for its hardest queries and overpays for the rest. We showed that a query's position predicts how each configuration will perform on it, and that this prediction must be kept per configuration. An estimator trained once on a labeled sample lets an optimizer choose configurations per query, under the requirement a static deployment is built to meet. Against the best static choice among the configurations we built, its estimated cost is a median of 0.59$\times$ with an MLP and 0.70$\times$ with a fidelity synopsis. An advisor modeled on relational index advisors keeps this saving with three configurations. Estimating fidelity before execution plays the role for vector search that cardinality estimation plays for relational plans. It is done once, offline, and consulted before any plan runs.

\clearpage
\bibliographystyle{plain}
\bibliography{mmir}

\end{document}